\documentclass[useAMS,usenatbib]{mn2e}
\usepackage{graphicx}

\newcommand{\tel}{$T_\mathrm{e}$} 		                    
\newcommand{\nel}{$N_\mathrm{e}$} 	                         
\newcommand{\chisq}{$\chi^{2}$}	                     
\newcommand{\etal}{\rm{et al.}}

\newcommand{\apj}{ApJ}

\newcommand{\apjss}{ApJS}
\newcommand{\ana}{A\&A}
\newcommand{\anass}{A\&AS}
\newcommand{\piaus}{Proc. IAU Symp.}

\newcommand{\pasp}{PASP}

\newcommand{\mn}{MNRAS}

\title[N~{\sc ii}~\& O~{\sc ii}~ORL Diagnostics]{Plasma Diagnostics for
Planetary Nebulae and H~{\sc ii}~Regions Using the N~{\sc ii}~and O~{\sc
ii}~Optical Recombination Lines}
\author[McNabb, Fang, Liu, Bastin \&~Storey]
 {I. A. McNabb$^{1}$\thanks{E-mail: imcnabb@pku.edu.cn (IAM)}, 
  X. Fang$^{2}$,
  X.-W. Liu$^{1,2}$,
  R.~J. Bastin$^{3}$\thanks{Current Address: Surbiton High School, Surbiton Crescent, Kingston upon Thames, Surrey, KT1 2JT, UK} \&  
  P. J. Storey$^{3}$\\
$^{1}$Kavli Institute for Astronomy and Astrophysics, Peking University, Beijing 100871, China\\
$^{2}$Department of Astronomy, School of Physics, Peking University, Beijing 100871, P.~R.~China\\
$^{3}$Department of Physics and Astronomy, University College London, London WC1E 6BT, UK}

\begin{document}

\date{Accepted 2012 October 24.  Received 2012 October 24; in original form 2012 August 10}

\maketitle

\label{firstpage}

\begin{abstract}
We carry out plasma diagnostic analyses for 123 planetary nebulae (PNe) and 
42 H~{\sc ii}~regions using the N~{\sc ii}~and O~{\sc ii}~optical recombination 
lines (ORLs).  New effective recombination coefficients for the N~{\sc ii}~and 
O~{\sc ii}~optical recombination spectra are used.  These data were calculated 
under the intermediate coupling scheme for a number of electron temperature 
($T_\mathrm{e}$) and density ($N_\mathrm{e}$) cases.  We used a new method to determine the $T_\mathrm{e}$'s
and $N_\mathrm{e}$'s for the nebular sample, combining the ORLs with the most reliable
measurements for each ion and the predicted intensities that are based on the
new atomic data.  Uncertainties of the derived $T_\mathrm{e}$~and $N_\mathrm{e}$~are estimated for
each object.

The diagnostic results from heavy element ORLs show reasonable agreement
with previous calculations in the literature.  We compare the electron
temperatures derived from the N~{\sc ii}~and O~{\sc ii}~ORLs, $T_\mathrm{e}$(ORLs), and
those from the collisionally excited lines (CELs), $T_\mathrm{e}$(CELs), as well as the
hydrogen Balmer jump, $T_\mathrm{e}$(H~{\sc i}~BJ), especially for the PNe with large
abundance discrepancies. Temperatures from He~{\sc i}~recombination lines,
$T_\mathrm{e}$(He~{\sc i}), are also used for comparison if available.  For all the
objects included in our sample, $T_\mathrm{e}$(ORLs) are lower than $T_\mathrm{e}$(H~{\sc i}~BJ),
which are in turn systematically lower than $T_\mathrm{e}$(CELs). Nebulae with
$T_\mathrm{e}$(He~{\sc i})~available show the relation $T_\mathrm{e}$(ORLs)~$\leq$~$T_\mathrm{e}$(He~{\sc
i})~$\leq$~$T_\mathrm{e}$(H~{\sc i}~BJ)~$\leq$~$T_\mathrm{e}$(CELs), which is consistent with
predictions from the bi-abundance nebular model postulated by
\citet{LSB00}.
\end{abstract}

\begin{keywords}
atomic data -- planetary nebulae: general -- H~{\sc ii}~regions
\end{keywords}

\section{Introduction}
Photoionized nebulae, such as planetary nebulae (PNe) and H~{\sc ii}~regions,
provide much of our knowledge of elemental abundances in the Milky Way and
other galaxies. Accurate measurements of electron temperatures (\tel) and
densities (\nel) are essential for reliable determination of elemental
abundances. Until recently, the principal means of determining elemental
abundances in nebulae has been from the measurement of collisionally 
excited lines (CELs). The emissivities of these lines relative to a hydrogen line are
very sensitive to \tel~under typical physical conditions of photoionized
nebulae, and thus are much affected by the errors in electron temperatures.  
These electron temperatures are in turn derived from the traditional method
based on the CEL ratios (\citealt{OF06}), such as the [O~{\sc iii}]
$\lambda$4363/($\lambda$4959~+~$\lambda$5007) and [N~{\sc ii}] 
$\lambda$5755/($\lambda$6548~+~$\lambda$6584) nebular-to-auroral line ratios.

An alternative method of nebular abundance determinations is to ratio the
intensity of an optical recombination line (ORL) of helium or a heavy
element with that of hydrogen. Unlike CELs, such as the [O~{\sc iii}]~
and [N~{\sc ii}]~nebular lines, whose emissivities relative to a hydrogen
recombination line increase exponentially with \tel, the emissivities of heavy
element ORLs relative to a hydrogen recombination line change weakly with both
\tel~and \nel, as the emissivities of recombination lines have only a similar,
power-law dependence on \tel~and \nel, apart from the parentage effects to be
discussed in Section\,2.2, and are essentially independent of \nel~under typical
nebular physical conditions (e.g. \citealt{L06a}, b). Consequently, the method based on
ORLs is much less affected by temperature measurement errors, and the results
in principal should be more conclusive. It is possible to use the intensity
ratio of two ORLs originating from the recombination of different ion parents
to estimate \nel~(\citealt{FSL11}), since the density sensitivity of the
parent populations is partially reflected in the resultant ORL
relative intensities. 

While emissivities of heavy element ORLs have
only a relatively weak, power-law dependence on \tel, this dependence varies
for lines originating from levels of different orbital angular momentum quantum
number $l$.  Therefore, the relative intensities of ORLs can be used to derive
electron temperature, provided that very accurate measurements can be secured 
(\citealt{L03}; \citealt{LLB04}; \citealt{TBL04}). Since the sensitivity of
the ORL ratios to electron temperature and density is very weak, in order to
obtain \tel's and \nel's of the nebular regions where ORLs arise, one needs
to acquire high-precision measurements of these ORLs.  
However, given the low nebular abundances of heavy elements
($\sim$10$^{-4}$--10$^{-3}$~or even lower relative to hydrogen) and the
relatively long time scales for a heavy element ion to recombine with an
electron under the physical conditions of gaseous nebulae (\citealt{OF06}),
the heavy element ORLs are very weak compared to the CELs and low-order 
hydrogen lines. Therefore, high-precision measurements of those ORLs are
required.

Although several deep spectroscopic surveys have been carried out during the past decade 
for several dozen Galactic disk and Bulge PNe (\citealt{TBL03b}, \citeyear{TBL04}; 
\citealt{LLB04}; \citealt{RG05}; \citealt{WLB05}; \citealt{WL07}) and for a number of
Galactic and extragalactic H~{\sc ii}~regions (\citealt{EPT02}, \citeyear{EPG04};
\citealt{TBL03a}; \citealt{PPR04}; \citealt{GEP04}, \citeyear{GEP05}, \citeyear{GPC06}),
most of the ORL measurements are still not accurate enough for nebular
analysis, due to either relatively low signal-to-noise ratios or line
blending. Currently only a limited number of objects (mainly nearby Galactic
PNe) have deep enough spectra for recombination line analysis
(e.g. \citealt{LSB95}, \citeyear{LSB00}, \citeyear{LLB01}; \citealt{EPP99}, \citeyear{EPG04}; 
\citealt{SBW04}; \citealt{ZLL05a}; \citealt{FL11}).

In nebular astrophysics, there has been a long-standing dichotomy in
abundance determinations and plasma diagnostics.  For a given PN, the
abundances of  heavy-element ions (C$^{2+}$, N$^{2+}$, O$^{2+}$, and
Ne$^{2+}$) relative to hydrogen derived from the nebular ORLs are all higher
than the corresponding abundance values derived from their CEL counterparts. 
The electron temperatures derived from the H~{\sc i}~Balmer jump are also 
systematically lower than those derived from the CELs. A number of mechanisms
have been proposed to explain this dichotomy (e.g. \citealt{P67};
\citealt{R89}; \citealt{VC94}), but have failed to provide a consistent
interpretation of all observations, especially for those PNe with dramatically
large abundance discrepancies (e.g. $>$10, 20). A bi-abundance nebular model
postulated by \citet{LSB00} provides a more natural explanation of this
dichotomy. In this model, the ORLs heavy element ions arise mainly from
``cold" H-deficient inclusions, while the strong CELs are emitted predominantly
from the warmer ambient ionised gas of ``normal" ($\sim$\,solar values)
chemical composition. Deep spectroscopic surveys (e.g. \citealt{TBL03b},
\citeyear{TBL04}; \citealt{LLB04}; \citealt{RG05}; \citealt{WLB05};
\citealt{WL07}) and ORL analysis of individual nebulae in the past decade
(e.g. \citealt{LSB95}, \citeyear{LSB00}, \citeyear{LLB01}, \citeyear{L06a})
have yielded strong evidence for the existence of such a ``cold" component
(Recent reviews on this topic are given by \citealt{L03}, \citeyear{L06b},
\citeyear{L11}). More recently, \citet{NDS12} proposed that a
$\kappa$-distribution for electron energies, which is a departure from the
conventional assumption of a Maxwell-Boltzmann equilibrium energy
distribution, can explain the dichotomy in both H~{\sc ii}~regions and PNe.

Despite the paucity of high-quality observational data of the heavy element
ORLs as well as the lack of the atomic data that are adequate enough for nebular
analysis, plasma diagnostics have been carried out for a number of PNe and
preliminary results have been obtained. Using the radiative recombination
coefficients of \citet{PPB91}, \citet{TBL04} derived electron temperatures
using the O~{\sc ii}~ORLs for 6 Galactic PNe, and found an average 
\tel(O~{\sc ii}~ORLs)~value of $2920\pm2690$~K.  \citet{LLB04} derived 
\tel(O~{\sc ii}~ORLs)~for 18 PNe, and found an average value of 
$4910\pm4060$~K. \citet{WLB05} derived electron temperatures for a
dozen Norther Galactic PNe using the O~{\sc ii}~ORL ratios
$\lambda4075/\lambda4089$, $\lambda4649/\lambda4089$,
$\lambda4072/\lambda4089$, $\lambda4414/\lambda4089$ and V1/$\lambda$4089.
\citet{WL07} determined \tel(O~{\sc ii}~ORLs) for 11 Galactic PNe using
the O~{\sc ii}~$\lambda4089/\lambda4649$~line ratio, and gave an average
value of $4370\pm5760$~K.  However, many of the recombination line intensities
used for diagnostics were not corrected for line blending, and the derived
electron temperatures could be problematic.

In this paper, we present a new approach to plasma diagnostics using the
N~{\sc ii}~and O~{\sc ii}~ORLs for a sample of PNe and H~{\sc ii}~regions,
which are selected from previous optical recombination line surveys and
individual-object studies. The most recent effective recombination
coefficients are utilised for the analysis. For each ion, we use one set of
detected ORLs with the most reliable measurements to constrain \tel~and
\nel~simultaneously, by comparing the observed intensities with the predicted
ones. The primary purpose of the current paper is to show that the plasma
diagnostic method employed here is applicable to all emission nebulae,
provided that more and more accurate measurements of the N~{\sc ii}~and
O~{\sc ii}~ORLs are available in the future.

\section[ORL diagnostics]{Optical recombination line diagnostics}

In this section, we compare the intensities of the observed N~{\sc ii}~and
O~{\sc ii}~lines with theoretical predictions over a wide range of electron
temperatures and densities. 

\begin{figure*}
\begin{minipage}{175mm}
\begin{center}
\vskip0.05truein
\rotatebox{-90}{\resizebox{10cm}{!}{\includegraphics{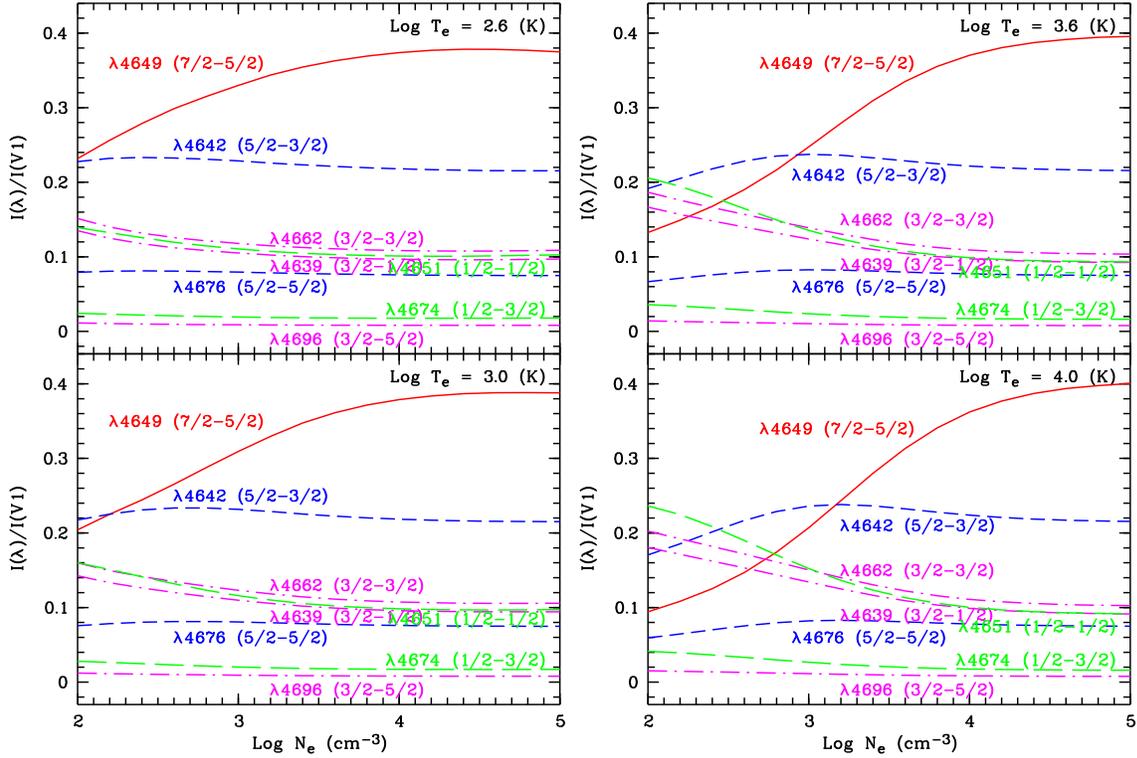}}}
\vskip-0.05truein
\caption[]{Theoretical fractional intensities of  O~{\sc ii}~Multiplet M1:
2s$^2$2p$^2$3p\,$^4$D$^{\rm o}$ -- 2s$^2$2p$^2$3s\,$^4$P as a function of
electron density.  The numbers in the brackets ($J_2 - J_1$) following the
line labels are the total angular momentum quantum numbers from the upper to
the lower levels.  Transitions from the upper levels with the same angular
momentum quantum number $J_2$ are represented by the curves with the same
colour and line type.  Four temperature cases, $\log~{T_\mathrm{e}}$~[K] = 2.5,
3.0, 3.5 and 4.0, are presented. The figure is based on the unpublished
calculations of Storey.}
\end{center}
\end{minipage}
\end{figure*}

\begin{figure}
\begin{minipage}{175mm}
\begin{center}
\vskip0.075truein
\rotatebox{-90}{\resizebox{10cm}{!}{\includegraphics{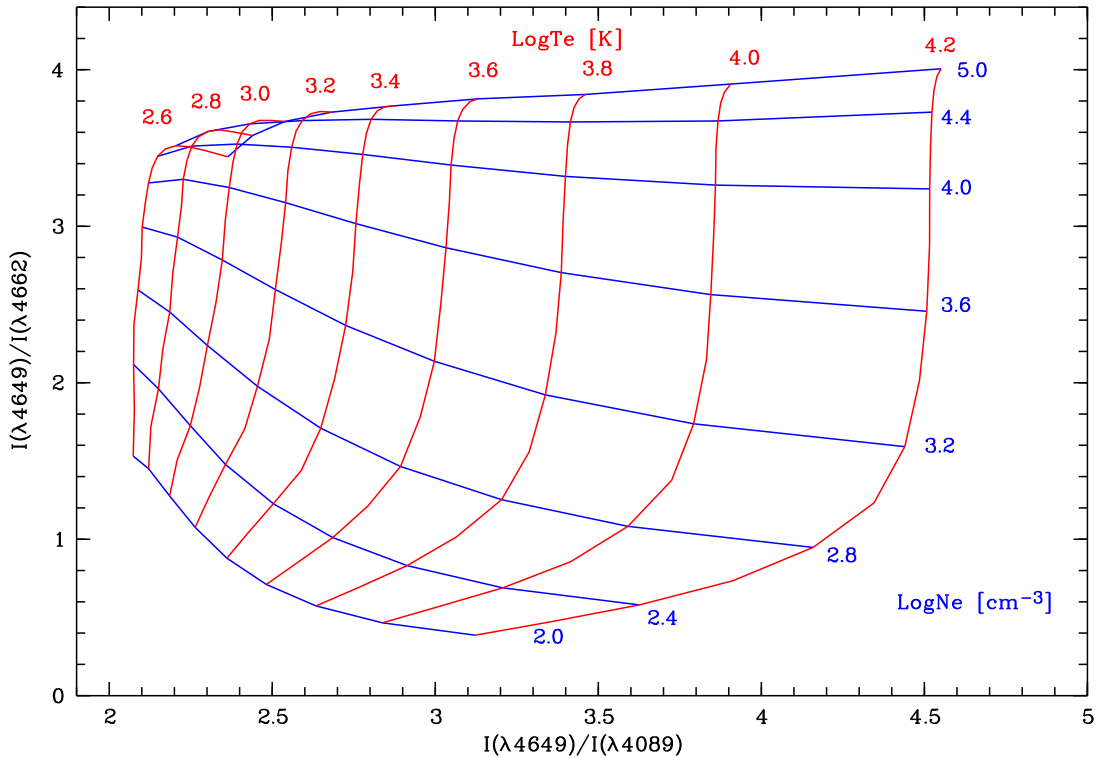}}}
\vskip0.1truein
\caption[]{Loci of the O II recombination line ratios
I($\lambda$4649)/I($\lambda$4662) and I($\lambda$4649)/I($\lambda$4089) for
different \tel's and \nel's.  The figure is based on the unpublished
calculations of Storey.}
\end{center}
\end{minipage}
\end{figure}

\subsection{Planetary nebula and H~{\sc ii}~region sample}

During the past two decades, several deep optical spectroscopic surveys have been 
carried out and published for several dozens of Galactic disk, Bulge and Halo PNe and for a 
number of Galactic and extragalactic H~{\sc ii}~regions.  These allow for detailed 
nebular plasma diagnostics and abundance analyses using the relatively weak hydrogen 
and helium recombination lines/continua and ORLs from heavy element ions.  In total, over 
100 PNe and 40 H~{\sc ii}~regions have been studied using ORLs.  The current paper 
makes use of the samples from the literature listed in Tables 1 and 2.

High spectral and spatial resolution spectroscopy of PNe has been published 
in the literature for detailed analysis 

\clearpage
\noindent 
of the physical condition and elemental 
abundance distributions across the nebulae.  \citet{TWP08} carried 
out a dedicated study of three Galactic PNe (NGC\,5882, NGC\,6153 and NGC\,7009) by
means of optical integral field spectroscopy using the Fibre Large Array Multi Element Spectrograph
on the Very Large Telescope (VLT/FLAMES). The ratio of
the abundances derived from ORLs and CELs, which is called the abundance 
discrepancy factor (ADF), was studied across the nebulae for doubly-ionised 
oxygen, O$^{2+}$.  Very small values of the temperature fluctuation parameter
$t^{2}$ (defined by \citealt{P67}, \citeyear{P71}) in the plane of the sky
were found in all three objects. Most recently, \citet{MNE12} published
results from integral field spectroscopy of a field located near the Trapezium
Cluster in the Orion Nebula using the Potsdam Multi-Aperture Spectrophotometer (PMAS). 
Detailed studies of one of the three
most prominent protoplanetary disks (proplyds) show that the ADF of O$^{2+}$
is close to one in the proplyd, while the background emission still yields
the typical ADF(O$^{2+}$) observed in the Orion nebula.

\subsection{Atomic data}

\begin{figure}
\begin{minipage}{85mm}
\begin{center}
\begin{tabular}{c}
\rotatebox{-90}{\resizebox{7.5cm}{!}{\includegraphics{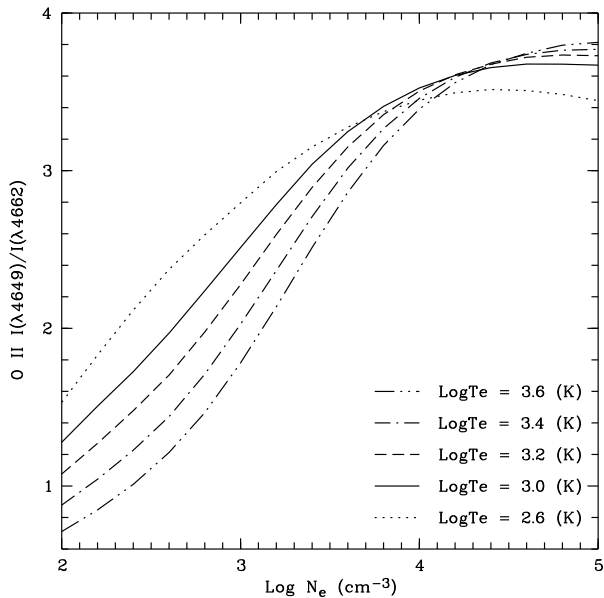}}}
\end{tabular}
\vskip0.1truein
\caption[]{The O~{\sc ii}~$\lambda$4649/$\lambda$4662 recombination line ratio
as a function of electron density. Different curves represent different
temperature cases. The plot is based on the unpublished calculations of P.~J.
Storey.}
\end{center}
\end{minipage}
\end{figure}

Since the 1990s, it has been possible to obtain reliable measurements of the
faint ORLs emitted by heavy element ions for bright nebulae. However, atomic
data are necessary to analyse these spectral features, in particular the
effective recombination coefficients. ORL ratios from states of different
orbital angular momenta do show some temperature dependence and therefore can
be used to measure the average temperature under which the lines are emitted
(e.g. \citealt{L03}).

Relative populations of the fine-structure levels of the ground term of a
recombining ion, such as N$^{2+}$~in the case of N~{\sc ii}~and O$^{2+}$~in
the case of O~{\sc ii}, deviate from the Boltzmann distribution and vary as a
function of electron density in low-density nebulae.  
This variation in the level population is reflected in the emissivities, and thus in the 
effective recombination coefficients of the recombination lines that are formed from 
different parent levels.  In essence, the intensity ratio of two such lines can be used 
to determine electron density (e.g. \citealt{FSL11}).

High-precision quantum mechanical calculations of the effective recombination
coefficients for N~{\sc ii}~(\citealt{FSL11}), and O~{\sc ii}~lines
(Storey, unpublished) have been completed. Both calculations improve those
from previous work (e.g. \citealt{NS84}; \citealt{EV90};
\citealt{PPB91}; \citealt{S94}; \citealt{LSB95};
\citealt{KSD98}; \citealt{KS02}). The strongest and
best-observed lines for N~{\sc ii}~so far have been those from multiplets V3
3p~$^{3}$D -- 3s~$^{3}$P$^{\rm o}$ and V39a,b 4f~G[7/2,9/2] --
3d~$^{3}$F$^{\rm o}$. For O~{\sc ii}, they are from multiplets V1 
3p~$^{4}$D$^{\rm o}$ -- 3s~$^{4}$P, V10 3d~$^{4}$F -- 3p~$^{4}$D$^{\rm o}$, 
and V48a,b 4f~G[5,4]$^{\rm o}$ -- 3d~$^{4}$F for O~{\sc ii}.  

Fig.\,1 shows the theoretical fractional intensities of the fine-structure
components of the O~{\sc ii}~Multiplet V1 2p$^2$3p\,$^4$D$^{\rm o}$ --
2p$^2$3s\,$^4$P~as a function of electron density at four temperature cases.
The same can be seen for the N~{\sc ii}~multiplet V3 2p3p\,$^{3}$D --
2p3s\,$^{3}$P$^{\rm o}$ in Fig.\,$5$ from \citet{FSL11}. Fig.\,2
shows the loci of the O~{\sc ii}~recombination line ratios
I($\lambda$4649)/I($\lambda$4662) and I($\lambda$4649)/I($\lambda$4089) for
different \tel's and \nel's. The same can be seen for N~{\sc ii}~recombination
line ratios I($\lambda$5679)/I($\lambda$5666) and
I($\lambda$5679)/I($\lambda$4041) in Fig.\,$8$ from \citet{FSL11}.
Using the accurate measurements of the N~{\sc ii}~and O~{\sc ii}~line ratios,
we can determine \tel's and \nel's simultaneously from the loci. Since
different ORL ratios of N~{\sc ii}~or O~{\sc ii}~have different sensitivity on
\tel~and \nel, this will affect the reliability of the results.

Given the different temperature-dependence of diagnostic lines, individual
diagnostic line ratios are expected to yield differing results if the nebulae
are not isothermal. While the small values of $t^2$ appear to suggest there
are no large temperature variations where the CELs are emitted, the
discrepancies between the CEL and ORL diagnostic results definitely exist,
thereby suggesting that the gas is not isothermal.
While the authors believe that the
bi-abundance nebular model is a satisfactory explanation of the abundance
(and temperature) discrepancies, an alternative explanation involving
non-equilibrium electron energies has also been proposed (\citealt{NDS12};
\citealt{OS83}). According to \citet{NDS12}, physical processes such as
magnetic reconnection, injection of high-energy electrons through
photoionization by a ''hard" photon spectrum, may generate such
non-equilibrium electrons. The role of shockwaves in generating the
discrepancies also needs to be explored. The choice of abundance discrepancy
model, however, does not impact on the validity of the ORL diagnostics
proposed here.

The traditional method, which uses one line ratio to determine an electron
temperature or a density, has the disadvantage that the density or temperature
is undefined. Although a temperature-sensitive recombination line ratio is
usually quite insensitive to electron density, some density-sensitive line
ratios, to a noticeable extent, may vary with electron temperature. As an
example, Fig.\,3 shows the O~{\sc ii}~$\lambda$4649/$\lambda$4662~ratio as
a function of electron density. From the
Figure one can see that at low densities ($\log~{T_\mathrm{e}}\,<$\,3.6), the
electron density yielded by the observed line ratio may differ by as much
as 0.6 dex when the logarithmic temperature $\log~{T_\mathrm{e}}$~increases
from 2.6 to 3.6.  Instead of using a single line ratio, we present our new method 
using one set of N~{\sc ii}~or O~{\sc ii}~ORLs to constrain the electron temperature and
density simultaneously.  Our method is based on the fact that the intensity ratio of two of these
selected ORLs is temperature-sensitive, while the ratio of the other (two)
lines of this group should be density-sensitive.

\begin{figure*}
\begin{minipage}{175mm}
\begin{center}
\vskip0.075truein
\begin{tabular}{ccc}
\resizebox{7.75cm}{!}{\includegraphics{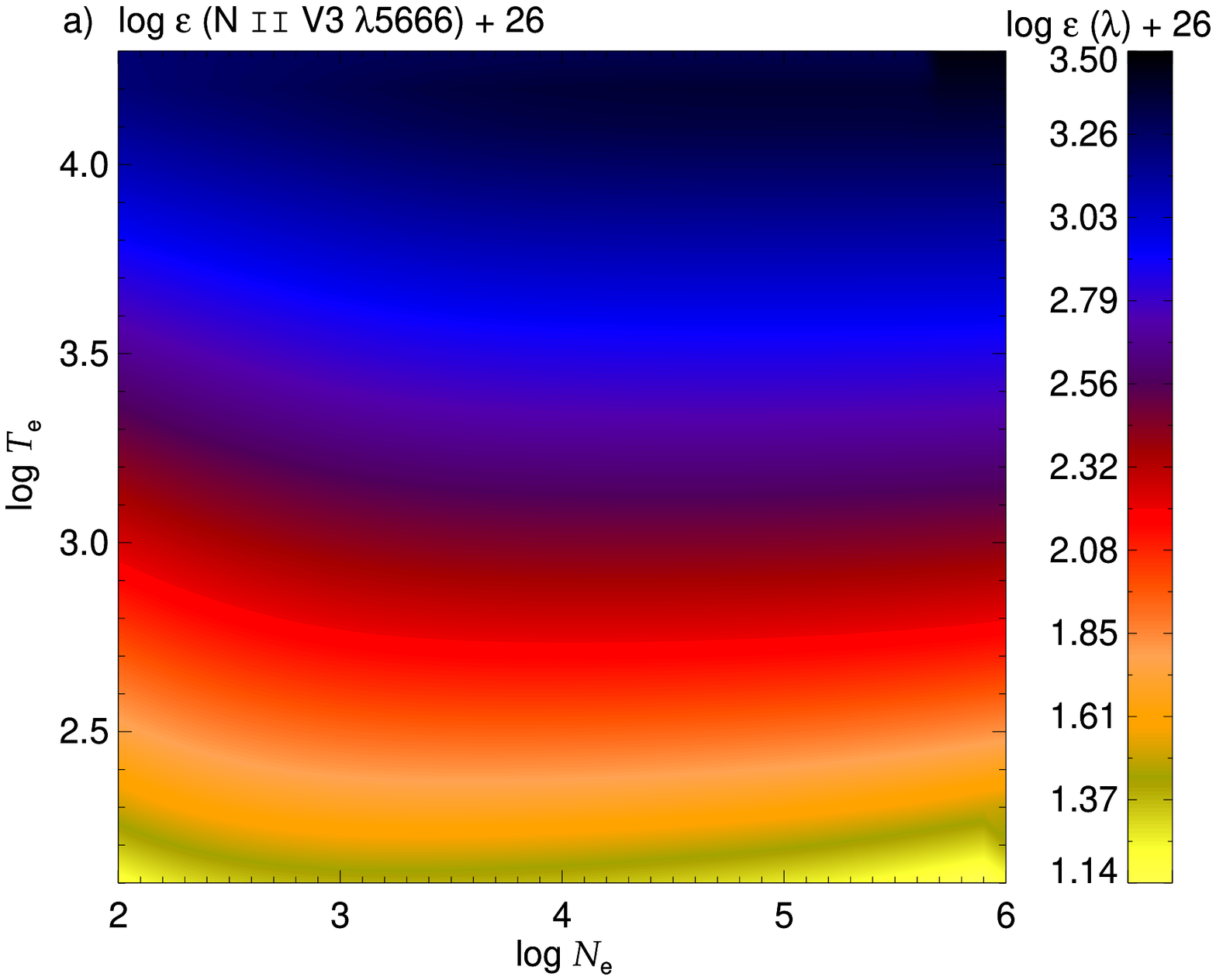}} & ~ & 
\resizebox{7.75cm}{!}{\includegraphics{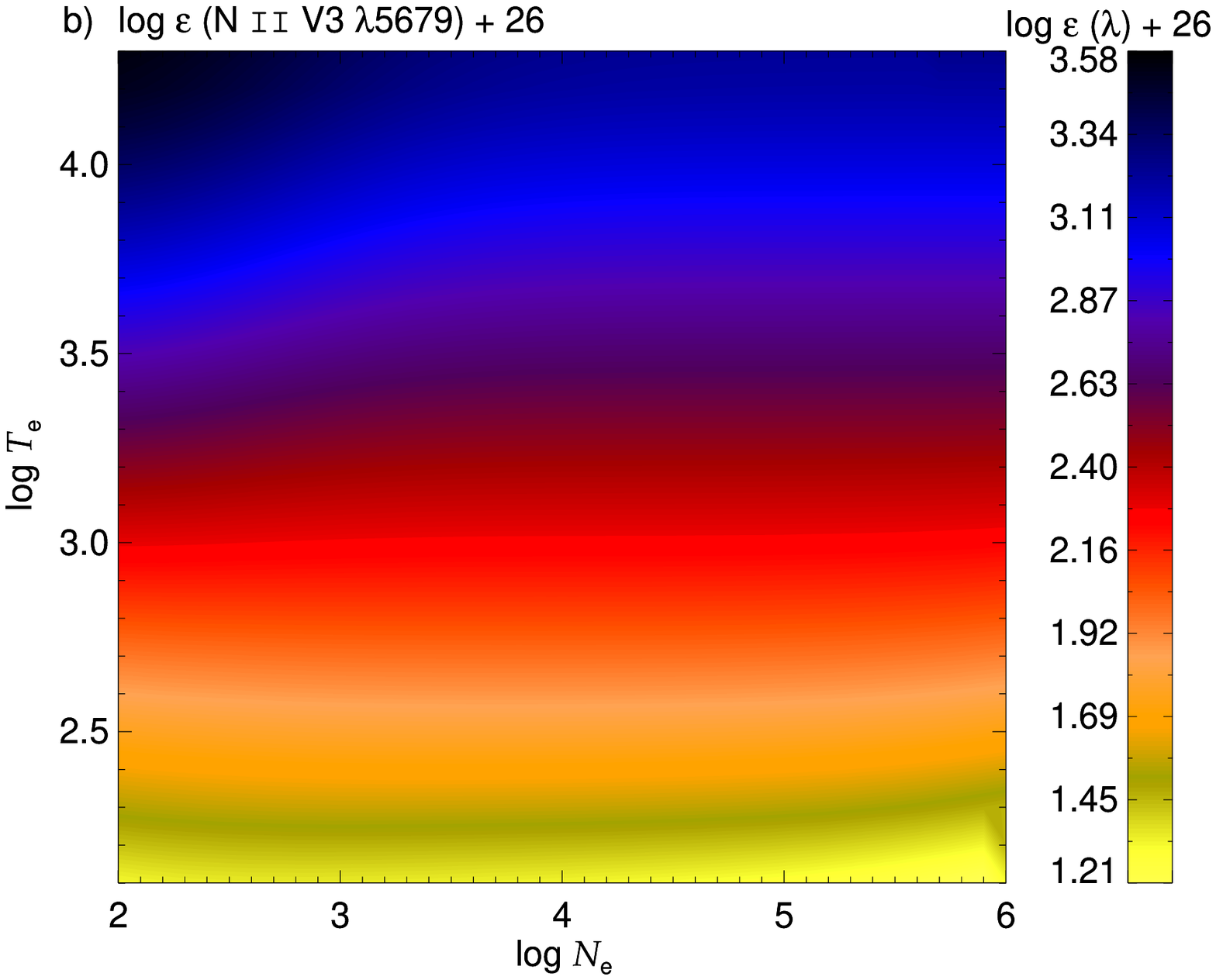}} \\ 
\end{tabular}
\vskip0.15truein
\begin{tabular}{ccc}
\resizebox{7.75cm}{!}{\includegraphics{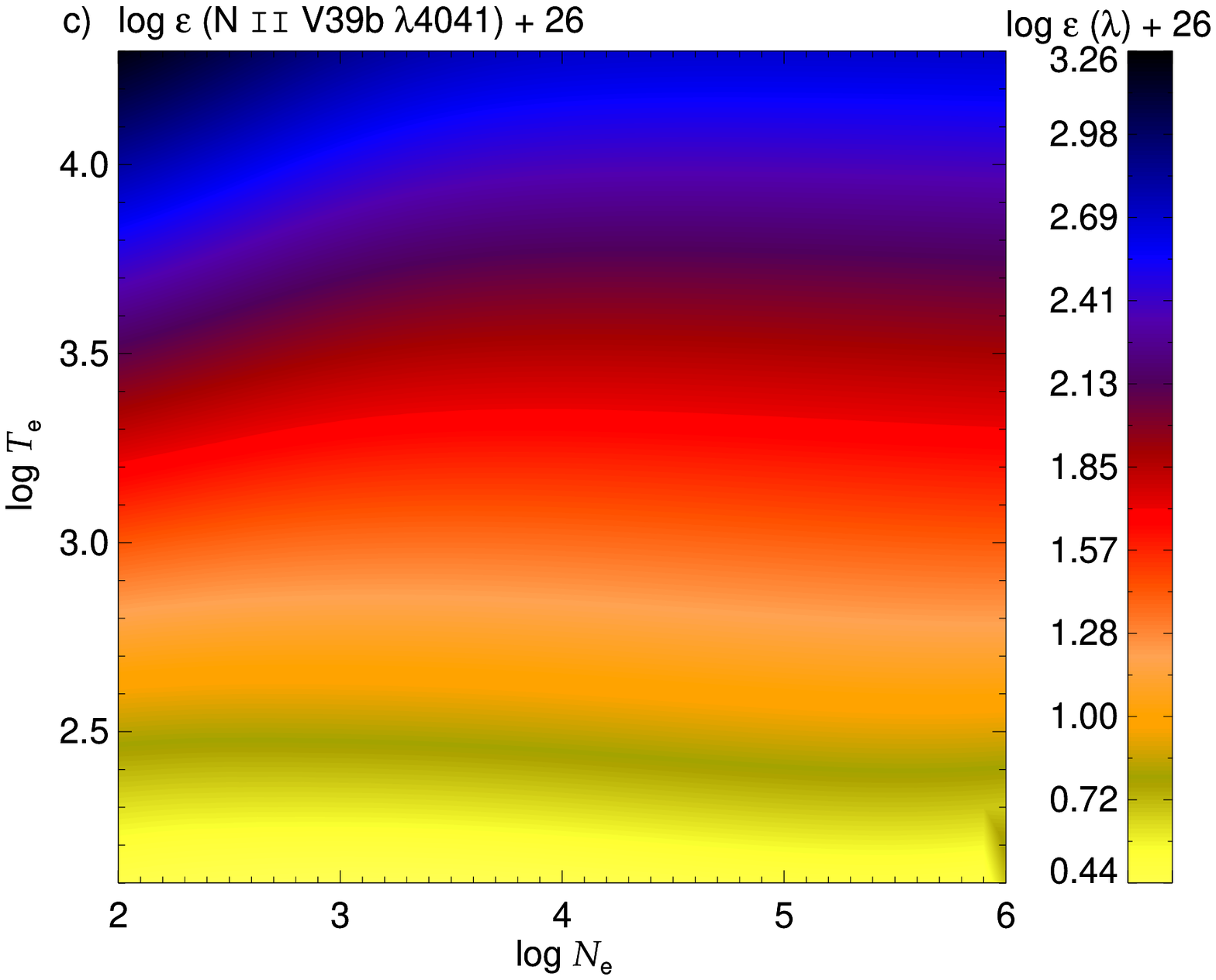}} & ~ & 
\resizebox{7.75cm}{!}{\includegraphics{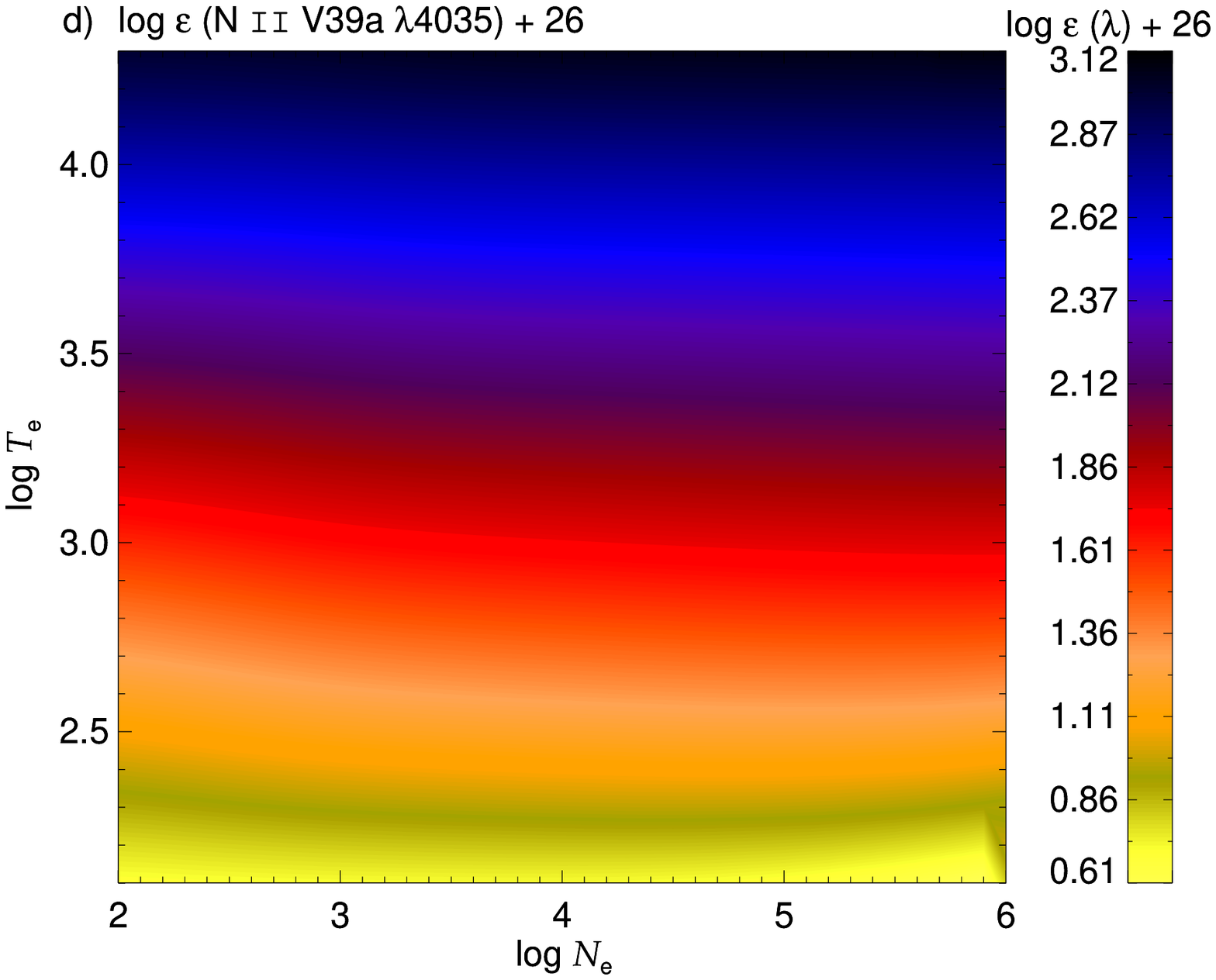}} \\ 
\end{tabular}
\vskip-0.1truein
\caption[]{Emissivities of the N~{\sc ii}~a) V3 $\lambda$5666, b) V3
$\lambda$5679, c) V39b $\lambda$4041 and d) V39a $\lambda$4035 lines as a
function of electron temperature and density, as derived by Eqn.\,(1). The
colour bar indicates the $\log~\varepsilon(\lambda)$ + 26 values in units of
ergs\,cm$^{+3}$\,s$^{-1}$.}
\end{center}
\end{minipage}
\end{figure*}

\begin{figure*}
\begin{minipage}{175mm}
\begin{center}
\vskip0.075truein
\begin{tabular}{ccc}
\resizebox{7.75cm}{!}{\includegraphics{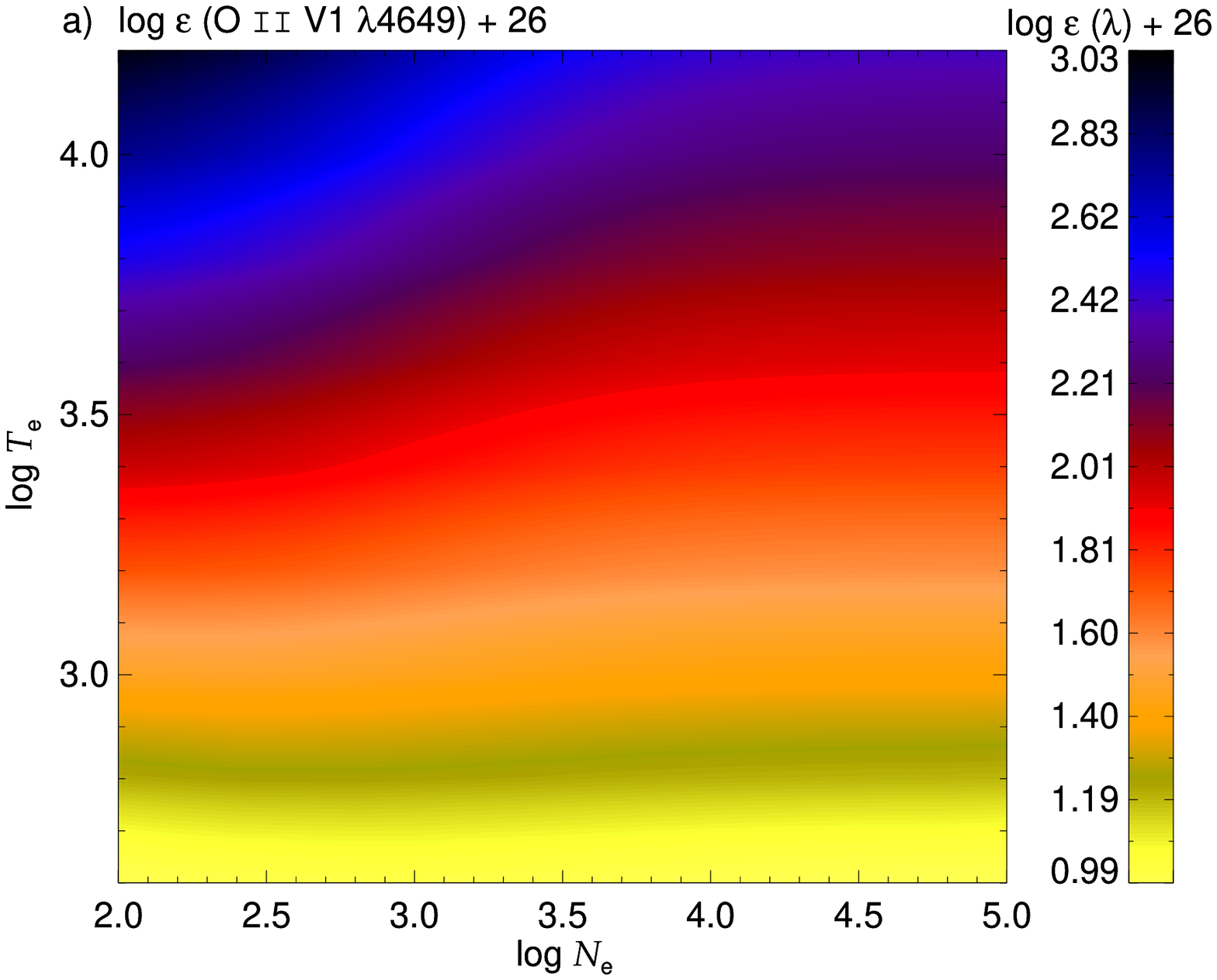}} & ~ & 
\resizebox{7.75cm}{!}{\includegraphics{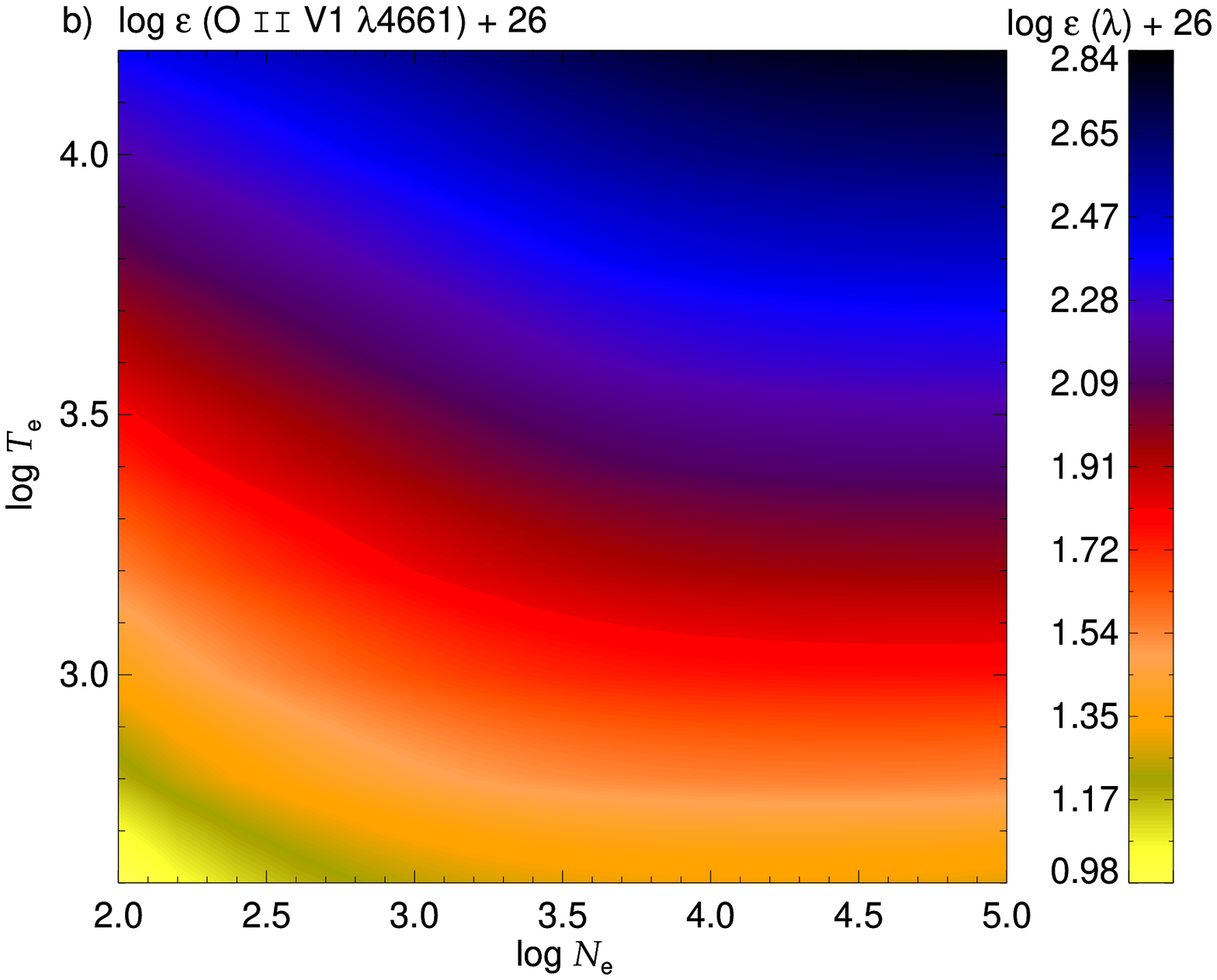}} \\ 
\end{tabular}
\vskip0.15truein
\begin{tabular}{ccc}
\resizebox{7.75cm}{!}{\includegraphics{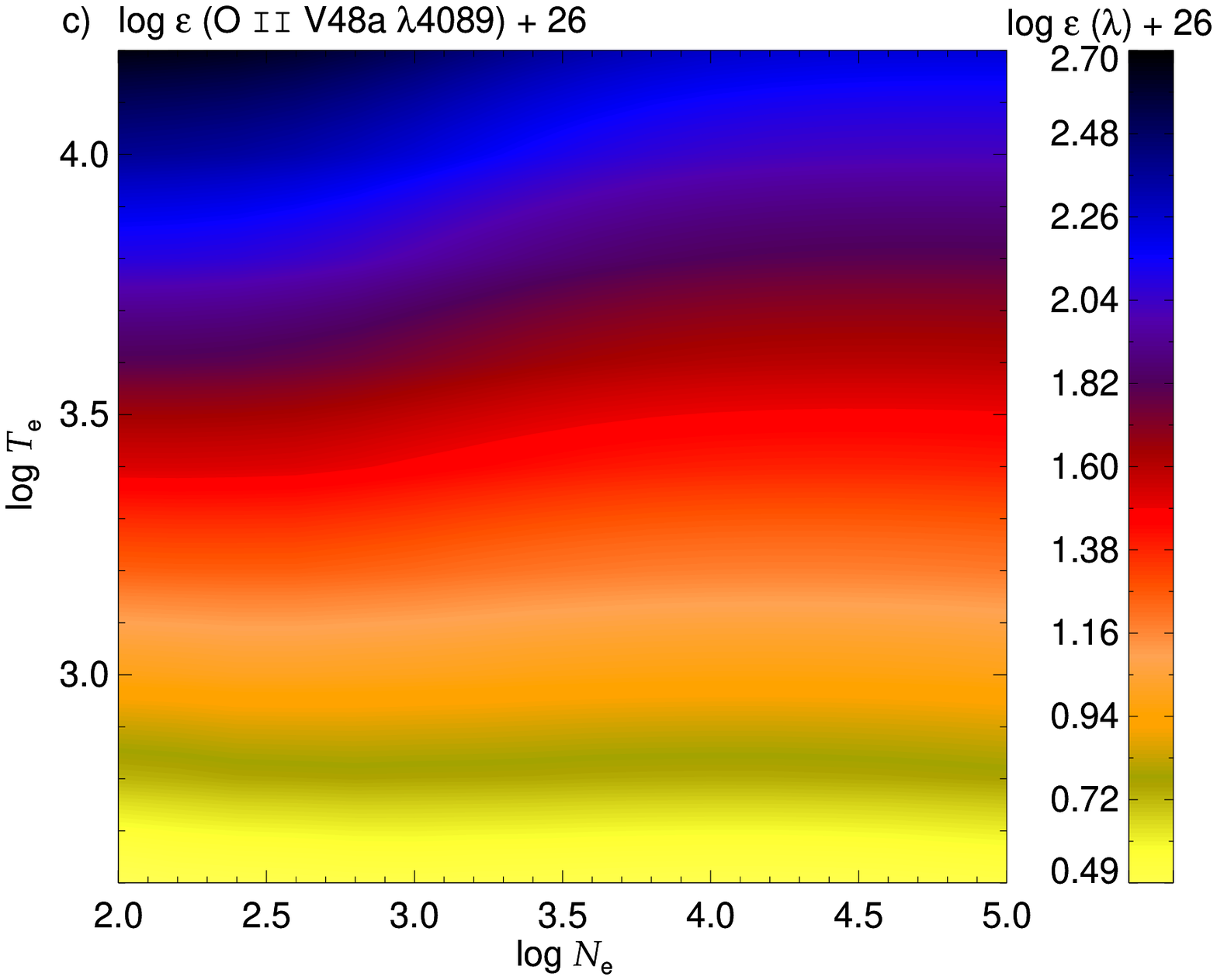}} & ~ & 
\resizebox{7.75cm}{!}{\includegraphics{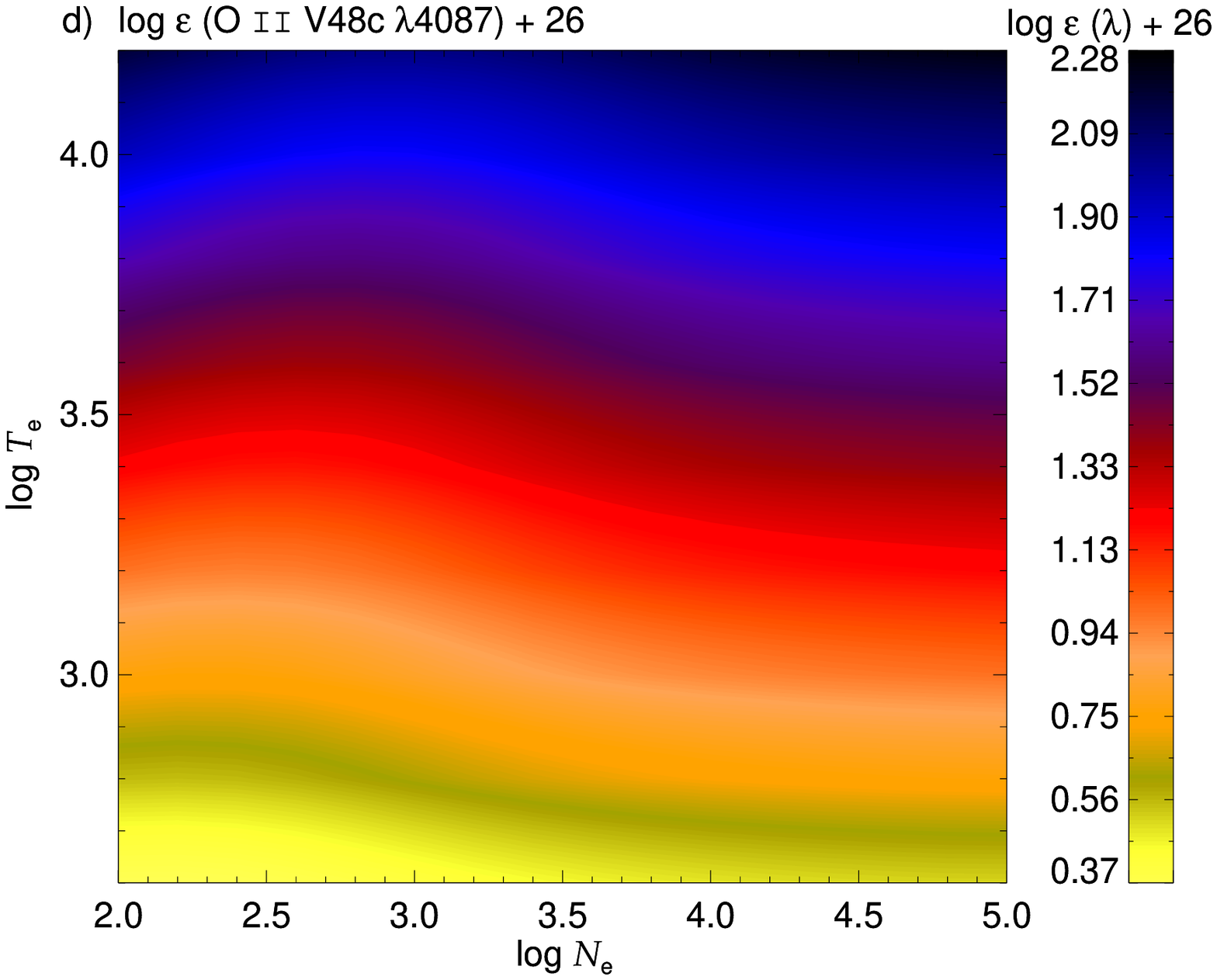}} \\ 
\end{tabular}
\vskip-0.1truein
\caption[]{The same as Figure 4, but for O~{\sc ii}~a) V1 $\lambda$4649, b)
V1 $\lambda$4662, c) V48a $\lambda$4089 and d) V48c $\lambda$4087 lines.}
\end{center}
\end{minipage}
\end{figure*}

\subsection{Theoretical intensities}

For the N~{\sc ii}~effective recombination coefficients (\citealt{FSL11}), 
$\log$~{\tel}~[K] covers a range of 2.1 to 4.3, with an incremental step
of 0.1, and $\log$~{\nel}~[cm$^{-3}$]~spans from 2.0 to 6.0, with an
incremental step of 0.1. For the O~{\sc ii}~effective recombination
coefficients (Storey, unpublished), $\log$~{\tel}~[K]~covers a range of 2.6 to
4.2, with an incremental step of 0.2, and $\log$~{\nel}~[cm$^{-3}$]~spans from
2.0 to 5.0, with an incremental step of 0.2. Both theoretical and observed
intensities are normalised such that H$\beta$ = 100.

For most objects, the other O~{\sc ii}~multiplets are not as strong as those
considered in Figs.\,1 and Fig.\,2, and are not suitable to constrain \tel~and
\nel, due to data quality. Figs.\,4 and 5 show the emissivities of N~{\sc ii}~
V3 and V39 ORLs (V3 $\lambda$5666, V3 $\lambda$5679, V39b $\lambda$4041 and
V39a $\lambda$4035) and O~{\sc ii}~V1 and V48 ORLs (V1 $\lambda$4649, V1
$\lambda$4662, V48a $\lambda$4089 and V48c $\lambda$4087), respectively, as a
function of \tel~and \nel, where the emissivity is derived from the effective
recombination coefficients as follows:

\vskip-0.15truein
\begin{equation}
\varepsilon (\lambda)~=~\alpha_{\rm eff}(\lambda)\times\frac{\it{hc}}{\lambda},
\end{equation}

\noindent
with $h$~as Planck's constant and $c$~as the speed of light, all in cgs units.  
Due to coarse grid resolution, we performed a bilinear interpolation on the 
effective recombination coefficients of the O~{\sc ii}~ORLs provided by 
Storey (unpublished).  Bilinear interpolation simply extends linear 
interpolation for functions of two variables, i.e. $\alpha_{eff}$~(\tel, \nel).

Based on the calculations of the effective recombination coefficients for the 
N~{\sc ii}~and O~{\sc ii}~lines, theoretical intensity for an N~{\sc ii}~or 
O~{\sc ii}~ORL relative to H$\beta$~can be calculated as a function of 
\tel~and \nel~as follows:

\vskip-0.15truein
\begin{equation}
I_{pred}\left(\lambda\right) = 
\frac{I\left(\lambda\right)}{I\left(\rm{H}\beta\right)} = 
\frac{\alpha_{\rm eff}\left(\lambda\right)}
{\alpha_{\rm eff}\left(\rm{H}\beta\right)}
\frac{4861}{\lambda}\frac{\rm{X}^+}{\rm{H}^+}\times{100},
\end{equation}

\noindent
where $I_{pred}$~(\tel,\nel)~is the theoretical predicted intensity of the 
transition $\lambda$, normalised such that $I(\rm{H}\beta)~=~100$.
$\alpha_{\rm eff}$($\lambda;\,$\tel,\nel)~is the effective recombination 
coefficient for the transition $\lambda$, and 
$\alpha_{\rm eff}$(\rm{H}$\beta;\,$\tel,\nel)~is the effective recombination 
coefficient for $\rm{H}\beta$, which is adopted from the hydrogenic 
calculations of \citet{SH95}.  The ionic abundances relative 
to hydrogen derived from ORLs, $\rm{X}^+/\rm{H}^+$, can be found in 
the literature for each nebula.

If the spectral resolution for the nebula is relatively poor, for instance 
FWHM~$<$~1.5\,{\AA}, then some blending may occur.  The theoretical 
intensities from the blended lines are then added together in the following manner:

\vskip-0.15truein
\begin{equation}
\sum_{i=1}^{n}I_{pred}\left(\lambda_i\right) = 
\sum_{i=1}^{n}\left(\frac{\alpha_{\rm eff}\left(\lambda_i\right)}{\lambda_i}\right)\frac{4861}{\alpha_{\rm eff}\left(\rm{H}\beta\right)}\frac{\rm{X}^+}{\rm{H}^+}\times{100},
\end{equation}

\noindent
where $i$~goes from 1 to the number of the blended lines.  

\subsection{Plasma diagnostics}

\subsubsection{Least-squares fit}

For each PN and H~{\sc ii}~region, the aforementioned theoretical intensities 
were compared with the observed intensities using a least-squares 
minimisation method as follows:

\vskip-0.15truein
\begin{equation}
\chi^2 =
\sum_{i=1}^{n}
\left(\frac{I_{obs}\left(\lambda_i\right) - I_{pred}\left(\lambda_i\right)}
{I_{pred}\left(\lambda_i\right)}\right)^2,
\end{equation}

\noindent
where \chisq~is the sum over the combination of $n$ lines used for 
either N~{\sc ii}~or O~{\sc ii}, as a function of \tel~and \nel.  $I_{obs}$~is the 
observed intensity for the transition $\lambda_i$~and $I_{pred}$~is the 
predicted intensity.  For a blended feature, the observed intensity is compared 
with the combined theoretical intensities of the blended lines, described in Eqn.\,(3).
As an example, Figs.\,6 and 7 show the $\log$~{\chisq}~distributions for 
4 PNe (Hf\,2-2, M\,1-42, NGC\,6153 and NGC\,7009) for N~{\sc  ii} and O~{\sc ii}, respectively.  
The location of the minimum $\log$~{\chisq}~value can be determined for each 
comparison, thus providing the optimal ($\log$~{\tel},\,$\log~${\nel}), as shown 
by the white cross hairs in both figures.  

\subsubsection{Error estimate}

For many objects, the resulting temperatures and densities are well confined.
However, it is difficult to decide how reliable these diagnostic results could
be. Not all of the surveys or individual detailed studies of PNe and/or H~{\sc
ii}~regions provided errors along with their observational intensities. This
makes estimating the uncertainties for the optimal ($\log$~{\tel},
$\log$~{\nel})~rather complicated. A few nebulae had observational errors
presented in the literature and are adopted in the current
analyses. For those nebulae whose measurement uncertainties are not presented
in the literature, we estimated errors according to their observational
conditions and data quality. In order to determine the uncertainties for the
optimal ($\log$~{\tel}, $\log$~{\nel}), we carried out an extra set of
calculations that begin with random number generation.

For each line in a combination of N~{\sc ii}~or O~{\sc ii}~ORLs (e.g.
V3~$\lambda$5666, V3~$\lambda$5679, V39a~$\lambda$4035, and V39b~$\lambda$4041
for N~{\sc ii}, and V1~$\lambda$4662, V1~$\lambda$4649, V48a~$\lambda$4089,
and V48c~$\lambda$4087 for O~{\sc ii})~that are used for plasma diagnostics,
we used {\sc idl}~to generate a number, $N_{\rm sim}$, of intensity values,
$I_{\rm sim}$($\lambda_i$), with a normal (Gaussian) distribution centred
around the original observational intensity, $I_{\rm obs}$($\lambda_i$), along
with a standard deviation obtained from the observational error,
$\sigma(I_{\rm obs}$), of this distribution.

The mean intensities of the simulations and standard deviations,
$I_{\rm sim}^{mean}$~and $\sigma(I_{\rm sim}^{mean})$, are calculated as
follows: 

\begin{equation}
I_{sim}^{mean} = \frac{\sum_{i=1}^{N_{\rm sim}}I_{\rm sim}^{i}(\lambda)}{N_{\rm sim}},
\end{equation}
and

\begin{equation}
\sigma(I_{\rm sim}^{mean}) = \sqrt{\frac{\sum_{i=1}^{N_{\rm sim}}\left(I_{\rm sim}^{i}(\lambda) - \left<I_{\rm sim}\right>\right)^2}{N_{\rm sim}}},
\end{equation}

\noindent
The standard deviation of these generated numbers also agree well with the 
observational errors.  Each combination of the randomly generated line 
intensities, $I_{\rm sim}^{i}$($\lambda$), where $i$~is from 1 to N$_{\rm sim}$, 
is used to determine a unique optimal ($\log$~{\tel},\,$\log$~{\nel}), following 
the method described in Section\,2.4.1. 

Errors of the optimal $\log$~{\tel}~and $\log$~{\nel}~of each nebula are then
calculated from these $N_{\rm sim}$~temperatures and densities, which can be
denoted as $T_\mathrm{e}^{i}$~and $N_\mathrm{e}^{i}$, where $i$~=~1, ...,
$N_{\rm sim}$.  For those objects whose $\log$~{\tel}~-~$\log$~{\nel}~diagram 
shows multiple peaks, standard deviations of \tel~and \nel~are calculated, 
which are centred on the mean values of these $N_{\rm sim}$~temperatures 
and densities.  The random intensities of each ORL are then normally 
distributed within the observational errors of the observed intensities.  
The program would then calculate the $\log$~\chisq~along the entire 
\tel~and \nel~grid for each combination of the random intensities and 
determine the minimum value, thus deriving the unique optimal 
($\log$~{\tel},\,$\log$~{\nel})~for that simulation.  

Figs.\,8 and 9 display the frequencies of these optimal $\log$~{\tel}~and
$\log$~{\nel}~distributions for the N~{\sc ii}~and O~{\sc ii}~ORLs,
respectively, for the same 4 PNe shown in Figs.\,6 and 7. The errors for
the optimal $\log$~{\tel}~and $\log$~{\nel}~are derived as follows: 

\begin{equation}
T_{\rm sim}^{mean}=\frac{\sum_{i=1}^{N_{\rm sim}}{T_\mathrm{e}^{i}}}{N_{\rm sim}},
\end{equation}
and

\begin{equation}
N_{\rm sim}^{mean}=\frac{\sum_{i=1}^{N_{\rm sim}}{N_\mathrm{e}^{i}}}{N_{\rm sim}},
\end{equation}

\noindent
with respective standard deviations calculated as follows:

\begin{equation}
\sigma({T_{\rm sim}^{mean}}) = \sqrt{\frac{\sum_{i=1}^{N_{\rm sim}}\left(T_\mathrm{e}^{i} - T_{\rm sim}^{mean} \right)^2}{N_{\rm sim}}},
\end{equation}
and

\begin{equation}
\sigma({N_{\rm sim}^{mean}}) = \sqrt{\frac{\sum_{i=1}^{N_{sim}}\left(N_\mathrm{e}^{i} - N_{\rm sim}^{mean} \right)^2}{N_{\rm sim}}}.
\end{equation}

\noindent
These values are represented in Figs.\,8 and 9 as green error bars centred at
$T_{\rm sim}^{mean}$~and $N_{\rm sim}^{mean}$, but projected on the white
crosshairs of the optimal $\log$~{\tel}~and $\log$~{\nel}~locations obtained
from observations, thus providing upper and lower limits.  The white solid
contour designates the 1-$\sigma$~level around the peak. We project the
standard deviations onto the optimal $\log$~{\tel}~and $\log$~{\nel}~locations 
obtained from observations because they are calculated using the 
1-$\sigma$~errors of the observed intensities.  

\begin{figure*}
\begin{minipage}{175mm}
\begin{center}
\vskip0.075truein
\begin{tabular}{cccccc}
\resizebox{7.75cm}{!}{\includegraphics{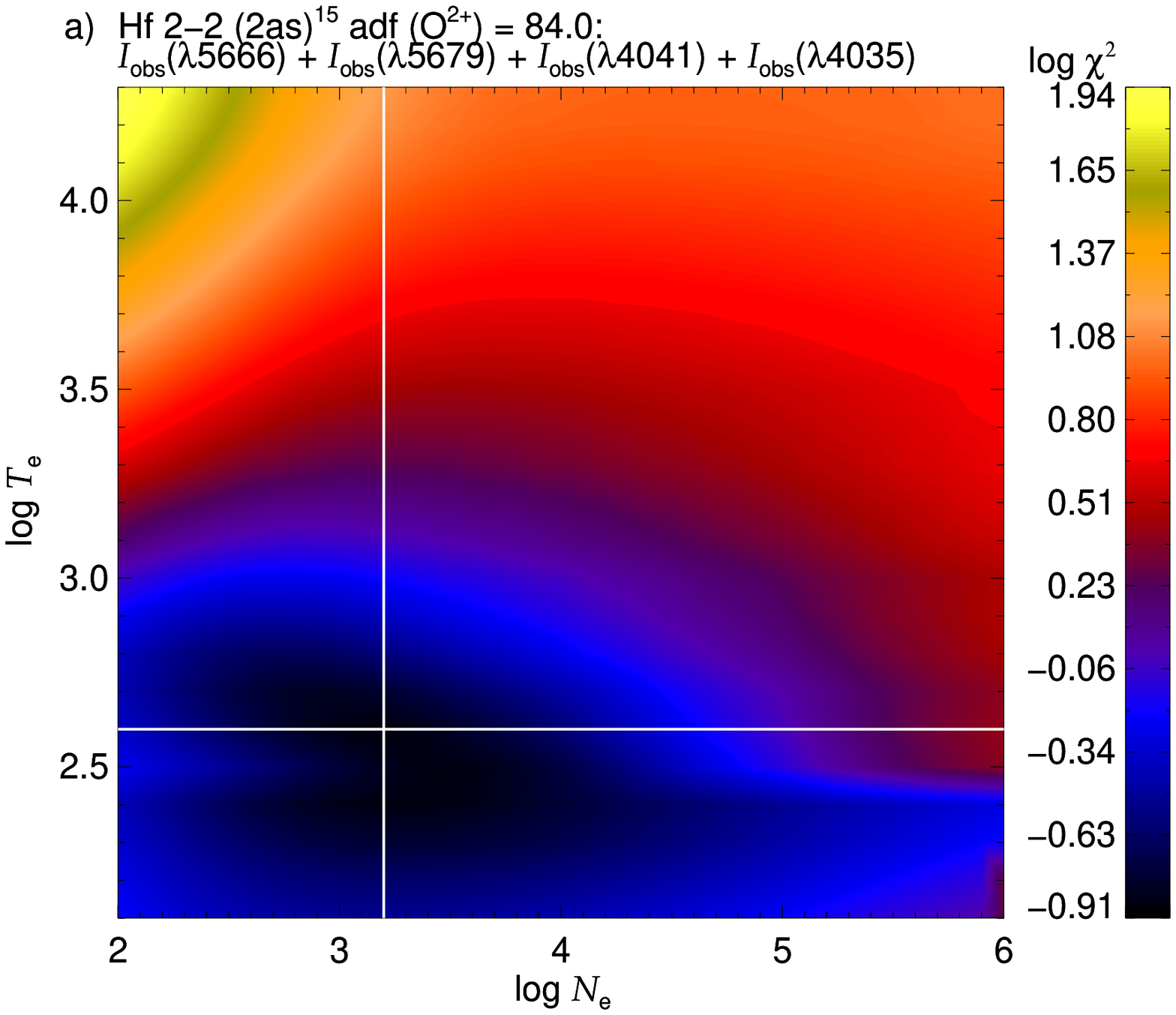}} & & 
\resizebox{7.75cm}{!}{\includegraphics{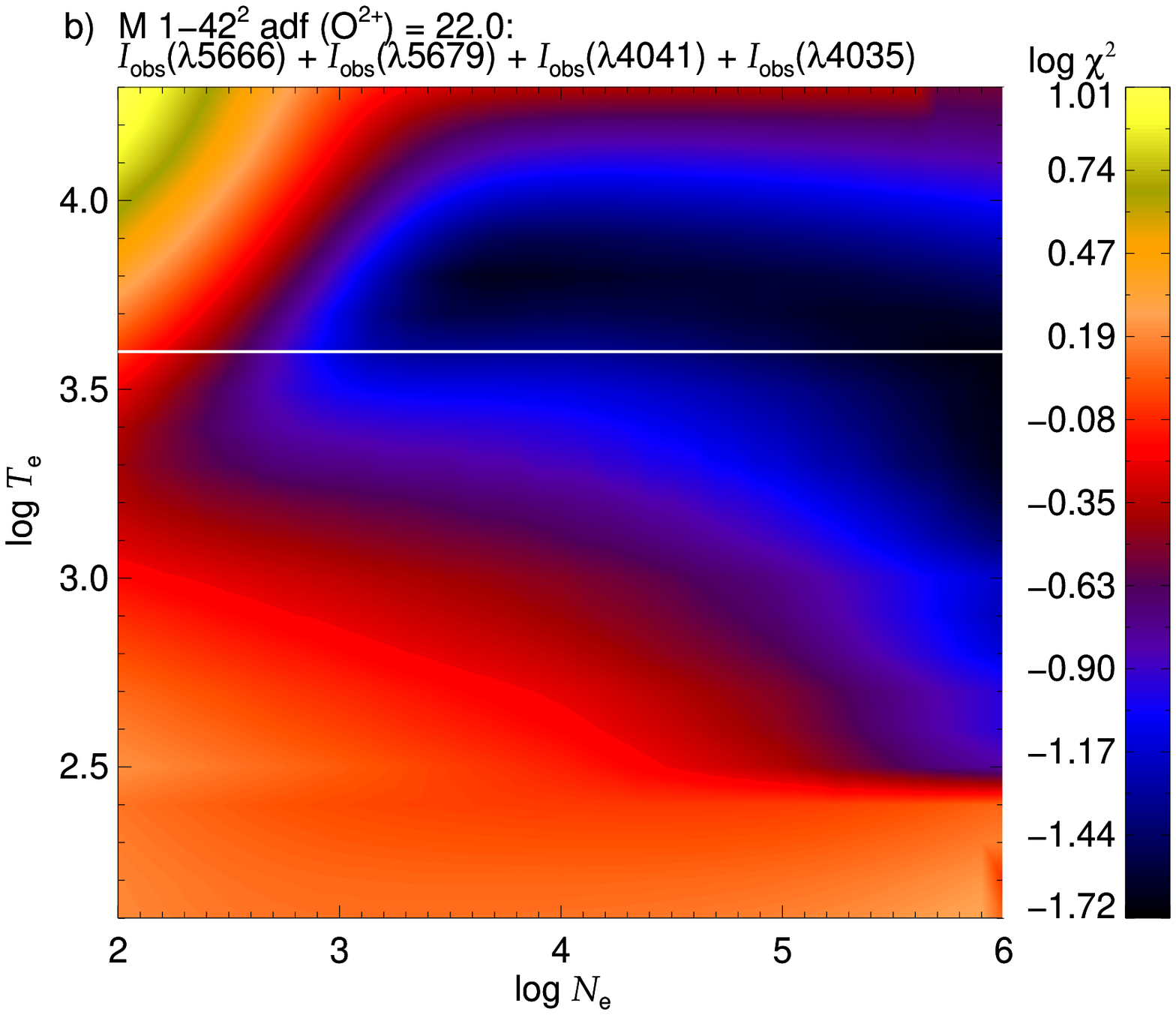}} & & & \\
\end{tabular}
\vskip0.2truein
\begin{tabular}{cccccc}
\resizebox{7.75cm}{!}{\includegraphics{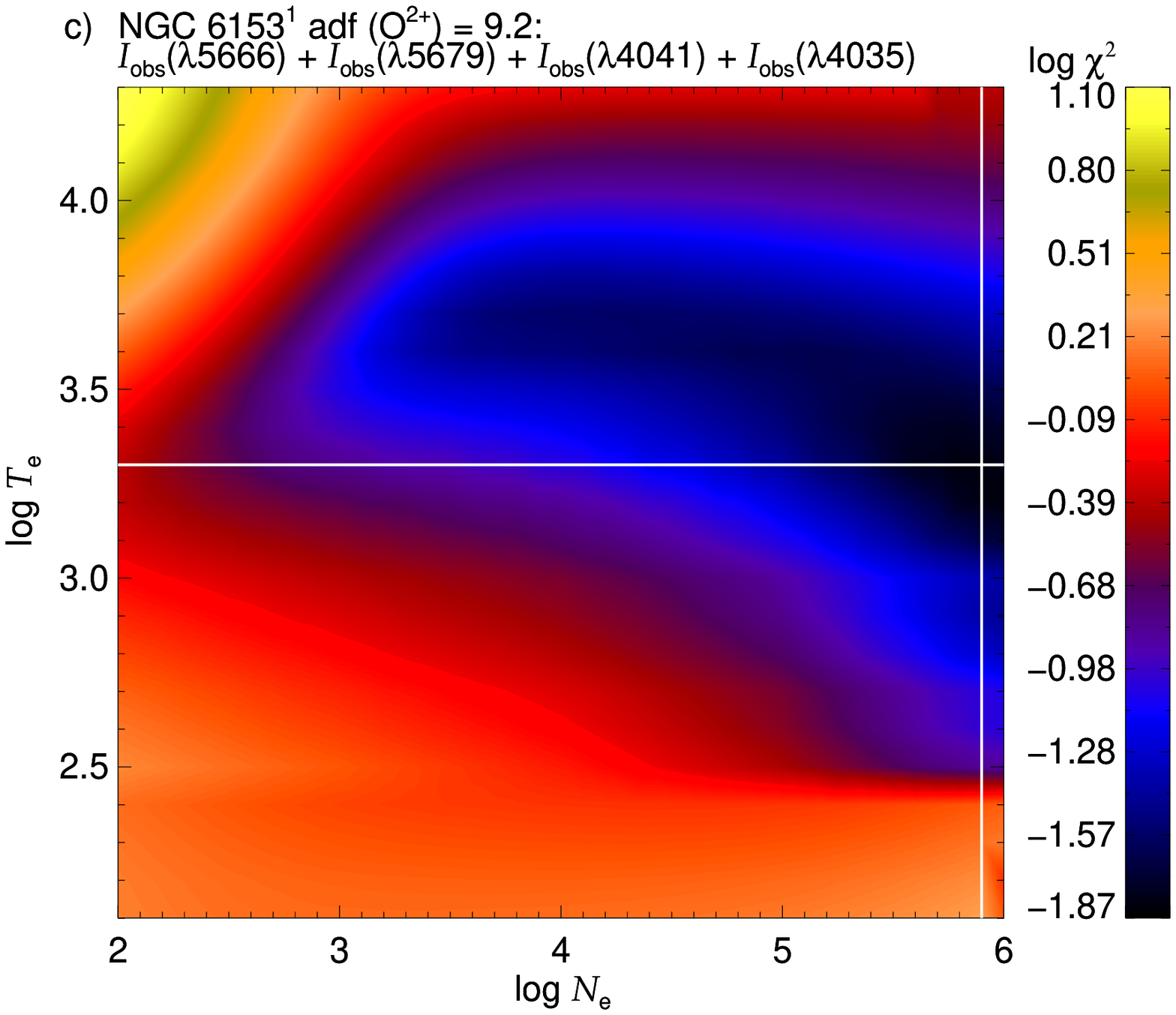}} & & 
\resizebox{7.75cm}{!}{\includegraphics{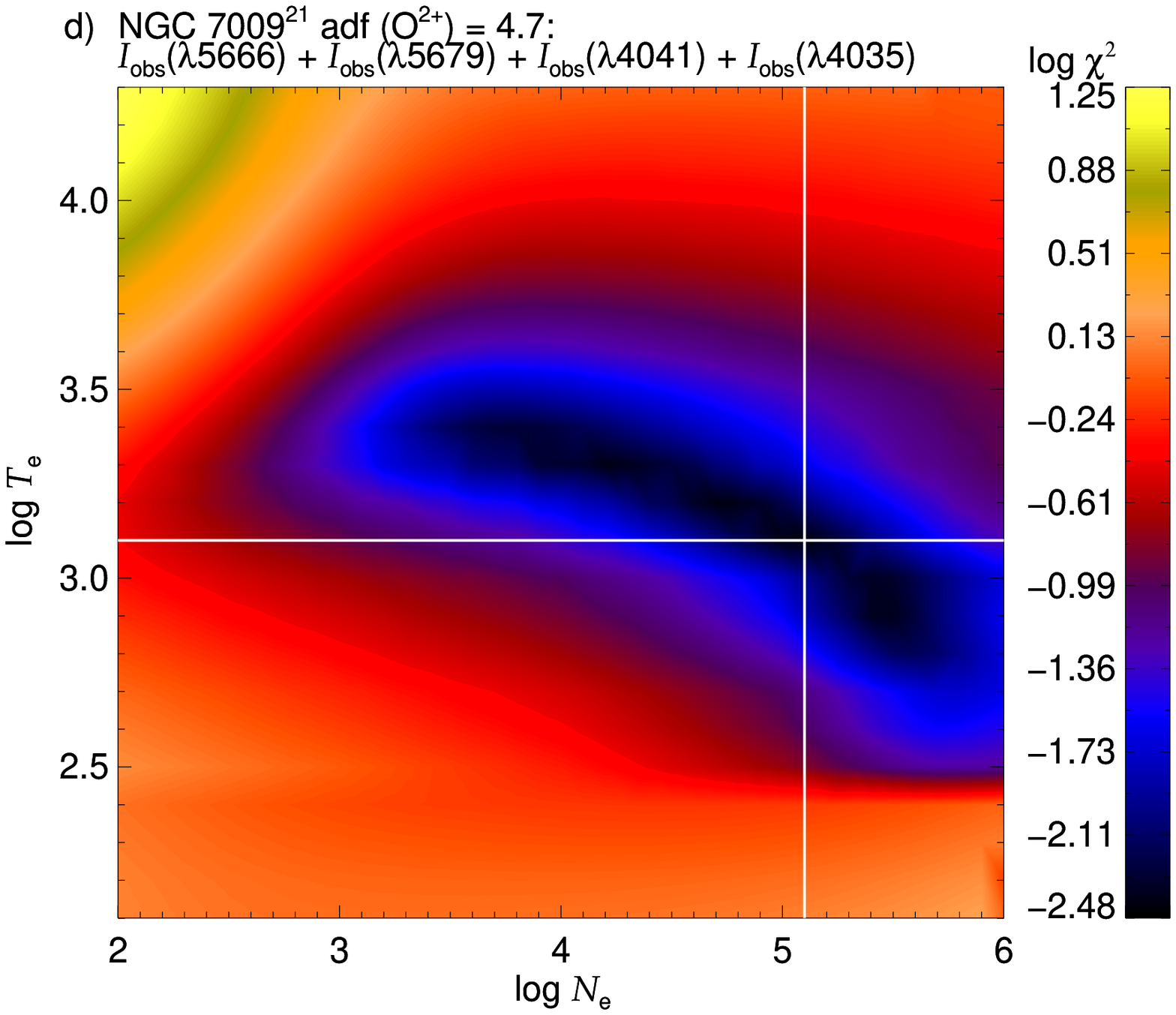}} & & & \\
\end{tabular}
\vskip-0.1truein
\caption[]{Log$\chi^2\left(T_{e},N_{e} \right)$ distributions for N~{\sc ii}~
V3 $\lambda$5666, V3 $\lambda$5679, V39b $\lambda$4041 and V39a $\lambda$4035
over the entire $\log$~\tel~-~$\log$~\nel~grid for 4 PNe.  The white crosshair
pinpoints the absolute minimum \chisq~value for the N~{\sc ii}~line
combinations indicating the optimal $\log$~{\tel} and $\log$~{\nel}.  The colour
bars indicate the $\log$~\chisq~values. ADF values are from literature: a)
Hf\,2-2\footnotemark[1]{(\citealt{LBZ06})},
b) M\,1-42\footnotemark[2]{(\citealt{LLB01})},
c) NGC\,6153\footnotemark[15]{(\citealt{LSB00})}, and 
d) NGC\,7009\footnotemark[21]{(\citealt{FL11})}.}
\end{center}
\end{minipage}
\end{figure*}

\begin{figure*}
\begin{minipage}{175mm}
\begin{center}
\vskip0.075truein
\begin{tabular}{cccccc}
\resizebox{7.75cm}{!}{\includegraphics{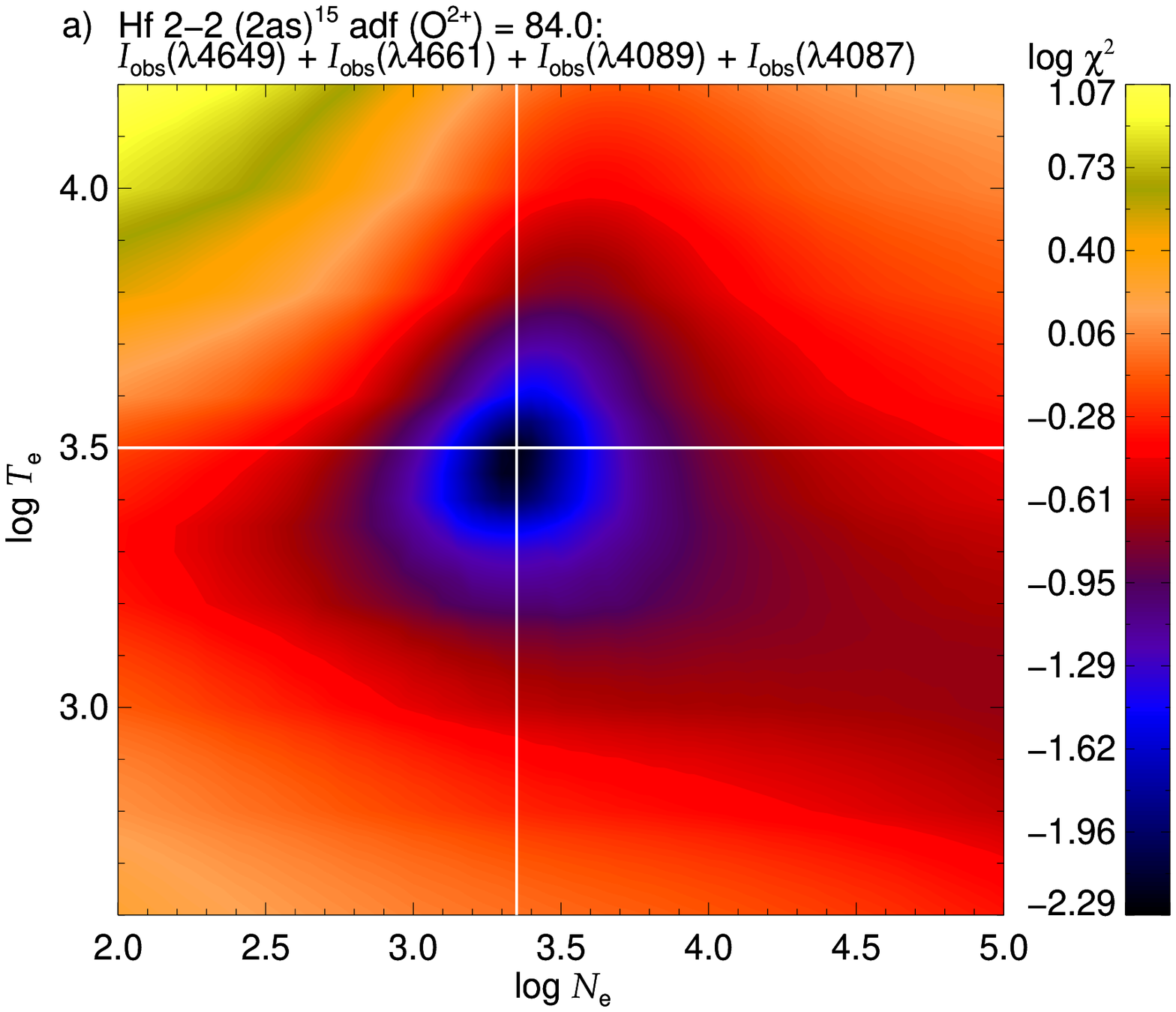}} & & 
\resizebox{7.75cm}{!}{\includegraphics{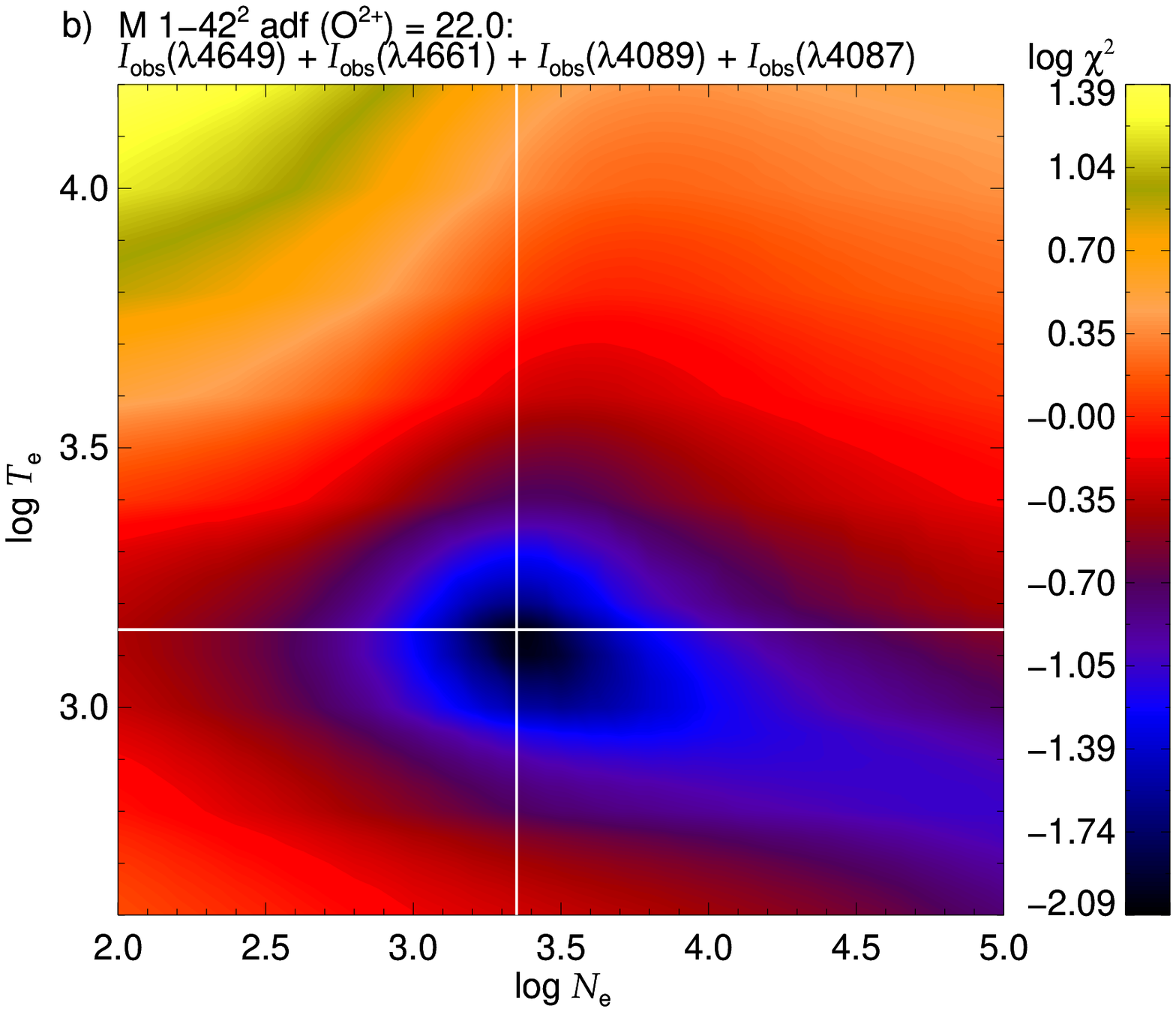}} & & & \\
\end{tabular}
\vskip0.2truein
\begin{tabular}{cccccc}
\resizebox{7.75cm}{!}{\includegraphics{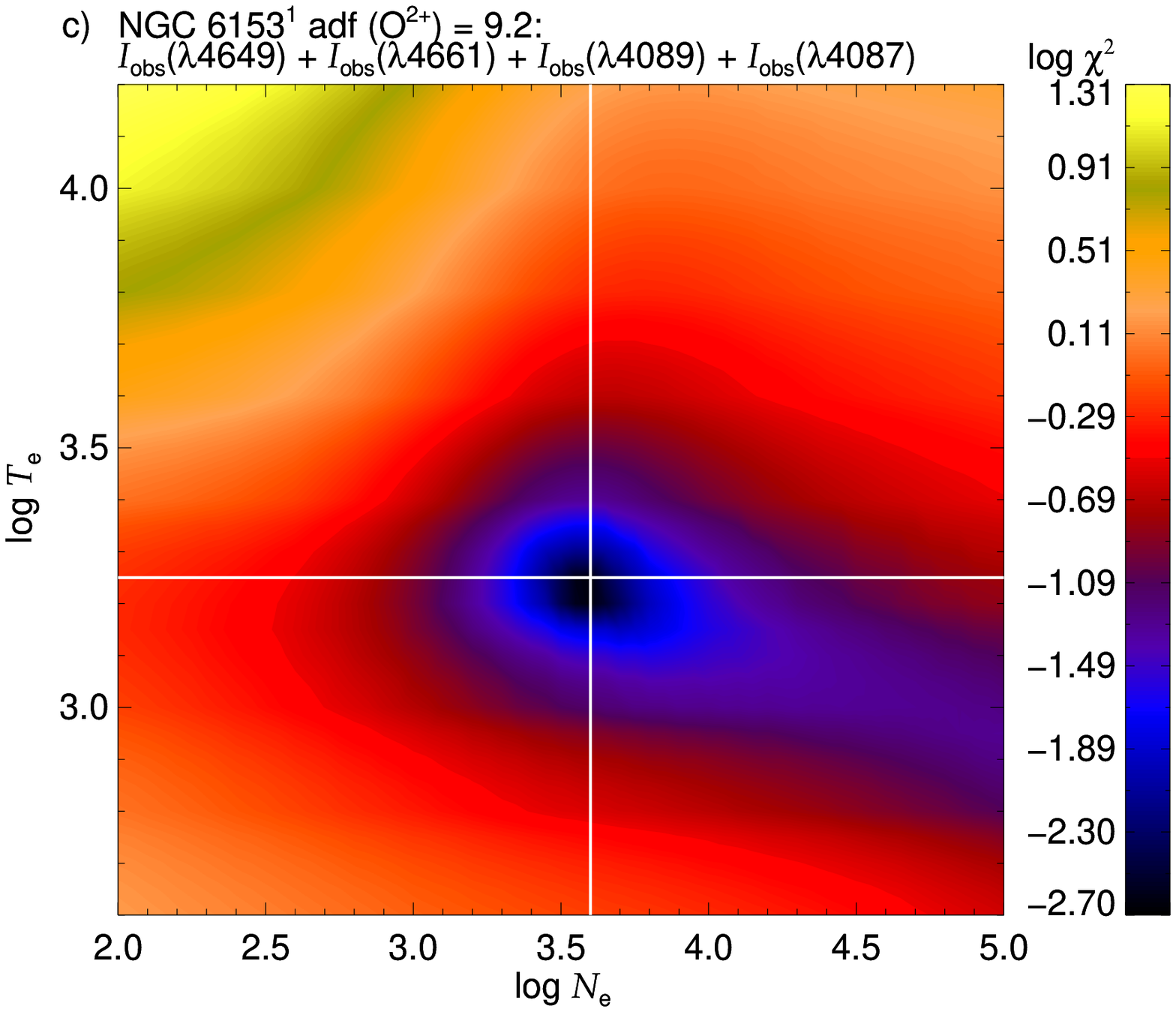}} & & 
\resizebox{7.75cm}{!}{\includegraphics{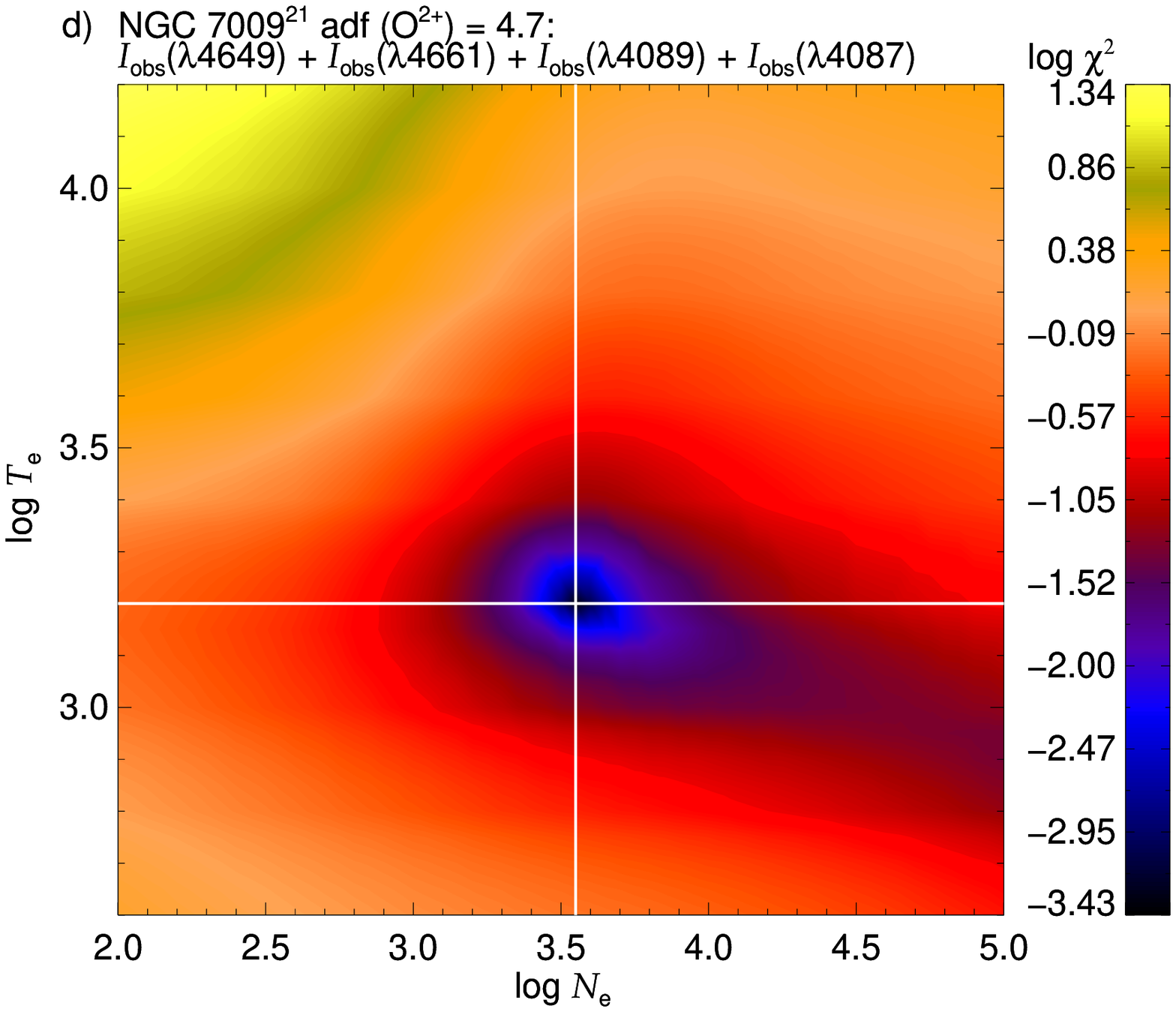}} & & & \\
\end{tabular}
\vskip-0.1truein
\caption[]{The same as Fig.\,6 but for O~{\sc ii}~V1 $\lambda$4649, V1
$\lambda$4662, V48a $\lambda$4089 and V48c $\lambda$4087 lines over the entire
$\log$~\tel~-~$\log$~\nel~grid for the same 4 PNe.}
\end{center}
\end{minipage}
\end{figure*}

\begin{figure*}
\begin{minipage}{175mm}
\begin{center}
\vskip0.075truein
\begin{tabular}{cccccc}
\resizebox{7.75cm}{!}{\includegraphics{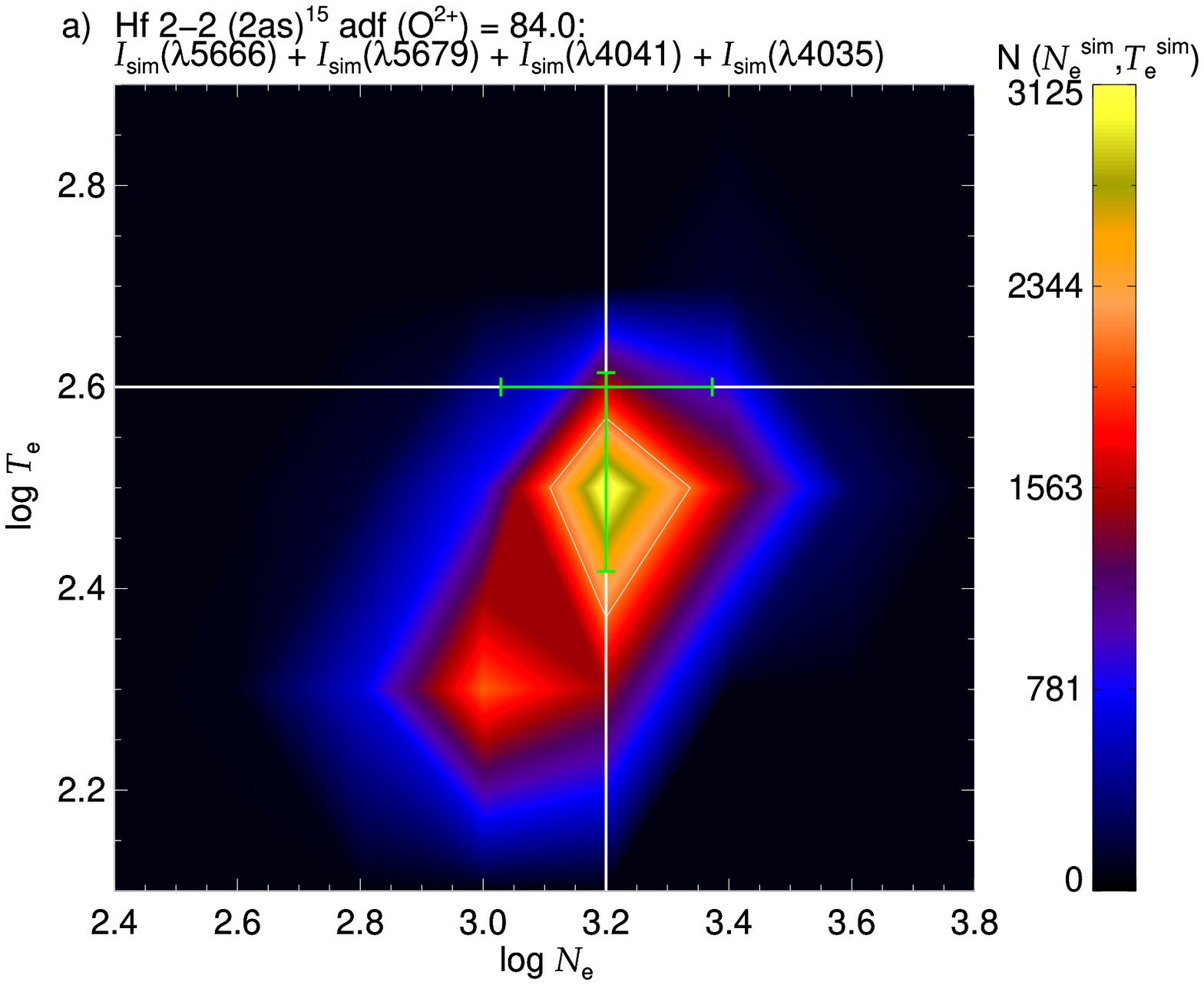}} & & 
\resizebox{7.75cm}{!}{\includegraphics{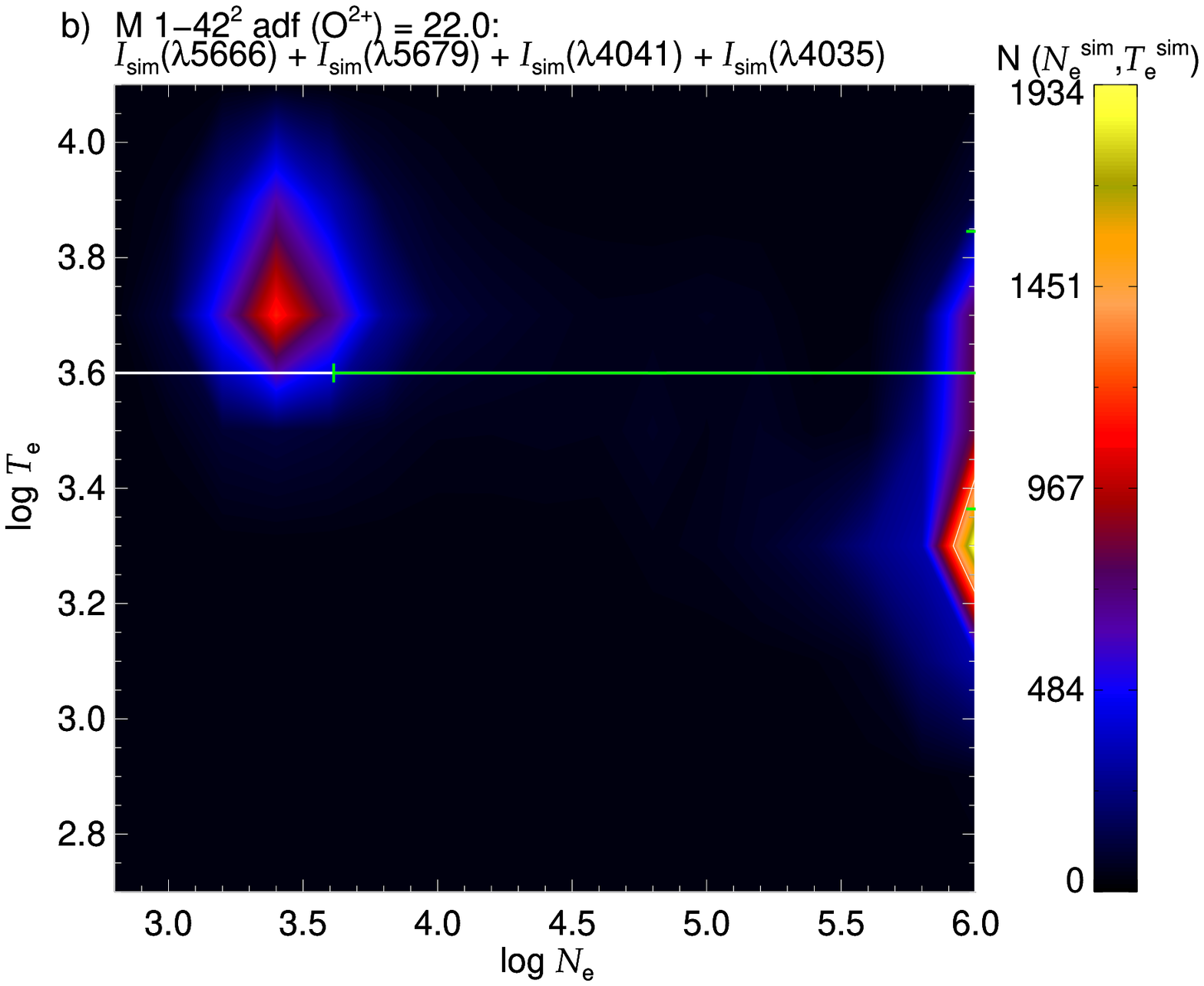}} & & & \\
\end{tabular}
\vskip0.2truein
\begin{tabular}{cccccc}
\resizebox{7.75cm}{!}{\includegraphics{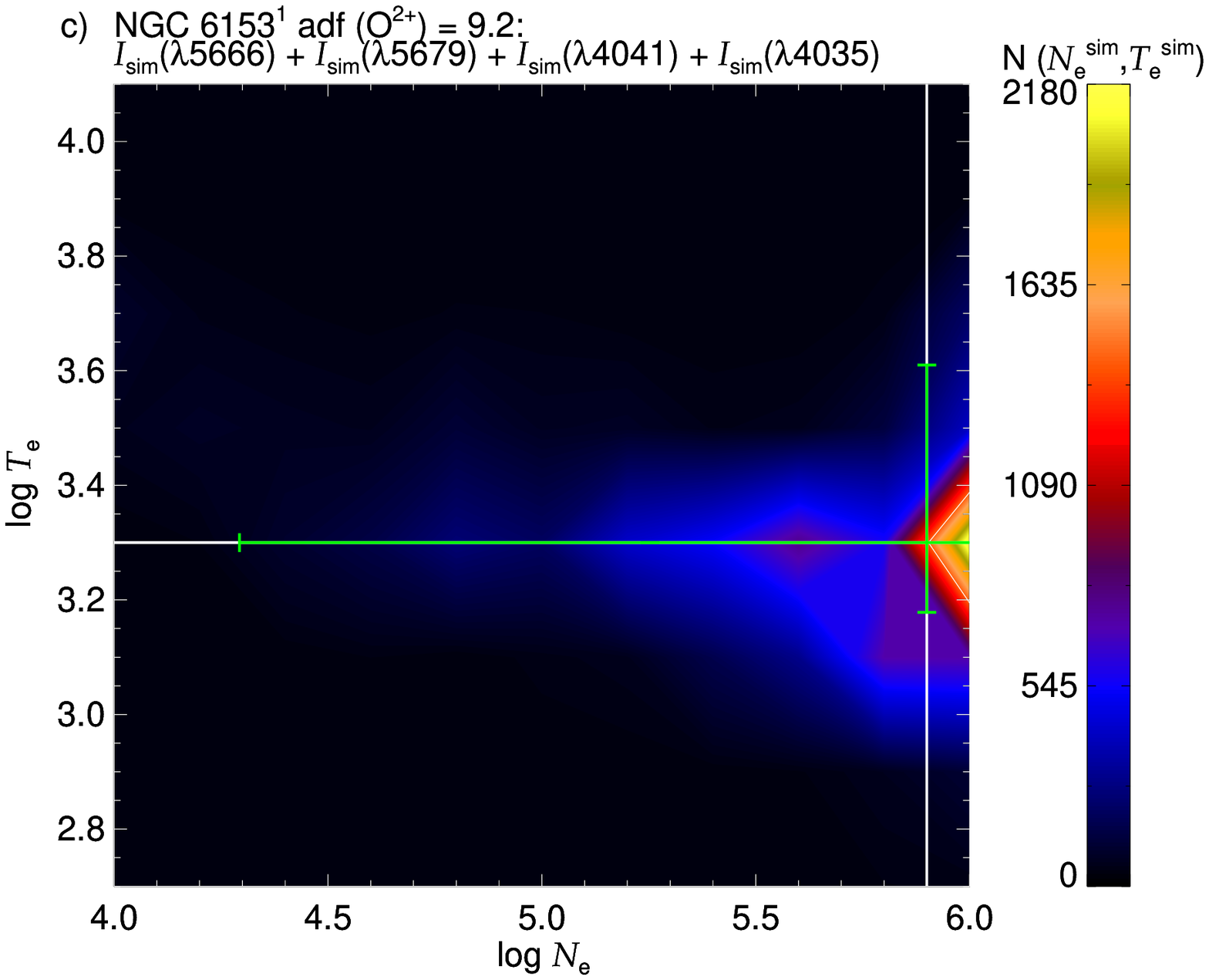}} & & 
\resizebox{7.75cm}{!}{\includegraphics{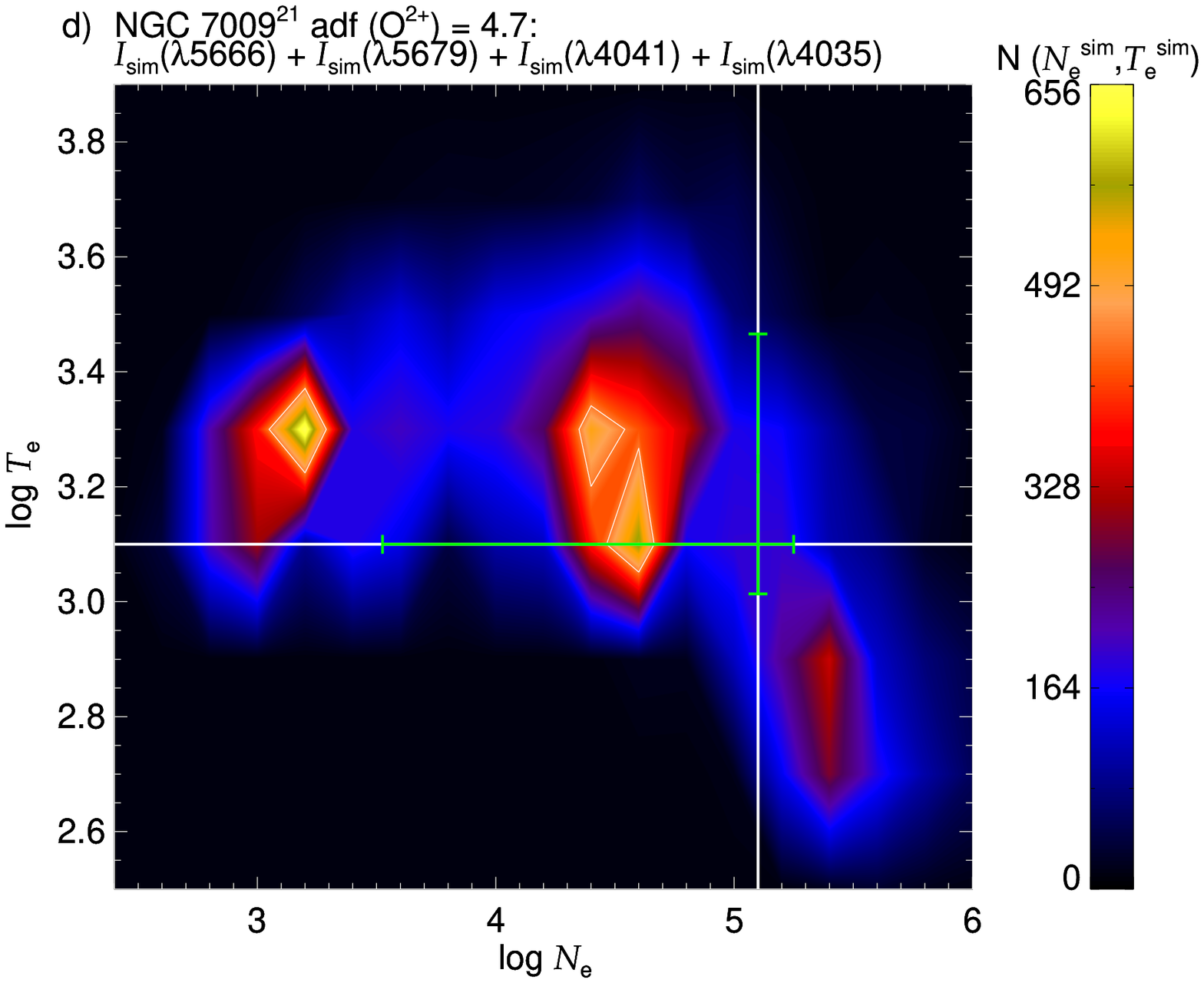}} & & & \\
\end{tabular}
\vskip-0.1truein
\caption[]{2-dimensional frequencies of the optimal $\log$~{\tel} and
$\log$~{\nel} locations calculated from 10\,000 simulations within a 1-$\sigma$
normal (Gaussian) distribution for N~{\sc ii}~V3 $\lambda$5666, V3
$\lambda$5679, V39b $\lambda$4041 and V39a $\lambda$4035 over a portion of the
$\log$\tel~-~$\log$~\nel~grid for the same 4 PNe as Figs\,5 and 6, sorted by
descending ADF. The white crosshair pinpoints the optimal $\log$~{\tel}~and
$\log$~{\nel}~location obtained from the observed intensities. The light-green
error bars also show the standard deviations as derived from Eqn.\,6. The ADF
values are adopted from literature:
a) Hf\,2-2\footnotemark[1] {(\citealt{LBZ06})},
b) M\,1-42\footnotemark[2] {(\citealt{LLB01})},
c) NGC\,6153\footnotemark[15] {(\citealt{LSB00})}, and
d) NGC\,7009\footnotemark[21] {(\citealt{FL11})}.}
\end{center}
\end{minipage}
\end{figure*}

\begin{figure*}
\vskip0.075truein
\begin{tabular}{cccccc}
\resizebox{7.75cm}{!}{\includegraphics{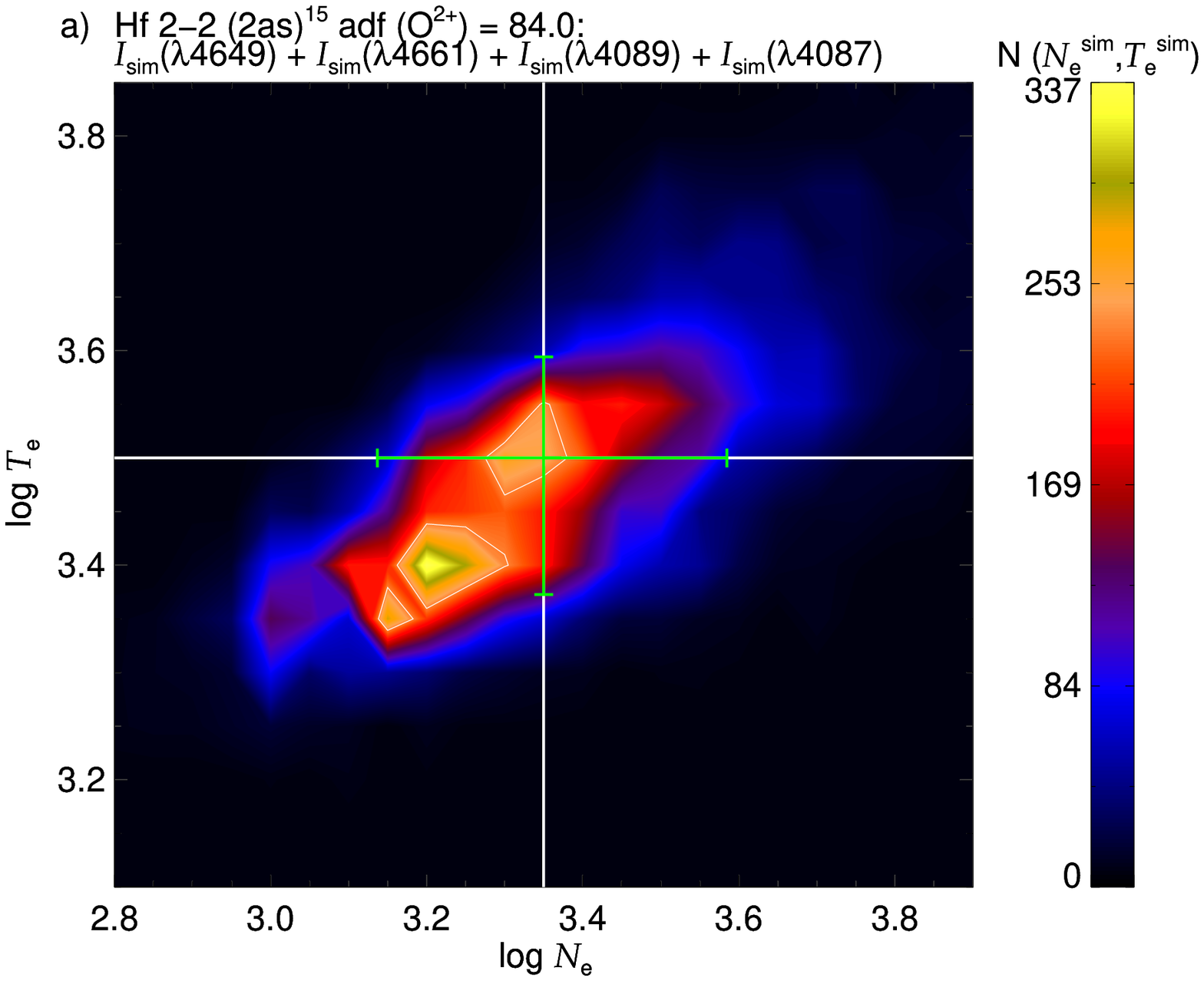}} & & 
\resizebox{7.75cm}{!}{\includegraphics{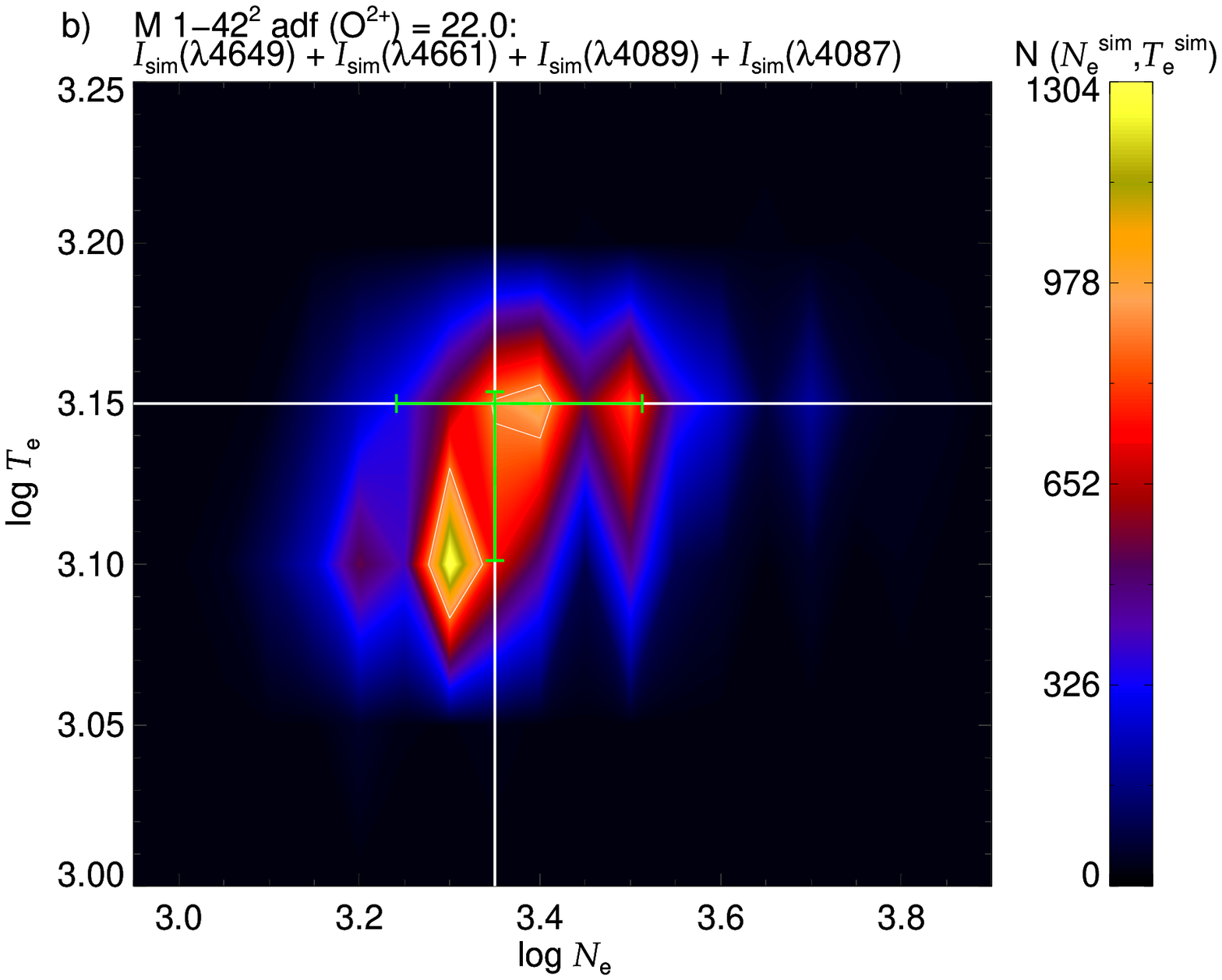}} & & & \\
\end{tabular}
\vskip0.2truein
\begin{tabular}{cccccc}
\resizebox{7.75cm}{!}{\includegraphics{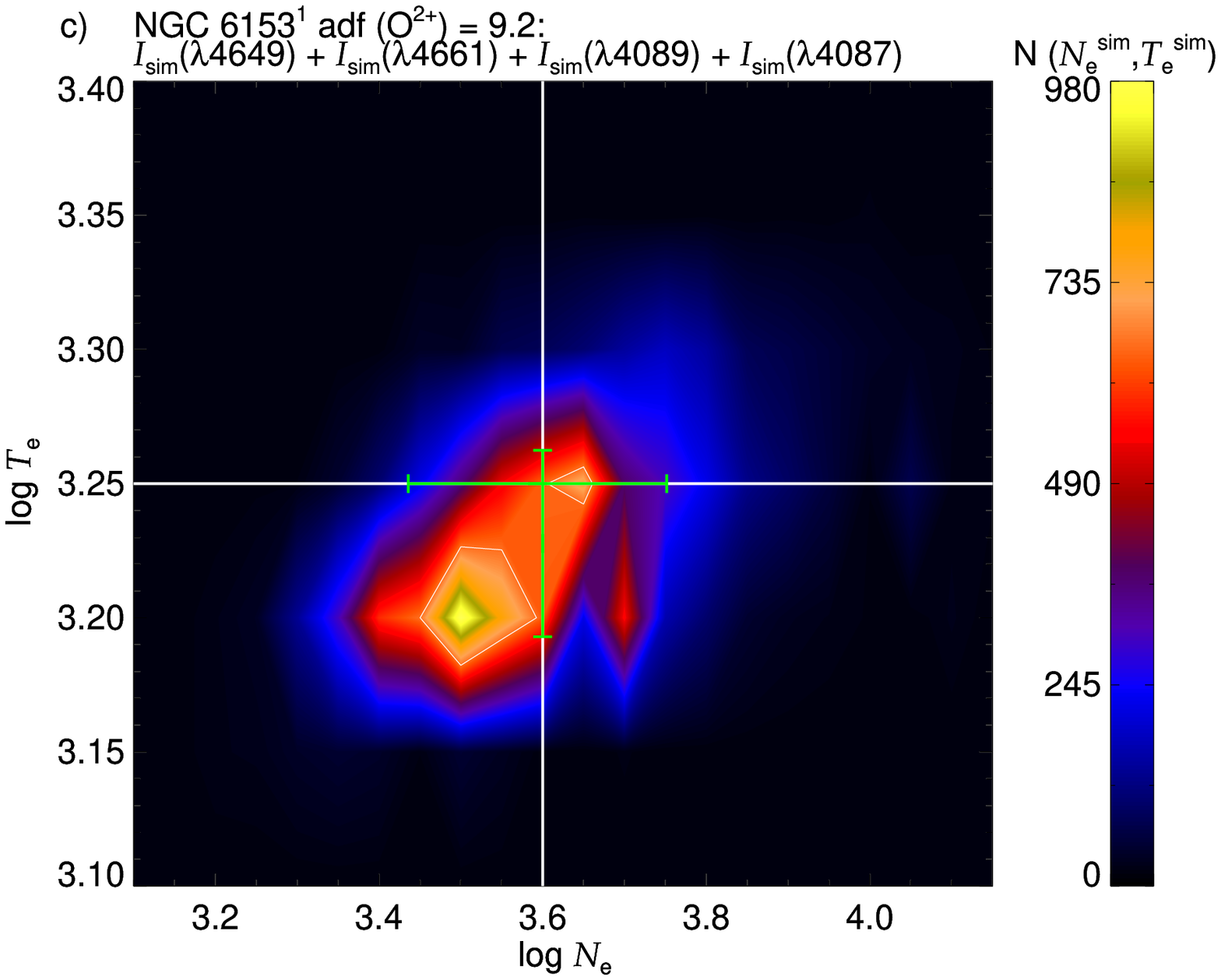}} & & 
\resizebox{7.75cm}{!}{\includegraphics{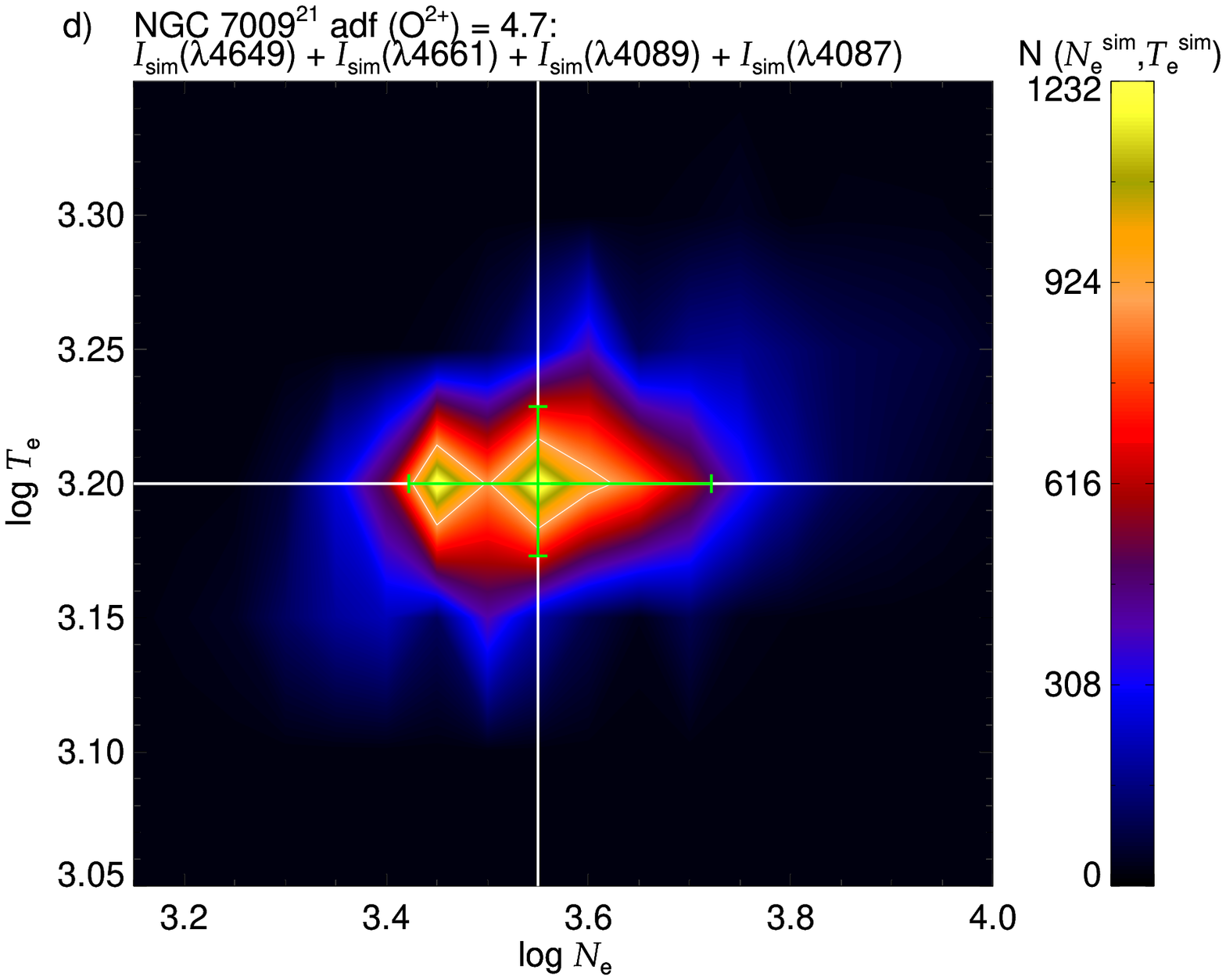}} & & & \\
\end{tabular}
\vskip-0.1truein
\caption[]{The same as Figure 8 but with O~{\sc ii}~V1 $\lambda$4649, V1
$\lambda$4662, V48a $\lambda$4089 and V48c $\lambda$4087 lines for the same 4
PNe.}
\end{figure*}

\begin{figure*}
\begin{minipage}{175mm}
\begin{center}
\vskip0.3truein
\begin{tabular}{c}
\rotatebox{90}{\resizebox{8.75cm}{!}{\includegraphics{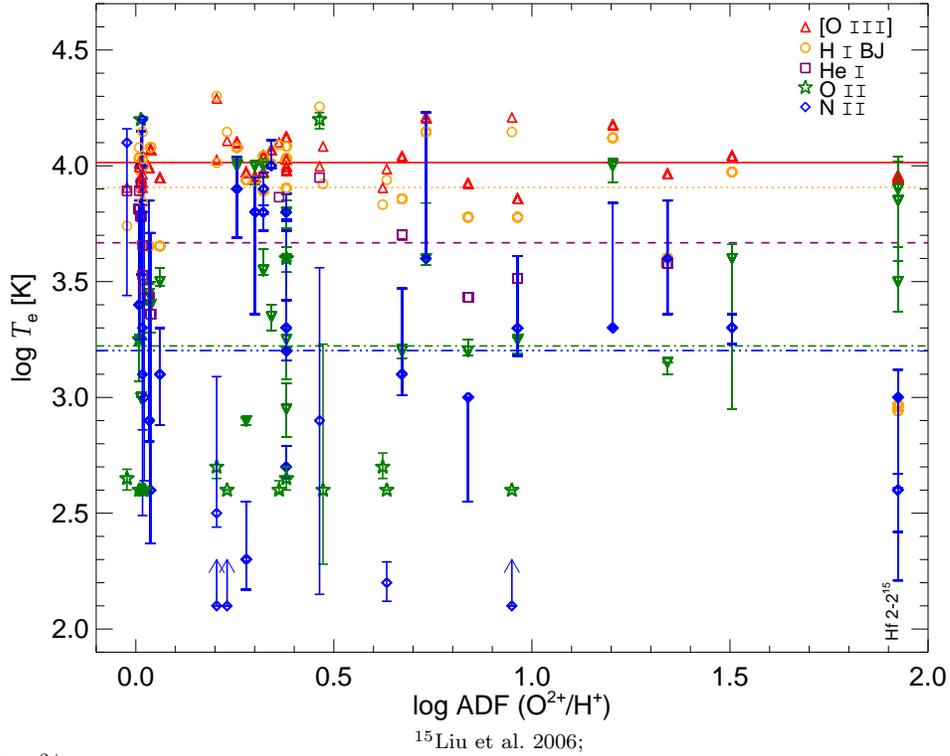}}} \\
\end{tabular}
\\
\footnotemark[15]{\citealt{LBZ06}}; 
\vskip-0.1truein
\caption[]{ADF(O$^{2+}$) values versus electron temperature using a variety of
diagnostics for 46 PNe.  Red triangles represent the log \tel([O~{\sc iii}]
values. Orange circles represent the log \tel(H~{\sc i}~BJ). Purple squares
represent the $\log$~\tel(He~{\sc i}~$\lambda$7281/$\lambda$6678). Green stars
represent the $\log$~\tel(O~{\sc ii})~from ORL diagnostics. Blue diamonds
represent the $\log$~\tel(N~{\sc ii})~from ORL diagnostics. The red solid line
indicates the mean $\log$~\tel([O~{\sc iii}])~value.  The orange dotted line
indicates the mean $\log$~\tel(H~{\sc i})~value. The purple dashed line
indicates the mean $\log$~\tel(He~{\sc i})~value. The green dot-dashed line
indicates the mean $\log$~\tel(O~{\sc ii})~value. The blue dot-dot-dot-dashed
line indicates the mean $\log$~\tel(N~{\sc ii}) value. The 3 PNe, Hu\,1-2 
(\citealt{LLB04}), NGC\,6790 (\citealt{RG05}) and NGC\,2440 (\citealt{TBL04}), 
that have $\log$~\tel(N~{\sc ii})~=~2.1 are lower limits due to
the poor quality of the data and the limited number of line intensities found
in their literature.}
\end{center}
\end{minipage}
\end{figure*}

\begin{figure*}
\begin{minipage}{175mm}
\begin{center}
\vskip0.25truein
\begin{tabular}{c}
\rotatebox{90}{\resizebox{8.75cm}{!}{\includegraphics{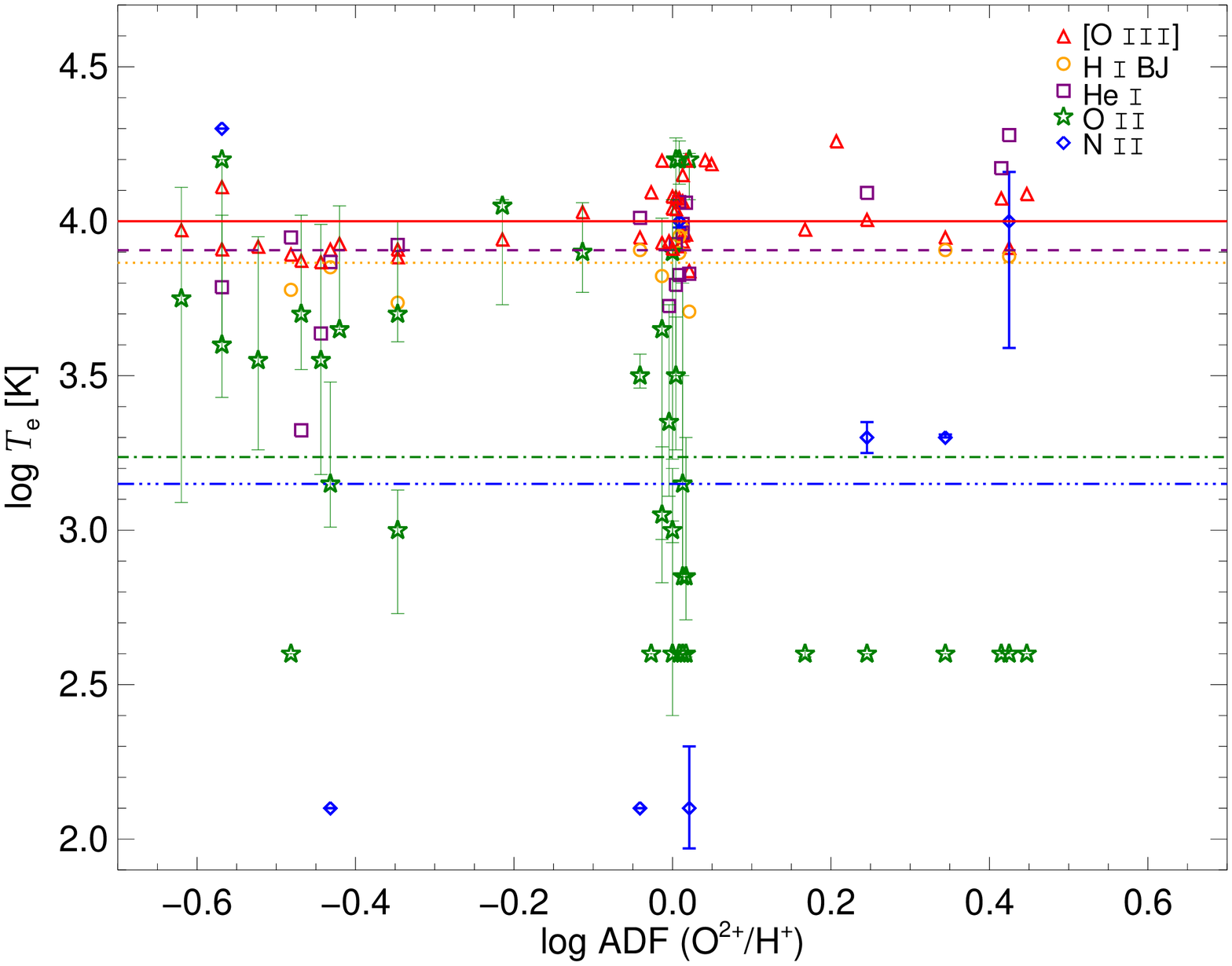}}} \\
\end{tabular}
\vskip0.05truein
\caption[]{The same as Figure\,10 but for 42 H~{\sc ii}~regions.}
\end{center}
\end{minipage}
\end{figure*}

\section{Results and Discussion}

\subsection{Temperature structure of photoionized nebulae}

Figs.\,10 and 11 show the electron temperatures from the literature and the
current ORL diagnostics for 46 PNe and for 42 H~{\sc ii}~regions, respectively.
The PNe in Fig.\,10 were chosen out of the whole list because they met the
following criteria: 1) \tel([O~{\sc iii}], \tel(H~{\sc i}) and \tel(He~{\sc
i}) are available from literature, 2) $2.6<\log$~\tel(O~{\sc ii}~ORLs)$<4.2$
and/or 3) $2.1<\log$~\tel(N~{\sc ii}~ORLs)$<4.3$.
For the PNe plotted in Fig.\,10, the mean value of $\log$~{\tel}([O~{\sc iii}])~
is 4.01$\pm$0.09\,[K]~(red solid line).  
The mean value of $\log$~{\tel}(H~{\sc i}~BJ)~is 3.91$\pm$0.29 [K]~(orange dotted line).  
The mean value of $\log$~{\tel}(He~{\sc i}~$\lambda$7281/$\lambda$6678)~is 3.67$\pm$0.19 [K]~(purple dashed line).  
The mean value of $\log$~{\tel}(O~{\sc ii}~ORLs)~is 3.22$\pm$0.52 [K]~(green dot-dashed line).  
The mean value of $\log$~{\tel}(N~{\sc ii}~ORLs)~is 3.20$\pm$0.59 [K]~(blue dot-dot-dot-dashed line).  
In the case of the PNe, the mean values clearly show the temperature sequence:
\tel(ORLs)~$\leq$~\tel(He~{\sc i})~$\leq$~\tel(H~{\sc i})~$\leq$~\tel(CELs),
which is consistent with predictions from the bi-abundance model
(\citealt{LSB00}; \citealt{L03}). While the physical conditions of the main
nebula and cold, H-deficient component may vary from nebula to nebula, over
all the nebulae there is a general trend that many contain a secondary cold,
H-deficient component.

For the H~{\sc ii}~regions plotted in Fig.\,11, the mean value of $\log$~{\tel}([O~{\sc iii}]) is 4.00$\pm$0.11 [K]~(red solid line).  
The mean value of $\log$~{\tel}(H~{\sc i}~BJ)~is 3.87$\pm$0.08 [K]~(orange dotted line).  
The mean value of $\log$~{\tel}(He~{\sc i}~$\lambda$7281/$\lambda$6678)~is 3.91 [K]~(purple dashed line).  
The mean value of $\log$~{\tel}(O~{\sc ii}~ORLs)~is 3.24$\pm$0.57 [K]~(green dot-dashed line).  
The mean value of $\log$~{\tel}(N~{\sc ii}~ORLs)~is 3.15$\pm$0.87 [K]~(blue dot-dot-dot-dashed line).  
In the case of the H~{\sc ii}~regions, the mean values show the temperature relation: 
\tel(ORLs)~$\leq$~\tel(H~{\sc i})~$\leq$~\tel(He{\sc i})~$\leq$~\tel(CELs).  
The difference in this relation are most likely due to the faint nature and 
large measurement errors of He~{\sc i}~lines in some of the H~{\sc ii}~regions.

\subsection{Physical evolution of photoionized nebulae}

Fig.\,12 shows histograms of \tel([O~{\sc iii}]), \tel(H~{\sc i}~BJ), \tel(He~{\sc i}), 
\tel(O~{\sc ii})~and \tel(N~{\sc ii}), for PNe (blue) and H~{\sc ii}~regions (red). 
The He~{\sc i}~ORL temperatures are derived from the line ratio
$\lambda$7281/$\lambda$6678 using the fitting functions of \citet{ZLL05b}.
Given the weakness of the N~{\sc ii}~and O~{\sc ii}~ORLs, the large scatter
seen in the \tel(O~{\sc ii})~and \tel(N~{\sc ii})~distributions are most likely 
due to observational uncertainties. The diagram
clearly shows the relation of the average temperatures for the PNe,
\tel(ORLs)~$\leq$ \tel(He~{\sc i})~$\leq$~\tel(H~{\sc i})~$\leq$~\tel(CELs),
which is also seen in Fig.\,10. The average \tel's and standard deviations
are labelled for PNe (in red) and H~{\sc ii}~regions (in blue). Many PNe and
H~{\sc ii}~regions in our analysis show \tel(O~{\sc ii})~lower than 2.6 and
\tel(N~{\sc ii})~lower than 2.1, the lowest \tel's for which the effective
recombination coefficients are available, respectively, although only a
handful of H~{\sc ii}~regions had N~{\sc ii}~ORLs listed in their literature 
for us to perform our diagnostics.  As can be seen in Fig.\,11, the temperature 
histograms of [O~{\sc iii}] CELs, He~{\sc i}, H~{\sc i}~BJ, O~{\sc ii}~and 
N~{\sc ii}~ORLs show an increase in the width of the distributions from 
[O~{\sc iii}]~through N~{\sc ii}.  This can be either caused by the
observational uncertainties arising from increasingly difficult measurements,
especially for the extremely faint ORLs or possibly due to variations of
physical properties of the cold H-deficient inclusions and main nebula from
object to object, causing the scatter in \tel(N~{\sc ii})~and \tel(O~{\sc ii}).

Fig.\,13 shows the ADF (O$^{2+}$/H$^{+}$) versus \tel([O~{\sc
iii}])\,--\,\tel(H~{\sc i}~BJ), for PNe (blue) and H~{\sc ii}~regions (red).
The 123 PNe analysed here are shown as the blue open circles; amongst them the
46 PNe plotted in Fig.\,10 are shown as the blue filled circles. A linear
least-squares fit to the 46 PNe from Fig.\,10 with reliable electron
temperatures indicates a positive correlation.
The PNe shown in open circles and H~{\sc ii}~regions are not considered 
for the linear fit.  This most likely indicates a tight correlation between the 
ADF of a nebulae and the temperature discrepancy. The
42 H~{\sc ii}~regions analysed here are shown as the red open squares with 40
H~{\sc ii}~regions shown as red closed squares if they had positive N~{\sc ii}~or 
O~{\sc ii}~$\log$~\tel~results. The H~{\sc ii}~regions plotted in Fig.\,13
just show large scatter, making it difficult to determine if there is a
correlation between the ADF value and temperature discrepancy.

The nebulae with lower ADF values might be relatively young in their
evolution processes when the main nebulae are very bright, and therefore the
cold, H-deficient components might be mixed with the main nebula, and the
effects of the cold, H-deficient components are insignificant. However, for the
nebulae with high ADF values, the nebula might have expanded such that the main
nebula has a relatively low surface brightness compared to the condensed, cold
H-deficient condensations, thus making it easier to distinguish the two based
on the temperatures derived from CELs and those derived from ORLs. Therefore
for most PNe, there may be an evolution from younger, more compact nebulae
where the cold, H-deficient component is embedded and almost indistinguishable
from the main bright nebula to older, more diffuse nebulae where the fainter
main nebula distinctly envelopes the cold, H-deficient condensations.

\subsection{Comments on several archetypal planetary nebulae}

\subsubsection{Hf\,2-2}

There are several well-studied planetary nebulae that have been observed 
for many decades.  Hf\,2-2 is a well--known southern PN, that has an unusually 
high ADF (O$^{2+}$) value of 84 (\citealt{LBZ06}).  They observed this PN with the 
ESO 1.52-m telescope with three separate slit widths of 2, 4, and 8 arcsec, 
using the 2-arcsec width for maximum spectral resolution.  Hf\,2-2 is known to have a 
close binary system with a very short orbital period (0.398 days, \citealt{LAB98}).  
It has also shown a very peculiar nature and may be included in the 'born-again' 
PNe scenario, relating the particularly large ADF to the phenomenon of novae, 
as stated in \citet{WLB03}.  From their 2-arcsec slit-width observations, 
\citet{LBZ06} derived the following electron temperatures: 
\tel([O~{\sc iii}])~=~8710, \tel(He~{\sc i}~$\lambda$6678/$\lambda$5876)~=~1570, and \tel(H~{\sc i})~=~933 K.  
From the plasma diagnostics based on the N~{\sc ii}~ORL lines of 
V3 $\lambda$5666, V3 $\lambda$5679, V39b $\lambda$4041, V39a $\lambda$4035, 
we derived an electron temperature of 398$^{+9}_{-135}$ K, as shown in the first panels of 
Fig.\,5 for the \chisq~minimisation and Fig.\,8 for the standard deviations.  
From the plasma diagnostics based on the O~{\sc ii}~ORLs of 
V1 $\lambda$4649, V1 $\lambda$4661, V48a $\lambda$4089, V48c $\lambda$4087, 
we derived an electron temperature of 3160$^{+730}_{-820}$ K, as shown in the first panels of 
Fig.\,7 for the \chisq~minimisation and Fig.\,9 for the standard deviations.

\subsubsection{M\,1-42}

Another PN is the Galactic bulge nebula M\,1-42, which was first discovered by
\citet{M46}. More recently, \citet{LLB01} observed M\,1-42 along with
M\,2-36 using the ESO 1.52-m telescope. For M\,1-42 they derived a fairly high
ADF of 22. \citet{LLB01} derived the following electron temperatures:
\tel([O~{\sc iii}])~=~9120~K, \tel(He~{\sc i}~$\lambda$7821/$\lambda$6678)~=
3790~K, and \tel(H~{\sc i})~=~3980~K. From our plasma diagnostics based on
the N~{\sc ii}~ORLs, V3 $\lambda$5666, V3 $\lambda$5679, V39b $\lambda$4041
and V39a $\lambda$4035, we derived an electron temperature of
3980$^{+3100}_{-1690}$~K, as shown in the second panel of Fig.\,5 for the
\chisq~minimisation and Fig.\,8 for the standard deviations. But as one can
see in the two figures, the electron density of 10$^{6}$ cm$^{-3}$ only
provides an upper limit for the diagnostics. From our plasma diagnostics based
on the O~{\sc ii}~ORLs, V1 $\lambda$4649, V1 $\lambda$4661, V48a $\lambda$4089
and V48c $\lambda$4087, we derived an electron temperature of
1410$^{~}_{-150}$~K, as shown in the second panel of Fig.\,7 for the 
\chisq~minimisation and Fig.\,9 for the standard deviations.

\subsubsection{NGC\,6153}

Over the last several decades, detailed studies have been put forth on the well-known 
planetary nebula NGC\,6153, a possibly super-metal-rich nebula and first noted by 
\citet{P84}.  \citet{LSB00} derived a moderately high ADF of 9.2.  
Deduced from their electron density distribution and the optical appearance of 
NGC\,6153, \citet{YLP11} suggested it is most likely a bipolar nebula probably 
with a central cavity and a density-enhanced waist, as viewed at a large angle to 
its polar axis.  However, they also argue that a convincing physical model accounting 
for the full range of behaviour NGC\,6153 exhibits is still to be found.  From the strong 
emission lines, they derived the following electron temperatures: 
\tel([O~{\sc iii}])~=~7240~K, \tel(He~{\sc i}~$\lambda$7821/$\lambda$6678)~=~3260~K, and \tel(H~{\sc i})~=~6030~K.  
Based on the N~{\sc ii}~ORLs of 
V3 $\lambda$5666, V3 $\lambda$5679, V39b $\lambda$4041 and V39a $\lambda$4035, 
our plasma diagnostics derived an electron temperature of 2000$^{+2080}_{-480}$~K, 
as shown in the third panel of Fig.\,6 for the \chisq~minimisation and Fig.\,8 for the standard deviations.
Based on the O~{\sc ii}~ORLs of 
V1 $\lambda$4649, V1 $\lambda$4661, V48a $\lambda$4089 and V48c $\lambda$4087, 
our plasma diagnostics yields an electron temperature of 1780$^{+40}_{-230}$~K, 
as shown in the third panel of Fig.\,7 for the \chisq~minimisation and Fig.\,9 for the standard deviations.  

\subsubsection{NGC\,7009}

NGC\,7009 has had hundreds of publications over the last 50 years or more.  
Also known as the Saturn Nebula, this large, double-ringed, high-surface-brightness nebula 
is particularly well known for having an unusually rich and strong O~{\sc ii}~optical permitted 
lines ever since the early high-resolution photographic spectroscopy observations of 
\citet{W42} and \citet{AK64}, having observed more than 100 O~{\sc ii}~permitted 
transitions (\citealt{LSB95}).  A much more recent work by \citet{FL11} has made use of 
very deep CCD spectrum of the Saturn Nebula covering from 3040 to 11\,000\,\AA.  They derived 
a reasonably high ADF of about 5.  They identified over 1000 emission lines with 81 per cent 
attributed to permitted lines and more than 200 O~{\sc ii}~permitted lines.  Their observations 
were made using the ESO 1.52 m telescope and the William Herschel Telescope (WHT), and securing all spectra with 
a long slit.  Table\,$3$ in their paper lists a very complete set of electron temperatures and 
densities derived from numerous line ratios.  They derived the following electron temperatures: 
\tel([O~{\sc iii}])~=~9810~K, \tel(He~{\sc i}~$\lambda$7821/$\lambda$6678)~=~5100~K, and \tel(H~{\sc i})~=~6420~K.
From our plasma diagnostics based on the N~{\sc ii}~lines of
V3 $\lambda$5666, V3 $\lambda$5679, V39b $\lambda$4041 and V39a $\lambda$4035,
we derived an electron temperature of 1260$^{+1690}_{-235}$~K, as shown in the
fourth panel of Fig.\,6 for the \chisq~minimisation and Fig.\,8 for the
standard deviations.  Due to the data quality, our \chisq~minimisation technique only 
provides an upper limit to the electron density for N~{\sc ii}~ORLs.  But judging from the 
2-D distributions of the results derived from the randomly generated simulated intensities, 
a secondary peak occurs at an electron density of 1580~cm$^{-3}$ and a temperature of 2000~K.  
From our plasma diagnostics based on the O~{\sc ii}~lines of 
V1 $\lambda$4649, V1 $\lambda$4661, V48a $\lambda$4089 and V48c $\lambda$4087, 
we derived an electron temperature of 1580$^{+113}_{-106}$ K, as shown in the fourth panel of 
Fig.\,7 for the \chisq~minimisation and Fig.\,9 for the standard deviations.

\begin{figure}
\begin{minipage}{85mm}
\vskip0.7truein
\begin{tabular}{c}
\resizebox{8cm}{!}{\includegraphics{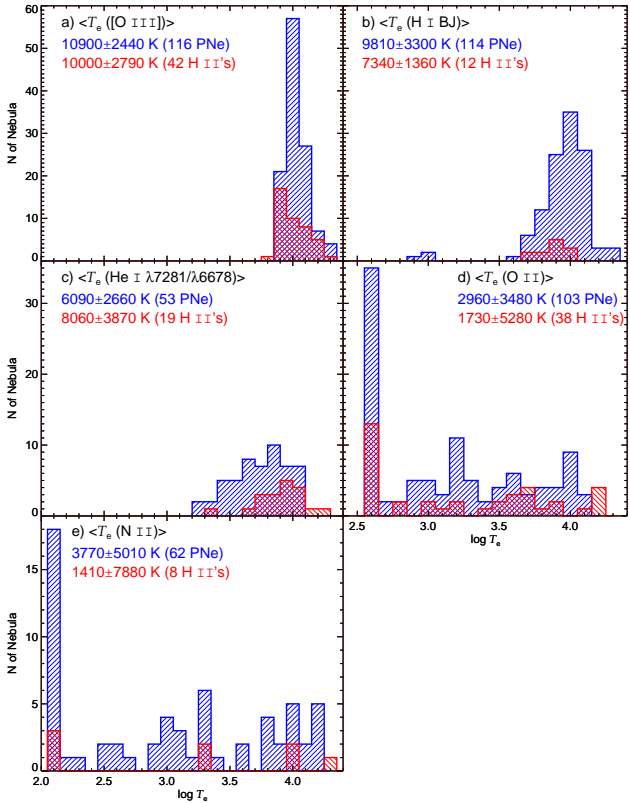}} \\
\end{tabular}
\vskip-0.1truein
\caption[]{Distributions of \tel~derived from a) [O~{\sc iii}] CELs; b) H~{\sc i}~Balmer Jump; c) He~{\sc i}~line ratio $\lambda$7281/$\lambda$6678 from literature and our own calculations; d) O~{\sc ii}~ORL diagnostics; and e) N~{\sc ii}~ORL diagnostics for PNe (blue) and H~{\sc ii}~regions (red).  For the distribution in each panel, the average \tel~and standard deviations are labelled for PNe (blue) and H~{\sc ii}~regions (red).}
\end{minipage}
\end{figure}

\begin{figure}
\begin{minipage}{85mm}
\begin{tabular}{c} 
\resizebox{7.5cm}{!}{\includegraphics{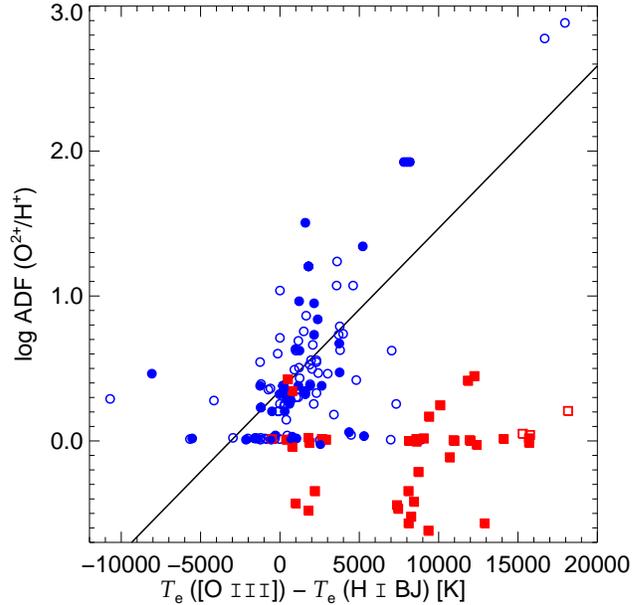}} \\
\end{tabular}
\vskip-0.1truein
\caption[]{Log ADF (O$^{2+}$/H$^{+}$) versus \tel([O~{\sc iii}]) - \tel(H~{\sc i}~BJ)~relation for PNe (blue) and H~{\sc ii}~regions (red).  The blue open circles are all 123 PNe listed in Table 1.  The blue closed circles are all the 46 PNe plotted in Fig.\,10.  The solid black line shows a least-squares fit to the 46 PNe plotted in Fig.\,10.  The red open squares are all 42 H~{\sc ii}~regions listed in Table 2.  The red closed squares are H~{\sc ii}~regions with either $\log$~\tel(O~{\sc ii})~or $\log$~\tel(N~{\sc ii})~greater than zero.}
\end{minipage}
\end{figure}

\subsubsection{Abell\,58 \& Abell\,30}

Another rather peculiar planetary nebula is Abell\,58, a ``born-again''
planetary nebula, known to contain an H-deficient knot surrounding V605\,Aql,
according to \citet{WBL08}. Hidden by a thick dusty torus, V605\,Aql
cannot be seen directly. However, from the surface abundances derived for the
star from observations of its scattered light, \citet{CKP06} found it
to be a typical Wolf-Rayet central star of PN. \citet{WBL08} did not
measure any N~{\sc ii}~ORLs in Abell\,58. From the plasma diagnostics based on the O~{\sc
ii} ORLs of V1 $\lambda$4649, V1 $\lambda$4661, V10 $\lambda$4075 and
V10 $\lambda$4069.62 (blended with V10 $\lambda$4069.89), we derived an upper
limit to the electron temperature ($\sim$15\,800~K).

\citet{WLB03} have provided detailed optical analysis of two of the
H-deficient knots (J1 \& J3) in another ``born-again'' PN Abell\,30. This
nebula consists of a rather large, $~$120 arcsec across, spherical shell of
low surface brightness and several bright clumps, first discovered by \citet{J79} 
and \citet{HTF80}, of material about 10 arcsec away from the
central star. For \citet{WLB03}, they observed Abell 30 with the 4.2~m
WHT at the Observatorio del Roque de los Muchachos, on La Palma, Spain, with
a 0.82 arcsec slit width. These two knots have extremely high ADF values of
598 and 766 for J1 and J3, respectively.

\citet{WLB03} derived the following electron temperatures for knot J1:
\tel([O~{\sc iii}])~=~20\,800~K, and
\tel(He~{\sc i}~$\lambda$5876/$\lambda$4471)~=~350~K. \citet{WLB03}
did not measure any N~{\sc ii}~ORLs in the knot J1. From the
plasma diagnostics based on the O~{\sc ii}~ORLs of V1 $\lambda$4649,
V1 $\lambda$4661, V10 $\lambda$4075 and V10 $\lambda$4069.62 (blended with
V10 $\lambda$4069.89), we derived an upper limit to the electron temperature
($\sim$15\,800~K). For knot J3, \citet{WLB03} derived the following
electron temperatures: \tel([O~{\sc iii}])~=~17\,960~K, \tel(He~{\sc i}~
$\lambda$6678/$\lambda$4471)~=~9\,240~K. They didn't measure any N~{\sc ii}~ORLs 
in this knot either. Plasma diagnostics based on the O~{\sc ii}~ORLs
yielded an upper limit to the electron temperature ($\sim$15\,800~K).

\subsection{Results and discussion}

Tables\,1 and 2 list the electron temperatures and densities for the 167
nebulae, including 123 PNe and 42 H~{\sc ii}~regions, respectively. These
objects are presented in descending order of ADF (O$^{2+}$/H$^{+}$).
Column\,1 is the nebula name.
Column\,2 contains the ADF values from literature.
Column\,3 is the $\log$~\tel([O~{\sc iii}])~[K]~value from literature.
Column\,4 is the $\log$~\tel(BJ)~[K]~value from literature.
Column\,5 is the $\log$~\tel(N~{\sc ii})~[K]~result from our plasma diagnostics.
Column\,6 is the $\log$~\nel(N~{\sc ii})~[cm$^{-3}$]~result from our plasma
diagnostics. Column\,7 is the $\log$~\tel(O~{\sc ii})~[K]~result from our
plasma diagnostics. Column\,8 is the $\log$~\nel(O~{\sc ii})~[cm$^{-3}$]~result
from our plasma diagnostics.

Detailed modelling for the He~{\sc i}~recombination lines have been computed
for a couple decades now. \citet{BSS99} combined the \citet{S96}
models with collisional transitions of \citet{SB93}
producing accurate emissivities. More recently, \citet{ZLL05b} applied
computational models for the He~{\sc i}~recombination lines for several dozen
PNe. They also provide fitting functions spanning from 5\,000 to 20\,000~K
with fitting parameters that are density-dependent.

Tables\,3 and 4 show the electron temperatures derived from the He~{\sc i}~ORLs 
for PNe and H~{\sc ii}~regions, respectively. The three He~{\sc i}~line
ratios $\lambda$7281/$\lambda$6678, $\lambda$6678/$\lambda$4471 and
$\lambda$6678/$\lambda$5876 are used, and the objects are in descending order
of the ADF values. The PN diagnostic results from \citet{ZLL05b} are
adopted. For those PNe that are not included in the sample of \citet{ZLL05b}, 
we derived electron temperatures using the fitting formula of \citet{ZLL05b}, 
and the He~{\sc i}~line intensities are adopted from
literature. Column\,1 is the nebula name. Column\,2 is the ADF value from
literature. Columns\,3 through \,5 are the \tel's derived from the He~{\sc i}~line 
ratios. Most literatures provided line intensities for the four He~{\sc i}~lines 
$\lambda$7281, $\lambda$6678, $\lambda$5876 and $\lambda$4471.

Resonance fluorescence may also have a minor effect on the strengths of ORLs.
\citet{G76} discussed in detail the effects of resonance fluorescence on
permitted transitions.
It has been known that the N~{\sc ii}~permitted lines from the
low-lying 3d\,--\,3p and 3p\,--\,3s triplet arrays, whose upper levels are
linked to the ground term 2p$^2$~$^3$P by resonance lines, can be enhanced by
fluorescence excitation. \citet{G76} used photoionization models to study
the excitation mechanisms of permitted transitions from common heavy element
ions observed in the spectra of the Orion nebula and two Galactic PNe
NGC\,7027 and NGC\,7662. 

\citet{G76} found that while the N~{\sc ii}~M28 3d~$^3$D$^{\rm o}$\,--\,3p~$^3$P 
multiplet is excited by both recombination and continuum
fluorescence of the starlight, emission of the N~{\sc ii}~M3
3p~$^3$D\,--\,3s~$^3$P$^{\rm o}$, M5 3p~$^3$P\,--\,3s~$^3$P$^{\rm o}$~and M30
4s~$^3$P$^{\rm o}$\,--\,3p~$^3$P~multiplets are dominated by fluorescence
excitation of the N~{\sc ii}~4s~$^3$P$^{\rm o}_{1}$~level by the 
He~{\sc i}~1s8p~$^1$p$^{\rm o}_{1}$\,--\,1s$^2$~$^1$S$_{0}$~$\lambda$508.643~resonance
line, which coincides in wavelength with the 
N~{\sc ii}~2p4s~$^3$P$^{\rm o}_{1}$\,--\,2p$^2$~$^3$P$_{0}$~$\lambda$508.668~line.  
Fluorescence excitation cannot excite the singlet transitions or transitions from the
3d\,--\,4f configuration. Given the fact that measurements of the N~{\sc ii}~ORLs 
for the majority of nebulae samples are of relatively large uncertainty
due to weakness and blend, the effects of fluorescence on the plasma
diagnostic results are insignificant. For those nebulae with good observations,
e.g., the several archetypal PNe discussed in Section\,3.3, detailed modelling
is needed to estimate the exact contribution of fluorescence mechanisms.
\citet{G76} showed that the dominant excitation mechanism of the strongest
O~{\sc ii}~permitted lines is recombination.

\section{Summary}

For nearly four decades, analyses of gaseous nebulae are mainly based on the
bright CELs, whose emissivities are acutely sensitive to electron temperature
(e.g. \citealt{OF06}). Thus the abundances derived from CELs
could be subject to the uncertainties caused by temperature fluctuations
if present in the nebulae. The ORLs of heavy element ions, although much
weaker by nature, are much less sensitive to temperature, and the resultant
abundances are supposed to be more reliable, provided that very accurate
measurements of the heavy element ORLs can be obtained. In the current paper,
we demonstrate a new plasma diagnostic method based on the N~{\sc ii}~and
O~{\sc ii}~optical recombination spectra, using the new effective
recombination coefficients. This is the first work devoted to nebular analysis
using heavy element ORLs, with a large sample of PNe and H~{\sc ii}~regions
considered. The results show systematic differences between the electron
temperatures derived from CELs, the optical recombination spectra of H~{\sc i}~
and He~{\sc i}, and the N~{\sc ii}~and O~{\sc ii}~ORLs. The observed
temperature sequence has been found in previous observations, and is in
agreement with the expectation of the bi-abundance nebular model (\citealt{LSB00}).
Although very deep, high-resolution spectra of gaseous nebulae, especially
accurate measurements of heavy element ORLs, are still rare, we need to
develop plasma diagnostic tools based on the heavy element ORLs for future
study, given that the most comprehensive treatment of the N~{\sc ii}~and
O~{\sc ii}~recombination under the physical conditions of gaseous nebulae
(i.e., the effective recombination coefficients for the N~{\sc ii}~and O~{\sc ii}~
recombination spectra) are now available. The main purpose of this paper
is to demonstrate the new method of nebular analysis and show its potential
application.

The method presented in the current work for determining \tel~and \nel~from
multiple transitions of N~{\sc ii}~and O~{\sc ii}~is general and quite
promising for the future deep spectroscopic studies of gaseous nebulae.
Transitions from multiplets V3 and V39 for N~{\sc ii}~and from V1, V10 and
V48 for O~{\sc ii}~tend to be the strongest and most reliable to constrain
the \chisq~distribution over the entire $\log$~\tel~-~$\log$~\nel~grid.
However, some other N~{\sc ii}~and O~{\sc ii}~transitions can be used as
well. Therefore, this method can potentially cover broad wavelength ranges
of a spectrum instead of just determining the \tel~and \nel~from just a
couple of lines, depending on the data quality. Given that the heavy element
ORLs are intrinsically faint, high signal-to-noise ratios are needed to
improve the data quality.

\section*{Acknowledgments}

IAM thanks M.~B.~N. Kouwenhoven and Rub\'{e}n Garc\'{i}a-Benito of KIAA for some helpful suggestions.

\clearpage
\setcounter{table}{0}
\begin{table*}
\centering
\setlength{\tabcolsep}{0.015in}
\caption{Electron temperatures and densities for PNe.}
\vskip-0.1truein
\begin{tabular}{lrcccccc}
\hline
& & \multicolumn{2}{l}{From Literature} & \multicolumn{4}{l}{From Current Work} \\
\multicolumn{1}{l}{Object} & ADF (O$^{2+}$) & $\log$~\tel([O~{\sc iii}]) & $\log$~\tel(BJ) & $\log$~\tel(N~{\sc ii}) & $\log$~\tel(N~{\sc ii}) & $\log$~\tel(O~{\sc ii}) & $\log$~\tel(O~{\sc ii}) \\
& ~ & [K] & [K] & [K] & [cm$^{-3}$] & [K] & [cm$^{-3}$] \\
\hline
       Abell 30 J3\footnotemark[20] & 766.00 & 4.25\footnotemark[20] & ~ & 4.30$^{0.01}_{0.01}$ & 2.00$^{~}_{~}$ & 4.20$^{~}_{~}$ & 2.00$^{~}_{~}$ \\ 
       Abell 30 J1\footnotemark[20] & 598.00 & 4.22\footnotemark[20] & ~ & 4.30$^{~}_{~}$ & 2.00$^{~}_{~}$ & 4.20$^{~}_{~}$ & 2.00$^{~}_{~}$ \\ 
          Abell 58\footnotemark[31] &  89.00 & ~ & ~ & ~$^{~}_{~}$ & ~$^{~}_{~}$ & 4.20$^{~}_{~}$ & 2.00$^{~}_{~}$ \\ 
        Hf 2-2 (2 arcsec)\footnotemark[15] &  84.00 & 3.94\footnotemark[15] & 2.97\footnotemark[15] & 2.60$^{0.01}_{0.18}$ & 3.20$^{0.17}_{0.17}$ & 3.50$^{0.09}_{0.13}$ & 3.35$^{0.23}_{0.21}$ \\ 
        Hf 2-2 (4 arcsec)\footnotemark[15] &  84.00 & 3.96\footnotemark[15] & 2.94\footnotemark[15] & 3.00$^{0.12}_{0.33}$ & 2.70$^{1.51}_{0.65}$ & 3.90$^{0.14}_{0.38}$ & 3.85$^{0.39}_{0.57}$ \\ 
        Hf 2-2 (8 arcsec)\footnotemark[15] &  84.00 & 3.95\footnotemark[15] & 2.96\footnotemark[15] & 3.00$^{~}_{0.79}$ & 2.30$^{3.59}_{~}$ & 3.85$^{0.17}_{0.20}$ & 5.00$^{0.10}_{0.93}$ \\ 
          NGC 1501\footnotemark[6] &  32.00 & 4.04\footnotemark[6] & 3.97\footnotemark[6] & 3.30$^{0.06}_{0.07}$ & 6.00$^{0.09}_{0.44}$ & 3.60$^{0.06}_{0.65}$ & 3.10$^{0.83}_{0.27}$ \\ 
            M 1-42\footnotemark[2] &  22.00 & 3.96\footnotemark[2] & 3.60\footnotemark[2] & 3.60$^{0.25}_{0.24}$ & 6.00$^{0.02}_{2.39}$ & 3.15$^{~}_{0.05}$ & 3.35$^{0.16}_{0.11}$ \\ 
            NGC 40\footnotemark[7] &  17.30 & 4.03\footnotemark[7] & 3.85\footnotemark[7] & ~$^{~}_{~}$ & ~$^{~}_{~}$ & 2.60$^{~}_{~}$ & 2.00$^{0.03}_{0.01}$ \\ 
            M 2-24\footnotemark[5] &  17.00 & ~ & 4.20\footnotemark[5] & 2.10$^{0.01}_{0.01}$ & 5.70$^{0.10}_{0.14}$ & 4.20$^{0.14}_{0.20}$ & 3.80$^{0.37}_{0.29}$ \\ 
          NGC 2022\footnotemark[4] &  16.00 & 4.18\footnotemark[4] & 4.12\footnotemark[4] & 2.10$^{0.01}_{0.01}$ & 5.90$^{0.05}_{0.06}$ & 2.60$^{~}_{~}$ & 2.00$^{~}_{~}$ \\ 
          NGC 2022\footnotemark[10] &  16.00 & 4.18\footnotemark[10] & 4.12\footnotemark[10] & 3.30$^{0.54}_{~}$ & 6.00$^{0.15}_{~}$ & 4.00$^{0.03}_{0.07}$ & 3.45$^{0.31}_{0.25}$ \\ 
            DdDm 1\footnotemark[13] &  11.80 & 4.09\footnotemark[13] & 3.94\footnotemark[13] & 4.30$^{~}_{~}$ & 2.00$^{~}_{~}$ & 3.65$^{0.20}_{0.12}$ & 5.00$^{0.01}_{0.02}$ \\ 
            Vy 2-2\footnotemark[13] &  11.80 & 4.14\footnotemark[13] & 3.97\footnotemark[13] & ~$^{~}_{~}$ & ~$^{~}_{~}$ & 4.20$^{~}_{~}$ & 2.90$^{0.02}_{0.04}$ \\ 
          NGC 6153\footnotemark[32] &  10.90 & ~ & ~ & ~$^{~}_{~}$ & ~$^{~}_{~}$ & 3.35$^{0.38}_{0.44}$ & 4.45$^{0.47}_{2.00}$ \\ 
          NGC 6153\footnotemark[1] &   9.20 & 3.86\footnotemark[1] & 3.78\footnotemark[1] & 3.30$^{0.31}_{0.12}$ & 5.90$^{0.31}_{1.61}$ & 3.25$^{0.01}_{0.06}$ & 3.60$^{0.15}_{0.16}$ \\ 
          NGC 2440\footnotemark[4] &   8.90 & 4.21\footnotemark[4] & 4.15\footnotemark[4] & 2.10$^{~}_{~}$ & 5.90$^{~}_{~}$ & 2.60$^{~}_{~}$ & 2.00$^{~}_{~}$ \\ 
           IC 2003\footnotemark[13] &   7.31 & 4.10\footnotemark[13] & 4.04\footnotemark[13] & ~$^{~}_{~}$ & ~$^{~}_{~}$ & 4.20$^{~}_{~}$ & 5.00$^{~}_{~}$ \\ 
            M 2-36\footnotemark[2] &   6.90 & 3.92\footnotemark[2] & 3.78\footnotemark[2] & 3.00$^{~}_{0.45}$ & 2.90$^{1.84}_{0.10}$ & 3.20$^{0.05}_{0.02}$ & 3.50$^{0.11}_{0.16}$ \\ 
            Vy 1-2\footnotemark[13] &   6.17 & 4.02\footnotemark[13] & 3.82\footnotemark[13] & ~$^{~}_{~}$ & ~$^{~}_{~}$ & 3.90$^{0.10}_{0.15}$ & 3.65$^{0.11}_{0.14}$ \\ 
          NGC 3242\footnotemark[4] &   5.70 & 4.07\footnotemark[4] & 4.01\footnotemark[4] & ~$^{~}_{~}$ & ~$^{~}_{~}$ & 2.60$^{~}_{~}$ & 2.00$^{~}_{~}$ \\ 
            M 3-27\footnotemark[13] &   5.48 & 4.11\footnotemark[13] & 3.96\footnotemark[13] & ~$^{~}_{~}$ & ~$^{~}_{~}$ & 4.10$^{0.09}_{0.03}$ & 3.45$^{0.08}_{0.06}$ \\ 
          NGC 2440\footnotemark[10] &   5.40 & 4.21\footnotemark[10]  & 4.15\footnotemark[10]  & 3.60$^{0.63}_{~}$ & 2.50$^{~}_{~}$ & 3.60$^{0.24}_{0.03}$ & 5.00$^{0.14}_{0.62}$ \\ 
          NGC 2440\footnotemark[16] &   5.40 & 4.17\footnotemark[16] & 4.04\footnotemark[16] & 2.50$^{1.57}_{0.12}$ & 6.00$^{0.08}_{1.74}$ & 4.20$^{~}_{~}$ & 5.00$^{0.12}_{0.99}$ \\ 
          NGC 7009\footnotemark[32] &   5.14 & ~ & ~ & ~$^{~}_{~}$ & ~$^{~}_{~}$ & 2.95$^{0.61}_{0.23}$ & 5.00$^{0.02}_{2.56}$ \\ 
          NGC 6818\footnotemark[4] &   4.90 & 4.12\footnotemark[4] & 4.08\footnotemark[4] & 2.10$^{~}_{~}$ & 5.90$^{~}_{~}$ & 2.60$^{~}_{~}$ & 2.00$^{~}_{~}$ \\ 
          NGC 7009\footnotemark[21] &   4.70 & 4.04\footnotemark[21] & 3.86\footnotemark[21] & 3.10$^{0.37}_{0.09}$ & 5.10$^{0.15}_{1.58}$ & 3.20$^{0.03}_{0.03}$ & 3.55$^{0.17}_{0.13}$ \\ 
           IC 3568\footnotemark[7] &   4.60 & 4.06\footnotemark[7] & 3.97\footnotemark[7] & 2.10$^{~}_{~}$ & 5.90$^{~}_{~}$ & 2.60$^{~}_{~}$ & 2.00$^{~}_{~}$ \\ 
          NGC 6210\footnotemark[7] &   4.30 & 3.99\footnotemark[7] & 3.94\footnotemark[7] & 2.20$^{0.09}_{0.08}$ & 5.90$^{0.61}_{1.24}$ & 2.60$^{~}_{~}$ & 2.00$^{~}_{~}$ \\ 
            M 3-34\footnotemark[13] &   4.23 & 4.09\footnotemark[13] & 3.93\footnotemark[13] & ~$^{~}_{~}$ & ~$^{~}_{~}$ & 4.00$^{0.01}_{0.01}$ & 5.00$^{~}_{~}$ \\ 
          NGC 6543\footnotemark[11] &   4.20 & 4.14\footnotemark[11] & 3.83\footnotemark[11] & 4.00$^{0.06}_{0.13}$ & 3.20$^{0.16}_{0.25}$ & 4.20$^{0.01}_{0.01}$ & 3.90$^{0.17}_{0.11}$ \\ 
          NGC 6543\footnotemark[12] &   4.20 & 3.90\footnotemark[12] & 3.83\footnotemark[12] & ~$^{~}_{~}$ & ~$^{~}_{~}$ & 2.70$^{0.06}_{0.05}$ & 2.80$^{0.26}_{0.21}$ \\ 
        NGC 6543 S\footnotemark[12] &   4.20 & 3.89\footnotemark[12] & 3.83\footnotemark[12] & ~$^{~}_{~}$ & ~$^{~}_{~}$ & 3.05$^{0.04}_{0.07}$ & 5.00$^{0.29}_{0.85}$ \\ 
            Hu 2-1\footnotemark[13] &   4.00 & 3.99\footnotemark[13] & 4.00\footnotemark[13] & 4.30$^{~}_{~}$ & 2.00$^{~}_{~}$ & 4.20$^{~}_{~}$ & 5.00$^{0.36}_{0.65}$ \\ 
            M 1-73\footnotemark[13] &   3.61 & 3.87\footnotemark[13] & 3.74\footnotemark[13] & ~$^{~}_{~}$ & ~$^{~}_{~}$ & 3.20$^{0.01}_{0.03}$ & 3.30$^{0.20}_{0.11}$ \\ 
          NGC 6302\footnotemark[10] &   3.60 & 4.26\footnotemark[10] & 4.21\footnotemark[10] & 2.10$^{~}_{~}$ & 5.90$^{~}_{~}$ & 4.10$^{0.16}_{0.21}$ & 5.00$^{0.08}_{0.17}$ \\ 
          NGC 3132\footnotemark[4] &   3.50 & 3.98\footnotemark[4] & 4.03\footnotemark[4] & 2.10$^{~}_{~}$ & 5.90$^{0.01}_{0.01}$ & 2.60$^{~}_{~}$ & 2.00$^{~}_{~}$ \\ 
          NGC 6302\footnotemark[4] &   3.50 & 4.26\footnotemark[4] & 4.21\footnotemark[4] & 4.30$^{~}_{~}$ & 2.00$^{~}_{~}$ & 2.60$^{~}_{~}$ & 2.00$^{~}_{~}$ \\ 
          NGC 7026\footnotemark[13] &   3.36 & 3.97\footnotemark[13] & 3.87\footnotemark[13] & 4.30$^{0.11}_{0.51}$ & 3.10$^{0.36}_{0.70}$ & 3.90$^{0.08}_{0.13}$ & 3.40$^{0.06}_{0.13}$ \\ 
           IC 1747\footnotemark[13] &   3.20 & 4.04\footnotemark[13] & 3.98\footnotemark[13] & ~$^{~}_{~}$ & ~$^{~}_{~}$ & 3.80$^{0.07}_{0.04}$ & 5.00$^{0.38}_{0.61}$ \\ 
            IC 351\footnotemark[13] &   3.14 & 4.12\footnotemark[13] & 4.04\footnotemark[13] & ~$^{~}_{~}$ & ~$^{~}_{~}$ & 3.90$^{~}_{0.13}$ & 3.20$^{0.42}_{0.27}$ \\ 
          NGC 6210\footnotemark[12] &   3.10 & 3.98\footnotemark[12] & 3.94\footnotemark[12] & ~$^{~}_{~}$ & ~$^{~}_{~}$ & 3.05$^{0.02}_{0.03}$ & 5.00$^{0.01}_{0.01}$ \\ 
            Hu 1-1\footnotemark[13] &   2.97 & 4.08\footnotemark[13] & 3.92\footnotemark[13] & ~$^{~}_{~}$ & ~$^{~}_{~}$ & 2.60$^{0.63}_{0.32}$ & 2.70$^{0.40}_{0.22}$ \\ 
            Sp 4-1\footnotemark[13] &   2.94 & 4.05\footnotemark[13] & 3.95\footnotemark[13] & ~$^{~}_{~}$ & ~$^{~}_{~}$ & 4.00$^{0.02}_{0.14}$ & 2.75$^{~}_{0.16}$ \\ 
           IC 4846\footnotemark[13] &   2.91 & 4.03\footnotemark[13] & 3.89\footnotemark[13] & ~$^{~}_{~}$ & ~$^{~}_{~}$ & ~$^{~}_{~}$ & ~$^{~}_{~}$ \\ 
           IC 4846\footnotemark[17] &   2.91 & 4.00\footnotemark[17] & 4.26\footnotemark[17] & 2.90$^{0.66}_{0.75}$ & 2.20$^{~}_{~}$ & 4.20$^{0.03}_{0.04}$ & 3.85$^{0.21}_{0.15}$ \\ 
          NGC 6803\footnotemark[13] &   2.71 & 3.99\footnotemark[13] & 3.93\footnotemark[13] & ~$^{~}_{~}$ & ~$^{~}_{~}$ & 3.70$^{0.04}_{0.07}$ & 4.20$^{0.37}_{0.27}$ \\ 
            BoBn 1\footnotemark[19] &   2.63 & 4.14\footnotemark[19] & 3.95\footnotemark[19] & 4.10$^{0.03}_{0.04}$ & 2.00$^{0.12}_{0.03}$ & 3.70$^{0.26}_{0.23}$ & 2.80$^{0.22}_{0.35}$ \\ 
          NGC 6833\footnotemark[13] &   2.47 & 4.11\footnotemark[13] & 4.15\footnotemark[13] & ~$^{~}_{~}$ & ~$^{~}_{~}$ & 4.20$^{0.21}_{0.64}$ & 3.80$^{0.17}_{0.40}$ \\ 
          NGC 6879\footnotemark[13] &   2.46 & 4.02\footnotemark[13] & 3.93\footnotemark[13] & ~$^{~}_{~}$ & ~$^{~}_{~}$ & 4.20$^{~}_{~}$ & 2.75$^{0.04}_{0.02}$ \\ 
     IC 4191 fixed\footnotemark[4] &   2.40 & 4.03\footnotemark[4] & 4.02\footnotemark[4] & 2.10$^{~}_{~}$ & 5.90$^{~}_{~}$ & 2.60$^{~}_{~}$ & 2.00$^{~}_{~}$ \\ 
  IC 4191 scanning\footnotemark[4] &   2.40 & 4.03\footnotemark[4] & 4.02\footnotemark[4] & 2.10$^{~}_{~}$ & 5.90$^{~}_{~}$ & 2.60$^{~}_{~}$ & 2.00$^{0.01}_{0.01}$ \\ 
          NGC 3132\footnotemark[10] &   2.40 & 3.98\footnotemark[10] & 4.03\footnotemark[10] & 3.30$^{0.12}_{0.14}$ & 6.00$^{0.49}_{0.76}$ & 3.20$^{0.34}_{0.12}$ & 5.00$^{0.15}_{0.99}$ \\ 
          NGC 6720\footnotemark[7] &   2.40 & 4.03\footnotemark[7] & 3.90\footnotemark[7] & 2.10$^{0.05}_{0.03}$ & 5.70$^{0.31}_{0.37}$ & 2.65$^{0.05}_{0.05}$ & 2.20$^{0.07}_{0.06}$ \\ 
\hline
\end{tabular}
\\
\begin{description}
\item [$^1$] \citet{LSB00};
$^2$ \citet{LLB01};
$^3$ \citet{RPP03};
$^4$ \citet{TBL03b};
$^5$ \citet{ZL03};
\item [$^6$] \citet{EWZ04};
$^7$ \citet{LLB04};
$^8$ \citet{PPR04};
$^9$ \citet{SBW04};
$^{10}$ \citet{TBL04};
\item [$^{11}$] \citet{WL04};
$^{12}$ \citet{RG05};
$^{13}$ \citet{WLB05};
$^{14}$ \citet{ZLL05a};
\item [$^{15}$] \citet{LBZ06};
$^{16}$ \citet{S07};
$^{17}$ \citet{WL07};
$^{18}$ \citet{GPP08};
$^{19}$ \citet{OTH10};
\item [$^{20}$] \citet{WLB03};
$^{21}$ \citet{FL11};
$^{22}$ \citet{EPT02};
$^{23}$ \citet{TBL03a};
$^{24}$ \citet{EPG04};
\item [$^{25}$] \citet{GEP04};
$^{26}$ \citet{GEP05};
$^{27}$ \citet{GPC06};
$^{28}$ \citet{GEP07};
\item [$^{29}$] \citet{LEG07};
$^{30}$ \citet{EBP09};
$^{31}$ \citet{WBL08};
$^{32}$ \citet{TWP08}.
\end{description}
\end{table*}

\clearpage
\setcounter{table}{0}
\begin{table*}
\centering
\setlength{\tabcolsep}{0.005in}
\caption{(continued)}
\vskip-0.1truein
\begin{tabular}{lrcccccc}
\hline
& & \multicolumn{2}{l}{From Literature} & \multicolumn{4}{l}{From Current Work} \\
\multicolumn{1}{l}{Object} & ADF (O$^{2+}$) & $\log$~\tel([O~{\sc iii}]) & $\log$~\tel(BJ) & $\log$~\tel(N~{\sc ii}) & $\log$~\tel(N~{\sc ii}) & $\log$~\tel(O~{\sc ii}) & $\log$~\tel(O~{\sc ii})  \\
& ~ & [K] & [K] & [K] & [cm$^{-3}$] & [K] & [cm$^{-3}$]  \\
\hline
          NGC 6818\footnotemark[10] &   2.40 & 4.12\footnotemark[10] & 4.08\footnotemark[10] & 3.20$^{0.52}_{~}$ & 6.00$^{0.37}_{2.52}$ & 2.95$^{0.11}_{0.12}$ & 4.75$^{0.09}_{0.15}$ \\ 
    IC 4191 nebula\footnotemark[10] &   2.40 & 4.03\footnotemark[10] & 4.02\footnotemark[10] & 3.80$^{0.08}_{0.03}$ & 2.70$^{0.12}_{0.10}$ & 3.80$^{0.05}_{0.07}$ & 3.90$^{0.15}_{0.14}$ \\ 
     IC 4191 fixed\footnotemark[10] &   2.40 & 4.03\footnotemark[10] & 4.02\footnotemark[10] & 3.80$^{0.08}_{0.04}$ & 2.70$^{0.15}_{0.13}$ & 3.25$^{0.05}_{0.03}$ & 4.05$^{0.36}_{0.24}$ \\ 
           IC 4191\footnotemark[16] &   2.40 & 4.00\footnotemark[16] & 3.90\footnotemark[16] & 2.70$^{0.09}_{0.05}$ & 2.20$^{0.02}_{0.16}$ & 3.60$^{0.05}_{0.01}$ & 5.00$^{~}_{~}$ \\ 
          NGC 3918\footnotemark[4] &   2.30 & 4.10\footnotemark[4] & 4.09\footnotemark[4] & 2.10$^{0.01}_{0.01}$ & 5.90$^{0.04}_{0.04}$ & 2.60$^{0.04}_{0.01}$ & 2.65$^{0.12}_{0.21}$ \\ 
          NGC 6884\footnotemark[7] &   2.30 & 4.04\footnotemark[7] & 4.06\footnotemark[7] & 2.10$^{0.05}_{0.03}$ & 5.90$^{0.21}_{0.24}$ & 2.60$^{~}_{~}$ & 2.40$^{0.01}_{0.02}$ \\ 
           IC 5217\footnotemark[13] &   2.26 & 4.05\footnotemark[13] & 4.08\footnotemark[13] & ~$^{~}_{~}$ & ~$^{~}_{~}$ & 3.50$^{0.08}_{0.04}$ & 5.00$^{0.19}_{0.45}$ \\ 
          NGC 3242\footnotemark[10] &   2.20 & 4.07\footnotemark[10] & 4.01\footnotemark[10] & 4.00$^{0.11}_{0.01}$ & 2.50$^{0.11}_{0.06}$ & 3.35$^{0.05}_{0.06}$ & 5.00$^{0.19}_{0.31}$ \\ 
          NGC 5882\footnotemark[4] &   2.20 & 3.97\footnotemark[4] & 3.89\footnotemark[4] & ~$^{~}_{~}$ & ~$^{~}_{~}$ & 2.60$^{~}_{~}$ & 3.10$^{0.18}_{0.28}$ \\ 
            M 1-74\footnotemark[13] &   2.14 & 4.01\footnotemark[13] & 3.89\footnotemark[13] & ~$^{~}_{~}$ & ~$^{~}_{~}$ & 3.75$^{0.30}_{0.32}$ & 4.45$^{0.53}_{0.32}$ \\ 
          NGC 5882\footnotemark[32] &   2.14 & ~ & ~ & ~$^{~}_{~}$ & ~$^{~}_{~}$ & 2.95$^{0.72}_{0.33}$ & 2.40$^{1.94}_{0.47}$ \\ 
          NGC 5882\footnotemark[10] &   2.10 & 3.97\footnotemark[10] & 3.89\footnotemark[10] & 3.90$^{0.07}_{0.11}$ & 3.00$^{0.60}_{0.43}$ & 3.55$^{0.09}_{0.02}$ & 5.00$^{0.10}_{0.67}$ \\ 
            Me 2-2\footnotemark[13] &   2.10 & 4.04\footnotemark[13] & 4.04\footnotemark[13] & 3.80$^{0.03}_{0.08}$ & 2.20$^{0.05}_{0.19}$ & 4.00$^{0.03}_{0.05}$ & 5.00$^{0.23}_{0.71}$ \\ 
          NGC 7662\footnotemark[7] &   2.00 & 4.13\footnotemark[7] & 4.09\footnotemark[7] & 2.10$^{~}_{~}$ & 5.90$^{~}_{~}$ & 2.60$^{~}_{~}$ & 2.00$^{~}_{~}$ \\ 
          NGC 5315\footnotemark[10] &   2.00 & 3.95\footnotemark[10] & 3.93\footnotemark[10] & 3.80$^{0.15}_{0.44}$ & 4.70$^{1.03}_{1.23}$ & 4.00$^{0.02}_{0.08}$ & 4.30$^{0.35}_{0.27}$ \\ 
          NGC 6807\footnotemark[13] &   2.00 & 4.04\footnotemark[13] & 4.00\footnotemark[13] & ~$^{~}_{~}$ & ~$^{~}_{~}$ & 4.00$^{0.04}_{0.04}$ & 3.15$^{0.43}_{0.31}$ \\ 
          NGC 5307\footnotemark[3] &   1.95 & ~ & 4.03\footnotemark[3] & ~$^{~}_{~}$ & ~$^{~}_{~}$ & 3.15$^{0.15}_{0.15}$ & 2.70$^{0.60}_{0.34}$ \\ 
           IC 4406\footnotemark[4] &   1.90 & 4.00\footnotemark[4] & 3.97\footnotemark[4] & 2.10$^{~}_{~}$ & 5.90$^{~}_{~}$ & 2.60$^{~}_{~}$ & 2.20$^{0.10}_{0.11}$ \\ 
           IC 4406\footnotemark[10] &   1.90 & 4.00\footnotemark[10] & 3.97\footnotemark[10] & 2.10$^{0.01}_{0.01}$ & 5.90$^{0.05}_{0.06}$ & 3.00$^{0.08}_{0.04}$ & 2.80$^{0.07}_{0.17}$ \\ 
          NGC 6741\footnotemark[7] &   1.90 & 3.99\footnotemark[7] & 4.15\footnotemark[7] & 4.20$^{0.34}_{0.45}$ & 2.00$^{0.84}_{0.12}$ & 2.60$^{~}_{~}$ & 2.20$^{~}_{~}$ \\ 
          NGC 6826\footnotemark[7] &   1.90 & 3.97\footnotemark[7] & 3.94\footnotemark[7] & 2.30$^{0.25}_{0.13}$ & 5.70$^{0.41}_{~}$ & 2.90$^{0.01}_{0.02}$ & 2.00$^{0.02}_{0.02}$ \\ 
          My Cn 18\footnotemark[4] &   1.80 & 3.86\footnotemark[4] & ~ & ~$^{~}_{~}$ & ~$^{~}_{~}$ & 2.60$^{~}_{~}$ & 2.15$^{~}_{0.15}$ \\ 
          My Cn 18\footnotemark[10] &   1.80 & ~ & ~ & 4.30$^{~}_{~}$ & 2.00$^{~}_{~}$ & 3.25$^{0.03}_{0.03}$ & 2.50$^{0.02}_{0.04}$ \\ 
          NGC 3918\footnotemark[10] &   1.80 & 4.10\footnotemark[10] & 4.09\footnotemark[10] & 4.00$^{0.21}_{0.22}$ & 2.70$^{0.28}_{0.39}$ & 2.60$^{~}_{~}$ & 2.00$^{0.09}_{0.07}$ \\ 
          NGC 7027\footnotemark[12] &   1.80 & 4.15\footnotemark[12] & 4.08\footnotemark[12] & ~$^{~}_{~}$ & ~$^{~}_{~}$ & 2.65$^{0.05}_{~}$ & 2.00$^{~}_{~}$ \\ 
          NGC 7027\footnotemark[14] &   1.80 & 4.10\footnotemark[14] & 4.08\footnotemark[14] & 3.90$^{0.14}_{0.21}$ & 3.10$^{1.43}_{0.65}$ & 4.00$^{0.03}_{0.10}$ & 5.00$^{~}_{~}$ \\ 
          NGC 5315\footnotemark[8] &   1.74 & 3.95\footnotemark[8] & 3.93\footnotemark[8] & 4.30$^{~}_{0.10}$ & 6.00$^{~}_{~}$ & 4.20$^{0.06}_{0.13}$ & 4.95$^{0.09}_{0.30}$ \\ 
          NGC 6790\footnotemark[7] &   1.70 & 4.11\footnotemark[7] & 4.15\footnotemark[7] & 2.10$^{~}_{~}$ & 5.90$^{~}_{~}$ & 2.60$^{~}_{~}$ & 2.20$^{~}_{~}$ \\ 
          NGC 6790\footnotemark[12] &   1.70 & 4.11\footnotemark[12] & 4.15\footnotemark[12] & ~$^{~}_{~}$ & ~$^{~}_{~}$ & 2.60$^{~}_{~}$ & 2.00$^{~}_{~}$ \\ 
            Hu 1-2\footnotemark[7] &   1.60 & 4.29\footnotemark[7] & 4.30\footnotemark[7] & 2.10$^{~}_{~}$ & 5.90$^{~}_{~}$ & 2.60$^{~}_{~}$ & 2.00$^{~}_{~}$ \\ 
          NGC 6572\footnotemark[7] &   1.60 & 4.03\footnotemark[7] & 4.01\footnotemark[7] & 2.50$^{0.59}_{0.06}$ & 5.40$^{0.29}_{~}$ & 2.70$^{0.01}_{0.05}$ & 3.35$^{0.07}_{~}$ \\ 
          NGC 6572\footnotemark[12] &   1.60 & 4.01\footnotemark[12] & 4.01\footnotemark[12] & ~$^{~}_{~}$ & ~$^{~}_{~}$ & 2.60$^{0.05}_{0.01}$ & 2.85$^{0.43}_{0.07}$ \\ 
        NGC 6572 S\footnotemark[12] &   1.60 & 4.01\footnotemark[12] & 4.01\footnotemark[12] & ~$^{~}_{~}$ & ~$^{~}_{~}$ & 2.60$^{1.62}_{~}$ & 2.30$^{~}_{~}$ \\ 
          NGC 6891\footnotemark[13] &   1.52 & 3.97\footnotemark[13] & 3.77\footnotemark[13] & ~$^{~}_{~}$ & ~$^{~}_{~}$ & 4.20$^{0.10}_{0.14}$ & 3.30$^{0.10}_{0.14}$ \\ 
          NGC 5315\footnotemark[4] &   1.40 & 3.95\footnotemark[4] & 3.93\footnotemark[4] & 2.10$^{0.01}_{0.01}$ & 5.90$^{~}_{~}$ & 2.90$^{0.10}_{0.06}$ & 5.00$^{0.03}_{0.04}$ \\ 
            M 3-32\footnotemark[17] &   1.15 & 3.95\footnotemark[17] & 3.65\footnotemark[17] & 3.10$^{0.20}_{0.22}$ & 5.30$^{0.47}_{0.47}$ & 3.50$^{0.06}_{0.02}$ & 3.35$^{0.11}_{0.11}$ \\ 
            M 3-33\footnotemark[17] &   1.10 & 4.02\footnotemark[17] & 3.77\footnotemark[17] & ~$^{~}_{~}$ & ~$^{~}_{~}$ & 3.60$^{0.03}_{0.06}$ & 3.25$^{0.08}_{0.11}$ \\ 
           IC 4699\footnotemark[17] &   1.09 & 4.07\footnotemark[17] & 4.08\footnotemark[17] & 2.60$^{1.11}_{0.23}$ & 3.40$^{1.41}_{0.93}$ & 3.40$^{0.10}_{0.12}$ & 3.40$^{0.23}_{0.16}$ \\ 
          NGC 6439\footnotemark[17] &   1.09 & 4.02\footnotemark[17] & 4.00\footnotemark[17] & 4.30$^{0.01}_{0.01}$ & 5.70$^{0.19}_{0.01}$ & 4.00$^{0.01}_{0.01}$ & 3.55$^{0.19}_{0.12}$ \\ 
            H 1-41\footnotemark[17] &   1.08 & 3.99\footnotemark[17] & 3.65\footnotemark[17] & 2.90$^{0.95}_{0.09}$ & 4.80$^{0.45}_{1.37}$ & 3.45$^{0.04}_{0.05}$ & 2.10$^{0.21}_{0.13}$ \\ 
             M 3-7\footnotemark[17] &   1.07 & 3.88\footnotemark[17] & 3.84\footnotemark[17] & 4.30$^{~}_{~}$ & 2.00$^{~}_{~}$ & 3.20$^{0.24}_{0.11}$ & 2.85$^{0.27}_{0.26}$ \\ 
          NGC 6620\footnotemark[17] &   1.06 & 3.98\footnotemark[17] & 4.00\footnotemark[17] & 4.20$^{0.04}_{0.52}$ & 3.00$^{0.05}_{0.74}$ & 3.35$^{0.08}_{0.02}$ & 3.75$^{0.19}_{0.16}$ \\ 
            H 1-50\footnotemark[17] &   1.05 & 4.04\footnotemark[17] & 4.10\footnotemark[17] & ~$^{~}_{~}$ & ~$^{~}_{~}$ & 3.15$^{0.02}_{0.04}$ & 2.85$^{0.16}_{0.07}$ \\ 
            H 1-54\footnotemark[17] &   1.05 & 3.98\footnotemark[17] & 4.10\footnotemark[17] & 4.30$^{~}_{~}$ & 2.00$^{~}_{~}$ & 3.25$^{0.32}_{0.20}$ & 3.75$^{0.32}_{0.24}$ \\ 
            M 1-29\footnotemark[17] &   1.05 & 4.03\footnotemark[17] & 4.00\footnotemark[17] & 3.00$^{0.80}_{0.36}$ & 3.40$^{1.45}_{0.57}$ & 2.60$^{~}_{~}$ & 2.00$^{~}_{~}$ \\ 
            M 3-21\footnotemark[17] &   1.05 & 3.99\footnotemark[17] & 4.04\footnotemark[17] & 4.30$^{0.02}_{0.02}$ & 3.80$^{1.41}_{0.36}$ & 2.60$^{~}_{~}$ & 2.00$^{~}_{~}$ \\ 
            H 1-35\footnotemark[17] &   1.04 & 3.96\footnotemark[17] & 4.00\footnotemark[17] & 4.30$^{~}_{~}$ & 2.00$^{~}_{~}$ & 3.20$^{0.02}_{0.01}$ & 4.75$^{0.10}_{0.05}$ \\ 
            H 1-42\footnotemark[17] &   1.04 & 3.99\footnotemark[17] & 4.00\footnotemark[17] & 4.30$^{~}_{~}$ & 2.00$^{~}_{~}$ & 3.60$^{0.03}_{0.07}$ & 3.75$^{0.43}_{0.30}$ \\ 
             M 2-6\footnotemark[17] &   1.04 & 4.00\footnotemark[17] & 4.07\footnotemark[17] & ~$^{~}_{~}$ & ~$^{~}_{~}$ & 3.80$^{0.03}_{0.15}$ & 5.00$^{~}_{~}$ \\ 
            M 2-27\footnotemark[17] &   1.04 & 4.08\footnotemark[17] & 4.15\footnotemark[17] & 4.30$^{~}_{~}$ & 2.00$^{~}_{~}$ & 3.30$^{0.29}_{0.17}$ & 2.90$^{0.18}_{0.20}$ \\ 
            M 2-33\footnotemark[17] &   1.04 & 3.91\footnotemark[17] & 3.85\footnotemark[17] & 3.30$^{0.53}_{0.44}$ & 2.50$^{1.25}_{0.59}$ & 2.60$^{0.02}_{0.01}$ & 3.60$^{0.14}_{0.22}$ \\ 
            M 2-42\footnotemark[17] &   1.04 & 3.93\footnotemark[17] & 4.15\footnotemark[17] & 4.20$^{~}_{1.71}$ & 4.80$^{~}_{1.88}$ & 2.60$^{~}_{~}$ & 2.00$^{~}_{~}$ \\ 
            M 3-29\footnotemark[17] &   1.04 & 3.96\footnotemark[17] & 4.03\footnotemark[17] & 3.10$^{0.15}_{~}$ & 5.30$^{0.32}_{~}$ & 2.60$^{~}_{~}$ & 2.00$^{~}_{~}$ \\ 
          NGC 6567\footnotemark[17] &   1.04 & 4.02\footnotemark[17] & 4.08\footnotemark[17] & 4.30$^{~}_{~}$ & 2.00$^{~}_{~}$ & 2.80$^{0.16}_{0.04}$ & 3.05$^{0.16}_{0.22}$ \\ 
\hline
\end{tabular}
\\
\begin{description}
\item [$^1$] \citet{LSB00};
$^2$ \citet{LLB01};
$^3$ \citet{RPP03};
$^4$ \citet{TBL03b};
$^5$ \citet{ZL03};
\item [$^6$] \citet{EWZ04};
$^7$ \citet{LLB04};
$^8$ \citet{PPR04};
$^9$ \citet{SBW04};
$^{10}$ \citet{TBL04};
\item [$^{11}$] \citet{WL04};
$^{12}$ \citet{RG05};
$^{13}$ \citet{WLB05};
$^{14}$ \citet{ZLL05a};
\item [$^{15}$] \citet{LBZ06};
$^{16}$ \citet{S07};
$^{17}$ \citet{WL07};
$^{18}$ \citet{GPP08};
$^{19}$ \citet{OTH10};
\item [$^{20}$] \citet{WLB03};
$^{21}$ \citet{FL11};
$^{22}$ \citet{EPT02};
$^{23}$ \citet{TBL03a};
$^{24}$ \citet{EPG04};
\item [$^{25}$] \citet{GEP04};
$^{26}$ \citet{GEP05};
$^{27}$ \citet{GPC06};
$^{28}$ \citet{GEP07};
\item [$^{29}$] \citet{LEG07};
$^{30}$ \citet{EBP09};
$^{31}$ \citet{WBL08};
$^{32}$ \citet{TWP08}.
\end{description}
\end{table*}

\clearpage
\setcounter{table}{0}
\begin{table*}
\centering
\setlength{\tabcolsep}{0.015in}
\caption{(continued)}
\vskip-0.1truein
\begin{tabular}{crcccccc}
\hline
& & \multicolumn{2}{l}{From Literature} & \multicolumn{4}{l}{From Current Work} \\
\multicolumn{1}{l}{Object} & ADF (O$^{2+}$) & $\log$~\tel([O~{\sc iii}]) & $\log$~\tel(BJ) & $\log$~\tel(N~{\sc ii}) & $\log$~\tel(N~{\sc ii}) & $\log$~\tel(O~{\sc ii}) & $\log$~\tel(O~{\sc ii})  \\
& ~ & [K] & [K] & [K] & [cm$^{-3}$] & [K] & [cm$^{-3}$]  \\
\hline
           IC 4593\footnotemark[12] &   1.03 & 3.92\footnotemark[12] & 3.92\footnotemark[12] & ~$^{~}_{~}$ & ~$^{~}_{~}$ & 2.80$^{0.17}_{0.14}$ & 2.00$^{0.47}_{0.32}$ \\ 
          He 2-118\footnotemark[17] &   1.03 & 4.10\footnotemark[17] & 4.26\footnotemark[17] & 4.30$^{~}_{~}$ & 2.00$^{~}_{~}$ & 3.05$^{0.05}_{~}$ & 2.45$^{0.21}_{0.18}$ \\ 
            M 1-61\footnotemark[17] &   1.03 & 3.95\footnotemark[17] & 3.98\footnotemark[17] & 4.00$^{0.15}_{0.47}$ & 2.30$^{~}_{0.76}$ & 3.00$^{~}_{~}$ & 2.25$^{0.02}_{0.06}$ \\ 
             M 2-4\footnotemark[17] &   1.03 & 3.93\footnotemark[17] & 3.90\footnotemark[17] & 4.00$^{0.01}_{0.01}$ & 2.00$^{~}_{~}$ & 4.20$^{~}_{~}$ & 3.45$^{0.04}_{0.07}$ \\ 
          NGC 6565\footnotemark[17] &   1.03 & 4.01\footnotemark[17] & 3.93\footnotemark[17] & 4.30$^{~}_{~}$ & 2.00$^{~}_{~}$ & 3.35$^{0.63}_{0.17}$ & 3.05$^{0.33}_{0.24}$ \\ 
            Vy 2-1\footnotemark[17] &   1.03 & 3.90\footnotemark[17] & 3.94\footnotemark[17] & 4.30$^{~}_{~}$ & 2.00$^{~}_{~}$ & 4.20$^{0.02}_{0.05}$ & 3.25$^{0.05}_{0.03}$ \\ 
            IC 418\footnotemark[9] &   1.02 & 3.95\footnotemark[9] & ~ & 4.20$^{0.01}_{0.02}$ & 2.30$^{0.06}_{0.14}$ & 4.20$^{~}_{~}$ & 3.85$^{0.28}_{0.16}$ \\ 
            Cn 1-5\footnotemark[17] &   1.02 & 3.94\footnotemark[17] & 4.00\footnotemark[17] & 3.30$^{0.09}_{0.07}$ & 6.00$^{0.42}_{0.54}$ & 2.60$^{~}_{~}$ & 2.00$^{~}_{~}$ \\ 
            Cn 2-1\footnotemark[17] &   1.02 & 4.01\footnotemark[17] & 4.03\footnotemark[17] & 3.40$^{0.45}_{0.15}$ & 6.00$^{~}_{~}$ & 3.25$^{0.15}_{0.18}$ & 3.75$^{0.78}_{0.29}$ \\ 
            M 1-20\footnotemark[17] &   1.02 & 3.99\footnotemark[17] & 4.08\footnotemark[17] & ~$^{~}_{~}$ & ~$^{~}_{~}$ & 2.60$^{0.01}_{0.01}$ & 4.00$^{0.13}_{0.15}$ \\ 
            M 2-23\footnotemark[17] &   1.02 & 4.08\footnotemark[17] & 3.70\footnotemark[17] & 4.30$^{~}_{~}$ & 2.00$^{~}_{~}$ & 4.15$^{0.06}_{0.07}$ & 5.00$^{0.05}_{0.94}$ \\ 
           IC 2501\footnotemark[16] &   1.01 & 3.98\footnotemark[16] & 3.85\footnotemark[16] & 4.20$^{0.03}_{0.08}$ & 6.00$^{0.31}_{1.88}$ & 4.00$^{0.06}_{0.03}$ & 5.00$^{0.02}_{0.02}$ \\ 
            M 2-39\footnotemark[17] &   0.95 & 3.91\footnotemark[17] & 3.74\footnotemark[17] & 4.10$^{0.06}_{0.66}$ & 5.80$^{~}_{~}$ & 2.65$^{0.04}_{0.05}$ & 2.60$^{0.05}_{0.06}$ \\ 
            Cn 3-1\footnotemark[13] &   ~ & 3.88\footnotemark[13] & 3.71\footnotemark[13] & 4.30$^{~}_{~}$ & 2.00$^{~}_{~}$ & ~$^{~}_{~}$ & ~$^{~}_{~}$ \\ 
            M 2-31\footnotemark[17] &   ~ & 3.99\footnotemark[17] & 4.15\footnotemark[17] & ~$^{~}_{~}$ & ~$^{~}_{~}$ & ~$^{~}_{~}$ & ~$^{~}_{~}$ \\ 
\hline
\end{tabular}
\\
\begin{description}
\item [$^{9}$] \citet{SBW04}; 
$^{13}$\citet{WLB05}; 
$^{16}$\citet{S07};
$^{17}$\citet{WL07}.
\end{description}
\end{table*}

\setcounter{table}{1}
\begin{table*}
\centering
\setlength{\tabcolsep}{0.015in}
\caption{Electron temperatures and densities for H~{\sc ii}~regions.}
\vskip-0.1truein
\begin{tabular}{lrcccccc}
\hline
& & \multicolumn{2}{l}{From Literature} & \multicolumn{4}{l}{From Current Work} \\
\multicolumn{1}{l}{Object} & ADF (O$^{2+}$) & $\log$~\tel([O~{\sc iii}]) & $\log$~\tel(BJ) & $\log$~\tel(N~{\sc ii}) & $\log$~\tel(N~{\sc ii}) & $\log$~\tel(O~{\sc ii}) & $\log$~\tel(O~{\sc ii})  \\
& ~ & [K] & [K] & [K] & [cm$^{-3}$] & [K] & [cm$^{-3}$]  \\
\hline
           SMC N87\footnotemark[1] &   2.80 & 4.09\footnotemark[1] & ~ & ~$^{~}_{~}$ & ~$^{~}_{~}$ & 2.60$^{~}_{~}$ & 2.20$^{~}_{~}$ \\ 
              M 17\footnotemark[4] &   2.66 & 3.91\footnotemark[4] & 3.89\footnotemark[4] & 4.00$^{0.16}_{0.41}$ & 2.40$^{~}_{0.79}$ & 2.60$^{~}_{~}$ & 2.00$^{~}_{~}$ \\ 
          LMC N141\footnotemark[1] &   2.60 & 4.07\footnotemark[1] & ~ & ~$^{~}_{~}$ & ~$^{~}_{~}$ & 2.60$^{~}_{~}$ & 2.00$^{~}_{~}$ \\ 
          NGC 3576\footnotemark[4] &   2.21 & 3.95\footnotemark[4] & 3.91\footnotemark[4] & 3.30$^{0.01}_{~}$ & 6.00$^{0.03}_{0.03}$ & 2.60$^{~}_{~}$ & 2.00$^{~}_{~}$ \\ 
        30 Doradus\footnotemark[4] &   1.76 & 4.00\footnotemark[4] & ~ & 3.30$^{0.05}_{0.05}$ & 6.00$^{0.21}_{0.29}$ & 2.60$^{~}_{~}$ & 2.00$^{~}_{~}$ \\ 
           LMC N66\footnotemark[1] &   1.61 & 4.26\footnotemark[1] & ~ & ~$^{~}_{~}$ & ~$^{~}_{~}$ & ~$^{~}_{~}$ & ~$^{~}_{~}$ \\ 
         LMC N11 B\footnotemark[4] &   1.47 & 3.97\footnotemark[4] & ~ & ~$^{~}_{~}$ & ~$^{~}_{~}$ & 2.60$^{0.01}_{0.01}$ & 2.00$^{0.02}_{0.01}$ \\ 
            PB 6-2\footnotemark[2] &   1.12 & 4.18\footnotemark[2] & ~ & ~$^{~}_{~}$ & ~$^{~}_{~}$ & ~$^{~}_{~}$ & ~$^{~}_{~}$ \\ 
            PB 6-1\footnotemark[2] &   1.10 & 4.20\footnotemark[2] & ~ & ~$^{~}_{~}$ & ~$^{~}_{~}$ & ~$^{~}_{~}$ & ~$^{~}_{~}$ \\ 
              PB 8\footnotemark[2] &   1.05 & 3.84\footnotemark[2] & 3.71\footnotemark[2] & 2.10$^{0.20}_{0.13}$ & 5.80$^{0.44}_{0.32}$ & 4.20$^{0.02}_{0.13}$ & 3.30$^{0.12}_{0.09}$ \\ 
 NGC 2366 NGC 2363\footnotemark[3] &   1.04 & 4.20\footnotemark[3] & ~ & ~$^{~}_{~}$ & ~$^{~}_{~}$ & 2.85$^{0.45}_{0.14}$ & 2.90$^{0.57}_{0.55}$ \\ 
          NGC 3603\footnotemark[8] &   1.04 & 3.96\footnotemark[8] & ~ & ~$^{~}_{~}$ & ~$^{~}_{~}$ & 2.60$^{0.01}_{0.01}$ & 2.35$^{0.02}_{0.09}$ \\ 
        NGC 2867-2\footnotemark[2] &   1.03 & 4.06\footnotemark[2] & 3.95\footnotemark[2] & ~$^{~}_{~}$ & ~$^{~}_{~}$ & 2.60$^{~}_{~}$ & 3.80$^{0.21}_{0.39}$ \\ 
             S 311\footnotemark[7] &   1.03 & 3.95\footnotemark[7] & 3.98\footnotemark[7] & ~$^{~}_{~}$ & ~$^{~}_{~}$ & 2.60$^{~}_{~}$ & 2.00$^{~}_{~}$ \\ 
     M101 NGC 5461\footnotemark[3] &   1.03 & 3.93\footnotemark[3] & ~ & ~$^{~}_{~}$ & ~$^{~}_{~}$ & 3.15$^{0.35}_{0.32}$ & 4.65$^{0.32}_{0.76}$ \\ 
     M101 NGC 5471\footnotemark[3] &   1.03 & 4.15\footnotemark[3] & ~ & ~$^{~}_{~}$ & ~$^{~}_{~}$ & 2.85$^{0.95}_{~}$ & 4.40$^{~}_{1.76}$ \\ 
              M 42\footnotemark[5] &   1.02 & 3.92\footnotemark[5] & 3.90\footnotemark[5] & 4.00$^{0.07}_{0.02}$ & 2.30$^{0.09}_{0.05}$ & 4.20$^{0.06}_{0.08}$ & 3.55$^{0.11}_{0.14}$ \\ 
        NGC 2867-1\footnotemark[2] &   1.02 & 4.07\footnotemark[2] & 3.95\footnotemark[2] & ~$^{~}_{~}$ & ~$^{~}_{~}$ & 2.60$^{~}_{~}$ & 3.75$^{0.21}_{1.40}$ \\ 
    NGC 5253 HII-2\footnotemark[10] &   1.01 & 4.08\footnotemark[10] & ~ & ~$^{~}_{~}$ & ~$^{~}_{~}$ & 3.50$^{0.19}_{0.24}$ & 3.20$^{0.33}_{0.33}$ \\ 
     NGC 5253 UV-1\footnotemark[10] &   1.01 & 4.04\footnotemark[10] & ~ & ~$^{~}_{~}$ & ~$^{~}_{~}$ & 4.20$^{0.07}_{0.70}$ & 2.85$^{~}_{0.55}$ \\ 
    NGC 5253 HII-1\footnotemark[10] &   1.00 & 4.08\footnotemark[10] & ~ & ~$^{~}_{~}$ & ~$^{~}_{~}$ & 3.90$^{0.17}_{0.67}$ & 3.50$^{0.42}_{0.57}$ \\ 
     NGC 5253 UV-2\footnotemark[10] &   1.00 & 4.04\footnotemark[10] & ~ & ~$^{~}_{~}$ & ~$^{~}_{~}$ & ~$^{~}_{~}$ & ~$^{~}_{~}$ \\ 
       M33 NGC 604\footnotemark[3] &   1.00 & 3.91\footnotemark[3] & ~ & ~$^{~}_{~}$ & ~$^{~}_{~}$ & 3.00$^{0.03}_{0.04}$ & 2.35$^{0.10}_{0.10}$ \\ 
       M33 NGC 604\footnotemark[11] &   1.00 & 3.91\footnotemark[11] & ~ & ~$^{~}_{~}$ & ~$^{~}_{~}$ & 2.60$^{0.60}_{0.20}$ & 2.00$^{0.15}_{0.09}$ \\ 
     M101 NGC 5461\footnotemark[11] &   0.99 & 3.93\footnotemark[11] & ~ & ~$^{~}_{~}$ & ~$^{~}_{~}$ & 3.35$^{0.38}_{0.24}$ & 2.65$^{0.21}_{0.11}$ \\ 
 NGC 2366 NGC 2363\footnotemark[11] &   0.97 & 4.20\footnotemark[11] & ~ & ~$^{~}_{~}$ & ~$^{~}_{~}$ & 3.65$^{0.36}_{0.68}$ & 2.00$^{0.26}_{0.12}$ \\ 
     M101 NGC 5447\footnotemark[11] &   0.97 & 3.93\footnotemark[11] & 3.82\footnotemark[11] & ~$^{~}_{~}$ & ~$^{~}_{~}$ & 3.05$^{0.22}_{0.22}$ & 2.15$^{0.52}_{0.22}$ \\ 
           SMC N66\footnotemark[4] &   0.94 & 4.09\footnotemark[4] & ~ & ~$^{~}_{~}$ & ~$^{~}_{~}$ & 2.60$^{~}_{~}$ & 2.00$^{~}_{~}$ \\ 
          NGC 3576\footnotemark[6] &   0.91 & 3.95\footnotemark[6] & 3.91\footnotemark[6] & 2.10$^{~}_{~}$ & 5.90$^{~}_{~}$ & 3.50$^{0.07}_{0.04}$ & 2.75$^{0.20}_{0.12}$ \\ 
   NGC 4395 Reg 70\footnotemark[11] &   0.77 & 4.03\footnotemark[11] & ~ & ~$^{~}_{~}$ & ~$^{~}_{~}$ & 3.90$^{0.16}_{0.13}$ & 2.45$^{0.17}_{0.22}$ \\ 
    NGC 2403 VS 38\footnotemark[11] &   0.61 & 3.94\footnotemark[11] & ~ & ~$^{~}_{~}$ & ~$^{~}_{~}$ & 4.05$^{0.02}_{0.32}$ & 2.70$^{0.02}_{0.46}$ \\ 
              M 16\footnotemark[8] &   0.45 & 3.88\footnotemark[8] & 3.74\footnotemark[8] & ~$^{~}_{~}$ & ~$^{~}_{~}$ & 3.00$^{0.13}_{0.27}$ & 2.60$^{1.00}_{0.41}$ \\ 
    NGC 2403 VS 24\footnotemark[11] &   0.45 & 3.91\footnotemark[11] & ~ & ~$^{~}_{~}$ & ~$^{~}_{~}$ & 3.70$^{0.30}_{0.09}$ & 2.40$^{0.36}_{0.19}$ \\ 
   NGC 1741 Zone C\footnotemark[11] &   0.38 & 3.93\footnotemark[11] & ~ & ~$^{~}_{~}$ & ~$^{~}_{~}$ & 3.65$^{0.40}_{~}$ & 2.05$^{0.69}_{~}$ \\ 
               M 8\footnotemark[9] &   0.37 & 3.91\footnotemark[9] & 3.85\footnotemark[9] & 2.10$^{~}_{~}$ & 5.90$^{~}_{~}$ & 3.15$^{0.33}_{0.14}$ & 2.50$^{0.12}_{0.26}$ \\ 
        M101 H1013\footnotemark[11] &   0.36 & 3.87\footnotemark[11] & ~ & ~$^{~}_{~}$ & ~$^{~}_{~}$ & 3.55$^{0.44}_{0.37}$ & 2.50$^{0.20}_{0.23}$ \\ 
       M33 NGC 595\footnotemark[11] &   0.34 & 3.87\footnotemark[11] & ~ & ~$^{~}_{~}$ & ~$^{~}_{~}$ & 3.70$^{0.32}_{0.18}$ & 2.55$^{0.34}_{0.28}$ \\ 
              M 20\footnotemark[8] &   0.33 & 3.89\footnotemark[8] & 3.78\footnotemark[8] & ~$^{~}_{~}$ & ~$^{~}_{~}$ & 2.60$^{~}_{~}$ & 2.00$^{~}_{~}$ \\ 
    NGC 2403 VS 44\footnotemark[11] &   0.30 & 3.92\footnotemark[11] & ~ & ~$^{~}_{~}$ & ~$^{~}_{~}$ & 3.55$^{0.40}_{0.29}$ & 2.00$^{0.18}_{0.09}$ \\ 
              M 17\footnotemark[9] &   0.27 & 3.91\footnotemark[9] & ~ & 4.30$^{~}_{~}$ & 2.00$^{~}_{~}$ & 3.60$^{0.42}_{0.17}$ & 2.30$^{0.52}_{0.27}$ \\ 
NGC 4861 BrightHII\footnotemark[11] &   0.27 & 4.11\footnotemark[11] & ~ & ~$^{~}_{~}$ & ~$^{~}_{~}$ & 4.20$^{~}_{0.59}$ & 2.90$^{~}_{0.64}$ \\ 
          M31 K932\footnotemark[11] &   0.24 & 3.97\footnotemark[11] & ~ & ~$^{~}_{~}$ & ~$^{~}_{~}$ & 3.75$^{0.36}_{0.66}$ & 2.00$^{0.36}_{0.13}$ \\ 
\hline
\end{tabular}
\\
\begin{description}
\item [$^1$] \citet{TBL03b}; 
$^{2}$ \citet{GPP08}; 
$^{3}$ \citet{EPT02}; 
$^{4}$ \citet{TBL03a}; 
$^{5}$ \citet{EPG04}; 
\item [$^{6}$] \citet{GEP04}; 
$^{7}$ \citet{GEP05}; 
$^{8}$ \citet{GPC06}; 
$^{9}$ \citet{GEP07}; 
\item [$^{10}$] \citet{LEG07}; 
$^{11}$ \citet{EBP09}. 
\end{description}

\end{table*}

\clearpage
\setcounter{table}{2}
\begin{table*}
\begin{minipage}{95mm}
\caption[]{The He~{\sc i}~diagnostics from literature and current analysis for PNe.}
\begin{tabular}{lrccc}
\hline
& ~ & \multicolumn{3}{c}{\tel(He~{\sc i})~[K]} \\
\cline{3-5}
\multicolumn{1}{l}{Object} & ADF(O$^{2+}$) &  \multicolumn{1}{c}{\underline{$\lambda$\rm{7281}}}  & \multicolumn{1}{c}{\underline{$\lambda$\rm{6678}}}  & \multicolumn{1}{c}{\underline{$\lambda$\rm{6678}}}  \\
& ~ &  \multicolumn{1}{c}{$\lambda$\rm{6678}}  & \multicolumn{1}{c}{$\lambda$\rm{4471}}  & \multicolumn{1}{c}{$\lambda$\rm{5876}}  \\
\hline
       Abell 30 J3\footnotemark[20] & 766.0 &     ~ &    9080 &    8230  \\
       Abell 30 J1\footnotemark[20] & 598.0 &     ~ &    5830 &    3910  \\
          Abell 58\footnotemark[31] &  89.0 &     ~ &        587 &       1250  \\
        Hf 2-2 (2 arcsec)\footnotemark[15] &  84.0 &     ~   &  1070 &     1570  \\
        Hf 2-2 (4 arcsec)\footnotemark[15] &  84.0 &     ~ &     ~ &      840  \\
        Hf 2-2 (8 arcsec)\footnotemark[15] &  84.0 &     ~ &     816 &     1460  \\
          NGC 1501\footnotemark[6] &  32.0 &    4800\footnotemark[6] & 5100\footnotemark[6]  &  4100\footnotemark[6]   \\
            M 1-42\footnotemark[2] &  22.0 &      3790 &       2500 &       2380  \\
            NGC 40\footnotemark[7] &  17.3 &     ~ &       7430 &   5470  \\
            M 2-24\footnotemark[5] &  17.0 &     ~ &     ~ &    4500  \\
          NGC 2022\footnotemark[4] &  16.0 &     ~ &      19700 &       6980  \\
          NGC 2022\footnotemark[10] &  16.0 &     ~ & 15900\footnotemark[10] &     ~  \\
            DdDm 1\footnotemark[13] &  11.8 &     ~ &     ~ &  3500\footnotemark[13]   \\
          NGC 6153\footnotemark[1] &   9.2 &    3260 &      10600 &       5570   \\
           IC 2003\footnotemark[13] &   7.3 &     ~ &  1600\footnotemark[13]  &  7670\footnotemark[13]   \\
            M 2-36\footnotemark[2] &   6.9 &       2710 &       3420 &      4370  \\
            Vy 1-2\footnotemark[13] &   6.2 &     ~ &  3550\footnotemark[13] &  4430\footnotemark[13]   \\
          NGC 3242\footnotemark[4] &   5.7 &       4620 &       9290 &      7120  \\
          NGC 2440\footnotemark[16] &   5.4 &      10400 &       7430 &       4500  \\
          NGC 6818\footnotemark[4] &   4.9 &      3420 &       5270 &       5880  \\
          NGC 7009\footnotemark[21] &   4.7 &    5030 &    9670 &  4310 \\
           IC 3568\footnotemark[7] &   4.6 &     ~ &     15900 &     10900 \\
          NGC 6210\footnotemark[7] &   4.3 &     ~ &       9470 &      10600  \\
            M 3-34\footnotemark[13] &   4.2 &     ~ & 17000\footnotemark[13]  &     ~  \\
          NGC 6543\footnotemark[11] &   4.2 &       6610 &    8440 &    5560 \\
            M 1-73\footnotemark[13] &   3.6 &     ~ &  8820\footnotemark[13]  &  7960\footnotemark[13]  \\
          NGC 6302\footnotemark[10] &   3.6 &     ~ & 15100\footnotemark[10]  &     ~ \\
          NGC 3132\footnotemark[4] &   3.5 &     10800 &      14500 &     10800  \\
          NGC 6302\footnotemark[4] &   3.5 &     ~ &     ~ &      5010  \\
          NGC 7026\footnotemark[13] &   3.4 &     ~ &  4050\footnotemark[13]  &  4120\footnotemark[13] \\
            IC 351\footnotemark[13] &   3.1 &     ~ &  3790\footnotemark[13]  &  4600\footnotemark[13]  \\
            Hu 1-1\footnotemark[13] &   3.0 &     ~ &  9550\footnotemark[13]  &  4740\footnotemark[13]  \\
            Sp 4-1\footnotemark[13] &   2.9 &     ~ &  3150\footnotemark[13]  &  2400\footnotemark[13]  \\
           IC 4846\footnotemark[13] &   2.9 &     ~ & 13400\footnotemark[13]  &     ~  \\
           IC 4846\footnotemark[17] &   2.9 &    8890 &    5880 &  2490  \\
          NGC 6803\footnotemark[13] &   2.7 &     ~ &  8100\footnotemark[13]  &  4840\footnotemark[13]  \\
            BoBn 1\footnotemark[19] &   2.6 &  9430\footnotemark[19]  &    5580 &      1590 \\
          NGC 6833\footnotemark[13] &   2.5 &     ~ & 14100\footnotemark[13]  &  2440\footnotemark[13] \\
          NGC 6879\footnotemark[13] &   2.5 &     ~ &  3750\footnotemark[13]  &  2340\footnotemark[13]  \\
     IC 4191 fixed\footnotemark[4] &   2.4 &       5850 &       3080 &       2050  \\
  IC 4191 scan\footnotemark[4] &   2.4 &       8340 &       2890 &       1920  \\
          NGC 3132\footnotemark[10] &   2.4 &     ~ & 13900\footnotemark[10]  &    ~  \\
          NGC 6720\footnotemark[7] &   2.4 &     ~ &      13300 &      10300  \\
          NGC 6818\footnotemark[10] &   2.4 &     ~ &  5000\footnotemark[10]  &    ~  \\
    IC 4191 neb\footnotemark[10] &   2.4 &     ~ &  3000\footnotemark[10]  &     ~ \\
     IC 4191 fixed\footnotemark[10] &   2.4 &     ~ &  2800\footnotemark[10]  &     ~  \\
          NGC 3918\footnotemark[4] &   2.3 &       7320 &      15700 &       7730  \\
          NGC 6884\footnotemark[7] &   2.3 &     ~ &     ~ &       8990 \\
           IC 5217\footnotemark[13] &   2.3 &     ~ &  5100\footnotemark[13]  &  3000\footnotemark[13]   \\
          NGC 3242\footnotemark[10] &   2.2 &     ~ & 10000\footnotemark[10]  &     ~  \\
          NGC 5882\footnotemark[4] &   2.2 &       5380 &      10300 &       6840  \\
            M 1-74\footnotemark[13] &   2.1 &     ~ &  9200\footnotemark[13]  &  3380\footnotemark[13]   \\
\hline
\end{tabular}
\end{minipage}
\setcounter{table}{2}
\begin{minipage}{80mm}
\caption[]{(Continued)}
\begin{tabular}{lrccc}
\hline
& ~ & \multicolumn{3}{c}{\tel(He~{\sc i})~[K]} \\
\cline{3-5}
\multicolumn{1}{l}{Object} & ADF(O$^{2+}$) &  \multicolumn{1}{c}{\underline{$\lambda$\rm{7281}}}  & \multicolumn{1}{c}{\underline{$\lambda$\rm{6678}}}  & \multicolumn{1}{c}{\underline{$\lambda$\rm{6678}}}  \\
& ~ &  \multicolumn{1}{c}{$\lambda$\rm{6678}}  & \multicolumn{1}{c}{$\lambda$\rm{4471}}  & \multicolumn{1}{c}{$\lambda$\rm{5876}}  \\
\hline
          NGC 5882\footnotemark[10] &   2.1 &     ~ & 10700\footnotemark[10]  &     ~  \\
          NGC 7662\footnotemark[7] &   2.0 &     ~ &       4830 &       3550  \\
          NGC 5315\footnotemark[10] &   2.0 &     ~ & 10000\footnotemark[10]  &     ~  \\
          NGC 6807\footnotemark[13] &   2.0 &     ~ &     ~ &  1000\footnotemark[13]   \\
          NGC 5307\footnotemark[3] &   1.9 &      8050 &       6040 &       4360  \\
           IC 4406\footnotemark[4] &   1.9 &       5260 &      7620 &     4600  \\
           IC 4406\footnotemark[10] &   1.9 &     ~ &  8000\footnotemark[10]  &     ~  \\
          NGC 6741\footnotemark[7] &   1.9 &     ~ &     19400 &      6720  \\
          NGC 6826\footnotemark[7] &   1.9 &     ~ &      8510 &    8310 \\
          My Cn 18\footnotemark[4] &   1.8 &       4740 &    ~ &      7670  \\
          NGC 3918\footnotemark[10] &   1.8 &    ~ & 12000\footnotemark[10]  &     ~  \\
          NGC 7027\footnotemark[14] &   1.8 & 10500\footnotemark[14]  &    8260 &      1970  \\
          NGC 5315\footnotemark[8] &   1.7 &     7910 &     7180 &       4350  \\
          NGC 6790\footnotemark[7] &   1.7 &     ~ &     ~ &     10600  \\
            Hu 1-2\footnotemark[7] &   1.6 &     ~ &     ~ &      12800 \\
          NGC 6572\footnotemark[7] &   1.6 &     ~ &     ~ &      7420  \\
          NGC 6891\footnotemark[13] &   1.5 &     ~ &  4100\footnotemark[13]  &  3660\footnotemark[13]   \\
          NGC 5315\footnotemark[4] &   1.4 &       6090 &      9220 &      5580  \\
            M 3-32\footnotemark[17] &   1.1 &  1710\footnotemark[17]   &  3620 &    2970  \\
            M 3-33\footnotemark[17] &   1.1 &    4960 &    4190 &    2970  \\
           IC 4699\footnotemark[17] &   1.1 &    2290 &    6540 &   5170  \\
          NGC 6439\footnotemark[17] &   1.1 &    5180 &    2270 &    2010  \\
            H 1-41\footnotemark[17] &   1.1 &    2700 &    6310 &    5100  \\
             M 3-7\footnotemark[17] &   1.1 &    2340 &    7390 &   2610  \\
          NGC 6620\footnotemark[17] &   1.1 &   3720 &    2570 &    2490 \\
            H 1-50\footnotemark[17] &   1.1 &    7990 &    8230 &    3140  \\
            H 1-54\footnotemark[17] &   1.1 &    6650 &  18800 &  5210 \\
            M 1-29\footnotemark[17] &   1.1 &    2940 &    5430 &       2750  \\
            M 3-21\footnotemark[17] &   1.1 &    6430  &  3740 &   2570  \\
            H 1-35\footnotemark[17] &   1.0 &   11000 &       3870 &     1830  \\
            H 1-42\footnotemark[17] &   1.0 &    7310 &   3130 &   1990  \\
             M 2-6\footnotemark[17] &   1.0 &    7870 &   4100 &    2870  \\
            M 2-27\footnotemark[17] &   1.0 &    2760 &    7630 &       3990  \\
            M 2-33\footnotemark[17] &   1.0 &    4530 &    7740 &   4880  \\
            M 2-42\footnotemark[17] &   1.0 &    3370 &   7240 &   5990  \\
            M 3-29\footnotemark[17] &   1.0 &  1800\footnotemark[17]  &   5990 &    4450 \\
          NGC 6567\footnotemark[17] &   1.0 &   10400 &    7100 &    2590  \\
          He 2-118\footnotemark[17] &   1.0 &   10600 &  15600 &   4360  \\
            M 1-61\footnotemark[17] &   1.0 &    6070 & 18500\footnotemark[17]  &     13200  \\
             M 2-4\footnotemark[17] &   1.0 &    8400 &      4060 &   4160 \\
          NGC 6565\footnotemark[17] &   1.0 &   4770 &    6470 &  5470  \\
            Vy 2-1\footnotemark[17] &   1.0 &    5000 &    5140 &    4330  \\
            IC 418\footnotemark[9] &   1.0 &     ~ &       4440 &       4040  \\
            Cn 1-5\footnotemark[17] &   1.0 &    2740 &  11600 &       7360  \\
            Cn 2-1\footnotemark[17] &   1.0 &    6500 &    5180 &       2210  \\
            M 1-20\footnotemark[17] &   1.0 &    7800 &       3850 &       1800  \\
            M 2-23\footnotemark[17] &   1.0 &   10300 &    4860 &    1830  \\
           IC 2501\footnotemark[16] &   1.0 &       9130 &       8400 &      4130  \\
            M 2-39\footnotemark[17] &   0.9 &   7790 &   6090 &    5600  \\
            Cn 3-1\footnotemark[13] &   0.0 &     ~ &  4700\footnotemark[13]  &  3400\footnotemark[13]   \\
            M 2-31\footnotemark[17] &   0.0 &    4390 &    2170 &    1770  \\
\hline
\end{tabular}
\\ \\
\end{minipage}
\begin{description}
\item [$^1$] \citet{LSB00};
$^2$ \citet{LLB01};
$^3$ \citet{RPP03};
$^4$ \citet{TBL03b};
$^5$ \citet{ZL03};
\item [$^6$] \citet{EWZ04};
$^7$ \citet{LLB04};
$^8$ \citet{PPR04};
$^9$ \citet{SBW04};
$^{10}$ \citet{TBL04};
\item [$^{11}$] \citet{WL04};
$^{12}$ \citet{RG05};
$^{13}$ \citet{WLB05};
$^{14}$ \citet{ZLL05a};
\item [$^{15}$] \citet{LBZ06};
$^{16}$ \citet{S07};
$^{17}$ \citet{WL07};
$^{18}$ \citet{GPP08};
$^{19}$ \citet{OTH10};
\item [$^{20}$] \citet{WLB03};
$^{21}$ \citet{FL11};
$^{22}$ \citet{EPT02};
$^{23}$ \citet{TBL03a};
$^{24}$ \citet{EPG04};
\item [$^{25}$] \citet{GEP04};
$^{26}$ \citet{GEP05};
$^{27}$ \citet{GPC06};
$^{28}$ \citet{GEP07};
\item [$^{29}$] \citet{LEG07};
$^{30}$ \citet{EBP09};
$^{31}$ \citet{WBL08};
$^{32}$ \citet{TWP08}.
\end{description}
\end{table*}

\clearpage
\setcounter{table}{3}
\begin{table*}
\caption[]{Electron temperatures of H~{\sc ii}~regions derived from the
He~{\sc i}~recombination lines.}
\begin{tabular}{lrccc}
\hline
& ~ & \multicolumn{3}{c}{\tel(He~{\sc i})~[K]} \\
\cline{3-5}
\multicolumn{1}{l}{Object} & ADF(O$^{2+}$) &  \multicolumn{1}{c}{\underline{$\lambda$\rm{7281}}}  & \multicolumn{1}{c}{\underline{$\lambda$\rm{6678}}}  & \multicolumn{1}{c}{\underline{$\lambda$\rm{6678}}}  \\
& ~ &  \multicolumn{1}{c}{$\lambda$\rm{6678}}  & \multicolumn{1}{c}{$\lambda$\rm{4471}}  & \multicolumn{1}{c}{$\lambda$\rm{5876}}  \\
\hline
              M 17\footnotemark[3] &   2.7 &     19000 &       7940 &     5050  \\
          LMC N141\footnotemark[1] &   2.6 &      14900 &       6570 &     ~  \\
          NGC 3576\footnotemark[3] &   2.2 &     ~ &      10500 &       6500  \\
        30 Doradus\footnotemark[3] &   1.8 &     12400 &       5350 &       5490  \\
           LMC N66\footnotemark[1] &   1.6 &     ~ &       3630 &     ~  \\
         LMC N11 B\footnotemark[3] &   1.5 &     ~ &       4740 &      1930  \\
            PB 6-2\footnotemark[2] &   1.1 &     ~ &      15200 &      15900  \\
            PB 6-1\footnotemark[2] &   1.1 &    ~ &     ~ &      15400  \\
              PB 8\footnotemark[2] &   1.1 &      6760 &      6940 &      11100  \\
          NGC 3603\footnotemark[7] &   1.0 &      11500 &      13900 &    8800  \\
        NGC 2867-2\footnotemark[2] &   1.0 &      9810 &       4750 &       3760  \\
             S 311\footnotemark[6] &   1.0 &       9190 &      7150 &      7480  \\
              M 42\footnotemark[4] &   1.0 &       6700 &     4860 &     2880  \\
        NGC 2867-1\footnotemark[2] &   1.0 &      11500 &       6550 &       6640  \\
    NGC 5253 HII-2\footnotemark[9] &   1.0 &       8440 &       6150 &       7440  \\
     NGC 5253 UV-1\footnotemark[9] &   1.1 &       6230 &       7990 &      5220  \\
    NGC 5253 HII-1\footnotemark[9] &   1.0 &     ~ &       6720 &       4320  \\
     NGC 5253 UV-2\footnotemark[9] &   1.0 &     ~ &      5210 &      2820  \\
       M33 NGC 604\footnotemark[10] &   1.0 &     ~ &   ~ &     17400  \\
     M101 NGC 5461\footnotemark[10] &   1.0 &       5320 &    ~ &     ~  \\
     M101 NGC 5447\footnotemark[10] &   1.0 &     ~ &      10400 &      4770  \\
           SMC N66\footnotemark[3] &   0.9 &     ~ &      12600 &    ~  \\
          NGC 3576\footnotemark[5] &   0.9 &      10300 &     ~ &     ~  \\
              M 16\footnotemark[7] &   0.4 &     8380 &     6350 &      5410  \\
               M 8\footnotemark[8] &   0.4 &       7380 &     6240 &     6790  \\
        M101 H1013\footnotemark[10] &   0.4 &       4330 &    ~ &     ~  \\
       M33 NGC 595\footnotemark[10] &   0.3 &      2110 &       6560 &     6280  \\
              M 20\footnotemark[7] &   0.3 &       8870 &      5660 &      6180  \\
              M 17\footnotemark[8] &   0.3 &      6120 &      5590 &       6690  \\
\hline
\end{tabular}
\\
\begin{description}
\item [$^1$] \citet{TBL03b}; 
$^{2}$ \citet{GPP08}; 
$^{3}$ \citet{TBL03a}; 
$^{4}$ \citet{EPG04}; 
$^{5}$ \citet{GEP04}; 
\item [$^{6}$] \citet{GEP05}; 
$^{7}$ \citet{GPC06}; 
$^{8}$ \citet{GEP07}; 
$^{9}$ \citet{LEG07}; 
\item [$^{10}$] \citet{EBP09}; 
\end{description}
\end{table*}

\label{lastpage}

\end{document}